\documentclass[11pt,a4paper]{article}
\usepackage[utf8]{inputenc}

\usepackage{jheppub} 

\usepackage[T1]{fontenc} 
\usepackage{CJK}
\usepackage{multirow}
\usepackage{longtable}
\usepackage{comment}
\usepackage{tikz}

\usepackage{amsmath,amsfonts,amssymb,amsthm}
\usepackage{hyperref}
\hypersetup{
    pdfstartview=FitH,
    bookmarksnumbered=true,
    colorlinks,
    breaklinks,
    urlcolor=blue,
    linkcolor=blue,
    citecolor=blue
}
\usepackage{graphicx}
\DeclareGraphicsExtensions{.pdf,.png,.jpg,.mps}
\usepackage{booktabs}
\usetikzlibrary{calc,matrix,decorations.pathmorphing,decorations.markings,arrows,positioning,intersections,mindmap,backgrounds,shapes.misc}
\usepackage{cancel}

\definecolor{w0}{RGB}{230,228,181}   
\definecolor{w1}{RGB}{190,206,147}   
\definecolor{w2}{RGB}{ 78,128,108}   
\definecolor{block}{RGB}{122,  0,  0} 

\usepackage{amsmath,amsthm,amsfonts,amssymb,amscd,mathtools
}
\usepackage{pifont}

\preprint{ USTC-ICTS/PCFT-25-18}
\title{\boldmath Differential Equations for Energy Correlators in Any Angle}

\author[a,b]{Rourou Ma,}
\author[c,d]{Jianyu Gong,}
\author[c,d]{Jingwen Lin,}
\author[c,d]{Kai Yan,}
\author[e,f,g]{Gang Yang,}
\author[a,h,i]{Yang Zhang}

\affiliation[a]{Interdisciplinary Center for Theoretical Study, University of Science and Technology of China, Hefei,
Anhui 230026, China}
\affiliation[b]{Max-Planck-Institut f\"ur Physik,  Werner-Heisenberg-Institut, Boltzmannstraße 8, 85748 Garching, Germany}
\affiliation[c]{State Key Laboratory of Dark Matter Physics, Shanghai Key Laboratory for Particle Physics and Cosmology,
Key Laboratory for Particle Astrophysics and Cosmology (MOE),
School of Physics and Astronomy, Shanghai Jiao Tong University, Shanghai 200240, China}
\affiliation[d]{Institute of Nuclear and Particle Physics (INPAC), Shanghai Jiao Tong University, Shanghai 200240, China}
\affiliation[e]{CAS Key Laboratory of Theoretical Physics, Institute of Theoretical Physics,\\
Chinese Academy of Sciences, Beijing 100190, China}
\affiliation[f]{School of Fundamental Physics and Mathematical Sciences, Hangzhou Institute for Advanced Study,
UCAS, Hangzhou 310024, China}
\affiliation[g]{International Centre for Theoretical Physics Asia-Pacific, Beijing/Hangzhou, China}
\affiliation[h]{Peng Huanwu Center for Fundamental Theory, Hefei, Anhui 230026, China}
\affiliation[i]{Center for High Energy Physics, Peking University, Beijing 100871, People’s Republic of China}

\emailAdd{marr21@mail.ustc.edu.cn}
\emailAdd{jianyu\_gong@sjtu.edu.cn}
\emailAdd{linjingwen@sjtu.edu.cn}
\emailAdd{yan.kai@sjtu.edu.cn}
\emailAdd{yangg@itp.ac.cn}
\emailAdd{yzhphy@ustc.edu.cn}

\abstract{Energy Correlators (EC) are the simplest IR finite observables, which connect theories and experiments. In this paper, we provide a systematic algorithm to calculate the canonical differential equations for energy correlators at generic angle in $\mathcal{N}=4$ super Yang-Mills theory. The integrand is obtained from the 5-point form factor square for scalar half-BPS operators. Applying the algorithm, we obtain the canonical basis for three-point EC and the full set of master integrals for four-point EC. We analyze the function space for four-point case. For multiple polylogrithmic (MPLs) integrals, we calculate their symbols, and for integrals beyond MPLs, we make further investigation by Picard-Fuchs operators. We find two elliptic curves and one genus 2 hyperelliptic curve. The results are achieved by means of integration by part (IBP) reduction and differential equations powered by computational algebraic geometry methods. We provide a package that implements the algorithm. The data is a valuable reference for exploring the structure of physical observables in perturbation theories.}

\begin{document} 
\maketitle
\flushbottom

\section{Introduction}
Physical observables are the connections between theories and experiments. Therefore, explicit calculations of physical observables in Quantum Field Theories (QFTs) are crucial, since they both provide theoretical predictions and reveal the hidden structures in field theories. Observables are constructed by scattering amplitudes. From the perturbative scattering amplitudes level, a lot of simplicity has been studied. However, observables or cross-section level attract less attention. An interesting and simplest class of observables are \textit{Energy Correlation Functions} or \textit{Energy Correlators} \cite{Basham:1977iq,Basham:1978bw,Basham:1978zq,Basham:1979gh}.

Energy correlators are the observables that measure the distribution and the correlation of energy flow in a particular direction. Unlike scattering amplitudes, the energy correlators are infrared finite objects \cite{Lee:1964is,Hofman:2008ar}, and this might mean that the energy correlators are the simplest observables to calculate analytically. In the experiment, they are reflected by deploying multiple detectors separated in specific angles. From a phenomenological aspect, they can be used as event shape or jet structures to further study properties of Quantum Chromodynamics (QCD) or even new physics. 

The n-point Energy Correlator ($\mathrm{E^n C}$) is defined by a weighted cross-section 
\begin{equation}
	\frac{d\sigma}{dx_{12}\cdots dx_{(n-1)n}} \equiv \sum_{m}\sum_{1\leq i_1,\cdots i_n \leq m}\int d\sigma_{m}\times  \prod_{1\leq k \leq n}\frac{E_{i_k}}{q^0} \prod_{1\leq j < l \leq n}\delta\left(x_{jl}-\frac{1-\cos\theta_{i_{j}i_{l}}}{2}\right)\,,
\end{equation}
where $m$ is the number of final-state particles, $d\sigma_m\equiv d\text
{PS}_m |\mathcal{M}|^2$ is the differential cross-section, $E_{i_k}$ and $q^0$ are the energy of final-state parton $i_k$ and total energy separately in the center-of-mass frame. $\theta_{i_ji_l}$ indicates the angle separation of partons $i_j$ and $i_l$.

\begin{figure}[h]
\centering
\begin{tikzpicture}[scale=2.5, line cap=round, line join=round, 
                    >=latex, 
                    font=\small] 
    \draw[thick] (0,0) circle(0.8);

    \begin{scope}[rotate=322]
   \fill[yellow!40, opacity=0.4] (0,0) ellipse (0.8cm and 0.25cm);
   \draw (0.8,0) arc [start angle=0, end angle=-180, x radius=0.8, y radius=0.25];
   \draw[dashed] (0.8,0) arc [start angle=0, end angle=180, x radius=0.8, y radius=0.25];
   \end{scope}
    
    \draw[->, thick] (0,0) -- (0.788,-0.139,0) node[right] {$\varepsilon(\vec{n}_4)$};
    \draw[->, thick] (0,0) -- (-0.139, 0.788, 0) node[above]{$\varepsilon(\vec{n}_1)$};
    \draw[->, thick] (0,0) -- (0.533, 0.205, 0) node[above]{$\varepsilon(\vec{n}_3)$};
    \draw[->, dashed,thick] (0,0) -- (0.185, 0.533, 0) node[above]{$\varepsilon(\vec{n}_2)$};
    
    \filldraw[fill=brown] (0,0,0) circle (1.0 pt); 

    \draw[dashed, thick] (0,0,0) -- (0.514,0.6128,0) node[above right]{$O$} ;
    \draw[dashed, thick] (0,0,0) -- (-0.514,-0.6128,0) node[below left]{$\infty$};
\end{tikzpicture}
\caption{Four-point energy correlators in any angle scattering}
\label{fig:enter-label}
\end{figure}
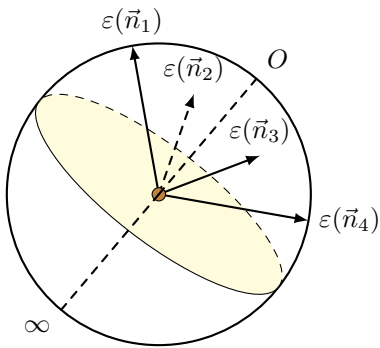

The $\mathrm{E^n C}$ can also be defined in terms of correlation functions of Average Null Energy Condition (ANEC) operators \cite{Sveshnikov:1995vi,Korchemsky:1999kt,Lee:2006nr,Hofman:2008ar,Belitsky:2013bja,Belitsky:2013xxa,Belitsky:2013ofa}, also called energy flux operators defined by
\begin{equation}
     \mathcal{E}=\int_{0}^{\infty}\mathrm{dt}\lim_{r\rightarrow\infty}r^2n^iT_{0i}(t,r\vec{n}).
\end{equation}
In the definition of the ANEC operator, the limit is taken with the retarded time $u=t-r$ fixed, and integration over the advanced time $v=t+r$. In terms of these operators, energy correlators can be expressed as

\begin{equation}\label{eq:ENCdef}
    \mathrm{E^n C}(\theta_{ij})=\int\prod_{i=1}^n \mathrm{d}\Omega_{\vec{n}_i}\prod_{i\neq j}\delta(\vec{n}_i \cdot \vec{n}_j-\cos\theta_{ij})\frac{\int d^4x e^{iqx}\langle0|\mathcal{O}^+(x)\mathcal{E}(\vec{n}_1) \cdots \mathcal{E}(\vec{n}_n)\mathcal{O}(0)|0\rangle}{(q^0)^n \int d^4x e^{iqx}\langle0|\mathcal{O}^+(x)\mathcal{O}(0)|0\rangle},
\end{equation}
where the integrations are over the solid angle and $\theta_{ij}$ is the angle separation of partons $i$ and $j$. The result should only depend on the angle between partons. Recall that $\theta_{ij}$s are not independent when $N\geq4$, and we will reparametrize them before concrete calculation. In the expression, the source operator $O$ and sink operator $O^\dagger$ create the initial state and the final state separately. These $O$ operators are chosen by the process to study.

Although people have calculated $\mathrm{E^n C}$ in strong coupling in $\mathcal{N}=4\,\, \mathrm{sYM}$ \cite{Hofman:2008ar}, little is known about the weak coupling aspect. In recent years, much progress has been made in calculating two-point, three-point energy correlators both in $\mathcal{N}=4\,\, \mathrm{sYM}$ and QCD even higher order \cite{Dixon:2018qgp,Luo:2019nig,Belitsky:2013ofa,Henn:2019gkr,Yan:2022cye} and their resummation \cite{chen2023nnllresummationprojectedthreepoint}, and four-point $\mathcal{N}=4$ energy correlators in collinear limits \cite{Chicherin:2024ifn}. However, differential equations \cite{Kotikov:1990kg} of energy correlators in any angle scattering have not been studied yet, even though differential equations are the main approach in studying integrals of observables like scattering amplitudes.  

Energy correlators are the integrals among phase space; more points, more complicated.
Even at leading order, sometimes, the difficulty of computation increases more dramatically of the energy correlators when the point increases than what happened to Feynman integrals when the loop increases. We will see that many new functions appear in the energy correlators for the 4-point case compared to the 3-point case. For traditional Feynman integrals, the most widely used analytical calculation method is the canonical differential equation \cite{Henn_2013} based on Integral by Parts (IBP) \cite{CHETYRKIN1981159}. People developed some advanced techniques for IBP based on algebraic geometry \cite{Gert_2008}, such as syzygy \cite{Gluza_2011} and lift \cite{Larsen_2018}.  Recently, thanks to the syzygy method, we have an IBP reduction package \textsc{NeatIBP} \cite{Wu:2025aeg} to generate a small size IBP system for traditional Feynman integrals. Until now, there has been no systematic method for generating an IBP system for high-point energy correlator integrals. This would also be a question of whether it is suitable to apply differential equations for energy correlators for higher points.

We present a new method for calculating the leading-order energy correlators in $\mathcal{N}=4$ Yang-Mills theory. They are infrared finite observables, and can be constructed from finite integrals. We aim to establish an integration by parts system and differential equations that exclude as many divergent integrals as possible. To avoid divergent integrals, from an algebra geometry point of view, with the help of a computer algebra system \textsc{Singular} \cite{DGPS}, we apply the \textit{syzygy} method to generate IBP systems  \cite{Henn:2022vqp,Wu:2023upw}, and the \textit{lift} method for differential equations. For these kinds of typical Feynman integrals, its canonical differential equation matrix is upper block triangular. Our new method can work smoothly for the three-point energy correlator. Even though there are elliptic integrals and hyperelliptic integrals for four-point energy correlators in arbitrary angle shape, our method can work well for the unitary transcendental sector. We are going to check how many integrals are elliptic and if there are more complex curves appearing in four-point energy correlators.

In this work, we use differential equation methods to obtain the master integrals of leading order $\mathrm{E^3C}$ and $\mathrm{E^4C}$ in $\mathcal{N}=4$ sYM. We will give the canonical differential equations of $\mathrm{E^3C}$ and the master integrals of $\mathrm{E^4C}$. We also introduce cuts and Picard-Fuchs operators \cite{Muller-Stach:2012tgj,Adams:2017tga,Dlapa:2020cwj,Dlapa:2022wdu} to classify these master integrals into Multiple Polylogarithms (MPL), elliptic integrals \cite{Bosma:2017ens,Primo:2016ebd,Primo:2017ipr,Henn:2020lye,Bourjaily:2021vyj,Frellesvig:2021hkr,Broedel:2017kkb,Broedel:2018qkq,Bourjaily:2017bsb,Broedel:2019hyg} and even hyperelliptic integrals \cite{Huang:2013kh,Georgoudis:2015hca}. For the MPL master integrals, we would like to give the preliminary symbols \cite{Duhr:2011zq,Duhr:2019tlz,Goncharov:2010jf} with the help of calculation packages like SOFIA \cite{Correia:2025yao} and \textsc{HyperInt} \cite{Panzer:2014caa}.

\subsection{Parametrization of Energy Correlators}

Before we move into the details of the algorithm, we first introduce the parametrization we use in the calculation. In $\mathcal{N}=4$ sYM, one can choose the source and sink operators as scalars that are the bottom component of the supermultiplet of conserved currents. The matrix element of producing a given super-state from vacuum is called form factor \cite{Bork:2010wf,Bork:2011cj,Penante:2014sza,Brandhuber:2011tv,vanNeerven:1985ja}
\begin{equation}
    \int \mathrm{d}^4x\,e^{iqx}  \langle X|\mathcal{O}(x)|0\rangle\equiv(2\pi)^4 \delta^4(q-q_X)\mathcal{F}_X.
\end{equation}
One can always insert unit operators in \eqref{eq:ENCdef} and $\mathrm{E^n C}$ can be obtained by performing a weighted sum over the onshell external states
\begin{equation}
    \mathrm{E^n C}=\frac{1}{\sigma_{\mathrm{tot}}}\sum_{(n_1,\cdots,n_n)\in X} \int \mathrm{d}\Pi_X \left(\prod_{i=1}^n\delta^2(\vec{n}_i-\hat{p}_{n_i})\frac{E_i}{q^0}\right)\big| \mathcal{F}_X\big|^2,
\end{equation}
where $\mathrm{d}\Pi_X$ is the on-shell phase-space of the final state.

The integrand of energy correlators depends on two types of variables. One is the energy fractions of the final particles and the other is the angles between every two final particles. We define energy fraction of a final state particle $x_i$ and angle parameter between two as following
\begin{align}
   &\text{energy fraction: } x_i=\frac{2q\cdot p_i}{q^2} \qquad i=1,\cdots ,n \,;\label{energyPara}\\
   &\text{angle parameter: } \zeta_{ij}=\frac{q^2 (p_i\cdot p_j)}{2(q\cdot p_i)(q\cdot p_j)} \qquad i, j=1,\cdots,n\,\label{anglePara};
\end{align}
where $q$ and $p$ are separately the momentum of incoming and outgoing particles. For massless particles in the final states with momentum $p_i$, on-shell conditions correspond to $\zeta_{ii}=0$. Leading order energy correlators for any angle scattering will only depend on angle parameters after integration over energy fraction
\begin{equation}\label{E^nC}
    \mathrm{E^n C}(\zeta_{ij})={\frac{1}{ \sigma_{\mathrm{tot}}}}\int_{0}^{1}\mathrm{d}x_1\cdots\mathrm{d}x_n\,(x_1\cdots x_n)^2\,\delta(1-Q_n) \big|\mathcal{F}_{n+1}^{(0)}\big|^2,
\end{equation}
where the $\delta$-function indicates momentum conservation
\begin{equation}
    \delta(1-Q_n)=\delta\left(1-\sum_i x_i+\sum_{1\leq i<j\leq n}\zeta_{ij}\,x_i\,x_j\right).
\end{equation}

In this work, we focus on three- and four-point energy correlators. For the three-point energy correlator, there are three $\zeta_{ij}$ s and are all independent. In order to calculate its canonical differential equations, we introduce two different parameterizations for different families:
\begin{align}
    &\text{Para. 1:}\qquad \zeta_{12}=\frac{z_2^2}{1+z_2^2},\,\zeta_{23}=\frac{(z_2-z_3)(z_2-\bar{z_3})}{(1+z_2^2)(1+z_3\bar{z_3})},\,\zeta_{13}=\frac{z_3\bar{z_3}}{1+z_3\bar{z_3}};
    \label{para 1}\\
    &\text{Para. 2:}\qquad \zeta_{12}=-\frac{s(1-x_2)^2}{(1+s)^2x_2},\,\zeta_{23}=-\frac{s(1-x_1x_2)^2}{(1+s)^2x_1x_2},\,\zeta_{13}=-\frac{s(1-x_1)^2}{(1+s)^2x_1}. \label{para 2}
\end{align}

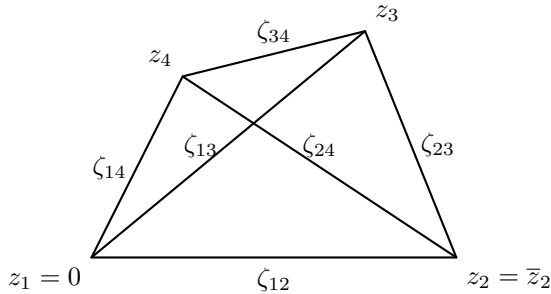
\begin{figure}[h]
\begin{center}
\begin{tikzpicture}[scale=1.2, line cap=round, line join=round, 
                    >=latex, 
                    font=\small] 
  \coordinate (Z1) at (0,0);     
  \coordinate (Z2) at (4,0);     
  \coordinate (Z4) at (1,2);     
  \coordinate (Z3) at (3,2.5);   

  \draw[thick] (Z1) -- (Z2) -- (Z3) -- (Z4) -- cycle;

  \draw[thick] (Z1) -- (Z3);
  \draw[thick] (Z2) -- (Z4);

  \node[below left]  at (Z1) {$z_{1} = 0$};
  \node[below right] at (Z2) {$z_{2} = \overline{z}_{2}$};
  \node[above right] at (Z3) {$z_{3}$};
  \node[above left]  at (Z4) {$z_{4}$};

  \node[below] at ($(Z1)!0.5!(Z2)$) {$\zeta_{12}$};
  \node[right] at ($(Z2)!0.5!(Z3)$) {$\zeta_{23}$};
  \node[above] at ($(Z3)!0.5!(Z4)$) {$\zeta_{34}$};
  \node[left]  at ($(Z4)!0.5!(Z1)$) {$\zeta_{14}$};

  \node[above] at ($(Z1)!0.4!(Z3)$) {$\zeta_{13}$};
  \node[above] at ($(Z2)!0.5!(Z4)$) {$\zeta_{24}$};

\end{tikzpicture}
\end{center}

 \caption{Parametrization of angle parameters for four-point energy correlators}
 \label{fig:E4Cpara}
\end{figure}

For the four-point energy correlators, there are six angle parameters. However, only five of them are independent. Therefore, $\zeta_{ij}$ can be parameterized by complex variable $z_i$ (as shown in Figure \ref{fig:E4Cpara})
\begin{equation}
    \zeta_{ij}=\frac{|z_i-z_j|^2}{(1+|z_i|^2)(1+|z_j|^2)},\quad i,j=1,\cdots,4,
    \label{4pointPara}
\end{equation}
where we set $z_1=0$ and $z_2$ real, which means $\bar{z_2}=z_2$.

\section{Form Factor Square}

%
An important ingredient in computing EEC is the module square of the form factor
\begin{equation}
\sum_{\textrm{physical states}} | \hat{\cal F}_{{\cal O},n} (1,2,...,n)|^2 \,,
\end{equation}
where the sum of physical states include all quantum numbers for the external states including helicities and color factors.
In ${\cal N}=4$ SYM, the chiral stress-tensor supermultiplet is a half-BPS operator with both the scalar primary operator ${\cal O}_{AB} = {\rm tr}(\phi_{AB}^2)$ and chiral Lagrangian ${\cal L}={\rm tr}(F_{\rm SD}^2)+ \cdots$ in the supermultiplet.
For convenience and without loss of generality, we consider the scalar primary ${\cal O}_{AB}$. The super MHV form factor take the compact expression \cite{Brandhuber:2011tv}:
\begin{equation}
{\cal F}_{{\rm tr}(\phi_{AB}^2),n}^{(0),{\rm MHV}}(1,2,...,n) = \frac{\delta^{(4)}(q-\sum_i \lambda_i\tilde\lambda_i) \delta_{AB}^{(4)}(\sum_i \lambda_i \eta_{i})} {\langle 12\rangle \langle 23\rangle \ldots  \langle n 1 \rangle} ,
\end{equation}
where fermionic $\eta$ variables characterize the physical states
\begin{align}\label{formula:superfield}
	\mathrm{\Phi}(p, \eta) = & g_{+}(p)+\eta^{A} \bar\psi_{A}(p)+\frac{1}{2!} \eta^{A} \eta^{B} \phi_{A B}(p) +\frac{1}{3!} \eta^{A} \eta^{B} \eta^{C} \varepsilon_{A B C D} \psi^{D}(p)\notag \\
	&+\frac{1}{4!} \eta^{A} \eta^{B} \eta^{C} \eta^{D} \varepsilon_{A B C D} g_{-}(p) \, .
\end{align}
The non-MHV tree-level form factors can be computed using MHV rules or BCFW recursion methods \cite{Brandhuber:2011tv}.
We denote $\hat{\cal F}$ for a full-color form factor and ${\cal F}$ for a color-ordered form factor. 
The sum of physical states is performed by summing over all MHV and non-MHV configurations and integrating out the $\eta$ variables.

A straightforward strategy to compute the form factor square is following. First, one can compute form factors with all possible helicity state configurations and sum over the square of all components. Next, it is important to rewrite the results in a representation in terms of Mandelstam variables that is convenient for the phase space integration. Since the super form factors are most conveniently computed in spinor-helicity variables, it is usually non-trivial to rewrite the first-step results in Mandelstam variables.

Here we introduce a more convenient method by utilizing the known Sudakov form factor results. This is based on a simple observation that the square of form factor is equivalent to a unitarity cut of the integrand of the two-point function in momentum space
\begin{equation}
\int d^4 x e^{i q\cdot x} \langle {\cal O}(x) {\cal O}(0) \rangle \big|_{n\textrm{-particle cut}} = \int d {\rm PS}_{n\textrm{-particle}} \sum_{\textrm{physical states}}  | \hat{\cal F}_{{\cal O},n} (1,2,...,n)|^2 .
\end{equation}
Furthermore, the integrand of the two-point function can be constructed from the Sudakov form factor by sewing the two external on-shell legs. 

For example, the one-loop Sudakov form factor is a scalar one-mass triangle integral. Sewing the two on-shell legs, we get a two-loop two-point integral as

\begin{figure}[h]
\centering
\begin{tikzpicture}[line width=1.2 pt,  
                    scale=1.2,       
                    >=stealth,       
                    font=\normalsize 
                   ]

  \draw[->] (-0.6,0) -- node[above] {$q$} (-0.2,0);
  \draw (-0.3,0) -- (0,0);

  \draw[->] (0,0) -- (1.2,0.6);
  \draw (1,0.5) -- (1.5,0.75)node[pos=1.0, right=-1 pt] {$1$};
  \draw[->] (0,0) -- (1.2,-0.6);
  \draw (1,-0.5) -- (1.5,-0.75)node[pos=1.0, right=-1 pt] {$2$};
  \draw (1,0.5) -- (1,-0.5);
  
  \node at (2.5,0) {\LARGE $\pmb{\Rightarrow}$};

  \draw[->] (3.9,0) -- node[above] {$q$} (4.3,0);
  \draw (4.2,0) -- (4.5,0);
  \draw[->] (6.5,0) -- node[above] {$q$} (6.9,0);
  \draw (6.8,0) -- (7.1,0);

  \draw (4.5,0) -- (5.5,0.5) -- (6.5,0) -- (5.5,-0.5) -- cycle;
  \draw (5.5,0.5) -- (5.5,-0.5);

\end{tikzpicture}
\caption{One-loop Sudakov form factor and its sewing}
\end{figure}
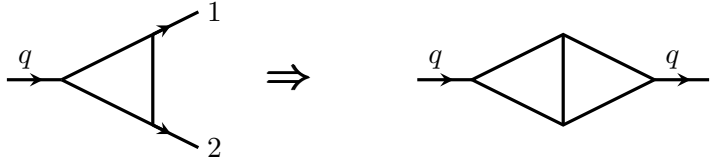

The two-point integral contains $q^2$ in the numerator. Considering the three-particle cuts of the two-point integral, one obtains the square of three-point form factor as
\begin{equation}
\sum_{\textrm{physical states}} | {\cal F}_{{\cal O},n} (1,2,3)|^2 =2 {q^2 \over s_{12} s_{23}} + 2{q^2 \over s_{12} s_{13}} + 2{q^2 \over s_{13} s_{23}} \,,
\end{equation}
which is equivalent to a direct form factor computation
\begin{equation}
\int \prod_{i=1}^3 d^4 \eta_i \, \Big[ {\cal F}_{{\cal O}_{12},3}^{ {\rm MHV}}(1,2,3)\, {\cal F}_{{\cal O}_{34},3}^{{\rm NMHV}}(3,2,1) + {\cal F}_{{\cal O}_{12},3}^{ {\rm NMHV}}(1,2,3) \, {\cal F}_{{\cal O}_{34},3}^{{\rm MHV}}(3,2,1) \Big] = 2{(q^2)^2 \over s_{12} s_{23} s_{31} }  \,.
\end{equation}
Here we consider color-ordered form factor which is sufficient as this order. We will comment more on the color factor in the end of this section.

The above strategy can be generalized to high loops.
From the two-loop Sudakov form factor \cite{vanNeerven:1985ja}, one has two three-loop two-point scalar integrals as

\begin{figure}[h]
\begin{center}
\begin{tikzpicture}[line width=1.2 pt,  
                    scale=1.2,       
                    >=stealth,       
                    font=\normalsize 
                   ]
\begin{scope}
  \draw[->] (-1.6,0) -- node[above] {$q$} (-1.2,0);
  \draw (-1.3,0) -- (-1,0);
  \draw (-1,0) -- (-0.5,0.5) -- (0.5,0.5) -- (1,0) -- (0.5,-0.5) -- (-0.5,-0.5) -- cycle;
  \draw (-0.5,0.5) -- (-0.5,-0.5);
  \draw (0.5,0.5) -- (0.5,-0.5);
  \draw[->] (1,0) -- node[above] {$q$} (1.4,0);
  \draw (1.3,0) -- (1.6,0);
\end{scope}
\begin{scope}[xshift =4 cm]
  \draw[->] (-1.6,0) -- node[above] {$q$} (-1.2,0);
  \draw (-1.3,0) -- (-1,0);
  \draw (-1,0) -- (-0.5,0.5) -- (0.5,0.5) -- (1,0) -- (0.5,-0.5) -- (-0.5,-0.5) -- cycle;
  \draw (-0.5,0.5) -- (0.5,-0.5);
  \draw (0.5,0.5) -- (0.1,0.1);
   \draw (-0.1,-0.1) -- (-0.5,-0.5);
  \draw[->] (1,0) -- node[above] {$q$} (1.4,0);
  \draw (1.3,0) -- (1.6,0);
\end{scope}
\end{tikzpicture}
\end{center}
    \caption{Three-loop two point scalar integrals}
\end{figure}
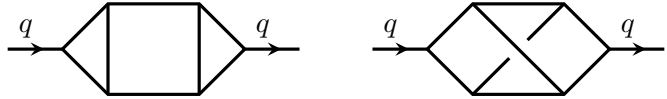

both with $(q^2)^2$ in the numerator.
Considering the four-particle cut of the two-point integrands, there are 26 cut contributions, which can be combined into following compact form:
\begin{equation}
{2 (q^2)^2 \over s_{12} s_{23} s_{34} s_{41}} + \bigg[ {(q^2)^2 \over s_{12} s_{34} s_{123} s_{341} } + {(q^2)^2 \over s_{12} s_{34} s_{234} s_{412} } + {2 (q^2)^2 \over s_{12} s_{34} s_{123} s_{234} } + { 2 (q^2)^2 \over s_{12} s_{23} s_{234} s_{412} }  + \textrm{cyclic perm.} \bigg] .
\label{F_4}
\end{equation}
We have checked that such obtained square of four-point form factor  is equivalent to a direct form factor computation
\begin{equation}
\int \prod_{i=1}^4 d^4 \eta_i \, \Big[ {\cal F}_{{\cal O}_{12},4}^{ {\rm MHV}}\, {\cal F}_{{\cal O}_{34},4}^{\textrm{N$^2$MHV}} + {\cal F}_{{\cal O}_{12},4}^{\textrm{N$^2$MHV}} \, {\cal F}_{{\cal O}_{34},4}^{{\rm MHV}} + {\cal F}_{{\cal O}_{12},4}^{ {\rm NMHV}} \, {\cal F}_{{\cal O}_{34},4}^{{\rm NMHV}} \Big] \,,
\end{equation}
where the color ordering is choose as ${\cal F}_{{\cal O}_{12},4}(1,2,3,4)$ and ${\cal F}_{{\cal O}_{34},4}(4,3,2,1)$.
Clearly, the result of the squre has cyclic permutational symmetry. 

For the five-point form factor square, we consider the three-loop Sudakov form factor. The three-loop form factor can be given by five trivalent topologies as given in \cite{Boels:2012ew}. By sewing the two external legs, there are five corresponding topologies for the two-point function:

\begin{figure}[h]
\begin{center}
\begin{tikzpicture}[line width=1.2 pt,  
                    scale=1.2,       
                    >=stealth,       
                    font=\normalsize 
                   ]\label{eq:2ptdiagram4loop}
\begin{scope}
  \draw[->] (-1.6,0) -- node[above] {$q$} (-1.2,0);
  \draw (-1.3,0) -- (-1,0);
  \draw (-1,0) -- (-0.5,0.5) -- (0.5,0.5) -- (1,0) -- (0.5,-0.5) -- (-0.5,-0.5) -- cycle;
  \draw (-0.5,0.5) -- (-0.5,-0.5);
  \draw (0,0.5) -- (0,-0.5);
  \draw (0.5,0.5) -- (0.5,-0.5);
  \draw[->] (1,0) -- node[above] {$q$} (1.4,0);
  \draw (1.3,0) -- (1.6,0);
\end{scope}

\begin{scope}[xshift =4 cm]
  \draw[->] (-1.6,0) -- node[above] {$q$} (-1.2,0);
  \draw (-1.3,0) -- (-1,0);
  \draw (-1,0) -- (-0.5,0.5) -- (0.5,0.5) -- (1,0) -- (0.5,-0.5) -- (-0.5,-0.5) -- cycle;
  \draw (-0.5,0.5) -- (0,-0.5);
  \draw (-0.5,-0.5) -- (-0.3,-0.1);
   \draw (-0.2,0.1) -- (0,0.5);
   \draw (0.5,0.5) -- (0.5,-0.5);
  \draw[->] (1,0) -- node[above] {$q$} (1.4,0);
  \draw (1.3,0) -- (1.6,0);
\end{scope}
\end{tikzpicture}
\end{center}
\begin{center}
\begin{tikzpicture}[line width=1.2 pt,  
                    scale=1.2,       
                    >=stealth,       
                    font=\normalsize 
                   ]
\begin{scope}
  \draw[->] (-1.6,0) -- node[above] {$q$} (-1.2,0);
  \draw (-1.3,0) -- (-1,0);
  \draw (-1,0) -- (-0.5,0.5) -- (0.5,0.5) -- (1,0) -- (0.5,-0.5) -- (-0.5,-0.5) -- cycle;
   \draw (-0.5,0.5) -- (-0.5,-0.5);
  \draw (0,-0.5) -- (0.5,0.5);
  \draw (0,0.5) -- (0.2,0.1);
   \draw (0.3,-0.1) -- (0.5,-0.5);
  \draw[->] (1,0) -- node[above] {$q$} (1.4,0);
  \draw (1.3,0) -- (1.6,0);
\end{scope}

\begin{scope}[xshift =4 cm]
  \draw[->] (-1.6,0) -- node[above left] {$q$} (-1.2,0);
  \draw (-1.3,0) -- (-1,0);
  \draw[->] (-1,0) -- node[above] {$l_1$} (-0.7,0.3);
  \draw (-0.8,0.2) -- (-0.5,0.5);
  \draw(-0.5,0.5) -- (0.5,0.5);
  \draw[->](1,0) -- node[above] {$l_3$} (0.7,0.3);
  \draw(0.8,0.2) -- (0.5,0.5);
  \draw[->](1,0) -- node[below] {$l_4$} (0.7,-0.3);
  \draw(0.8,-0.2) -- (0.5,-0.5);
   \draw[->](-1,0) -- node[below] {$l_2$} (-0.7,-0.3);
  \draw(-0.8,-0.2) -- (-0.5,-0.5);
  \draw(-0.5,-0.5) -- (0.5,-0.5);
  \draw (-0.5,0.5) -- (-0.5,-0.5);
   \draw (-0.5,0) -- (0.5,0);
   \draw (0.5,0.5) -- (0.5,-0.5);
  \draw[->] (1,0) -- node[above] {$q$} (1.4,0);
  \draw (1.3,0) -- (1.6,0);
\end{scope}

\begin{scope}[xshift =8 cm]
  \draw[->] (-1.6,0) -- node[above] {$q$} (-1.2,0);
  \draw (-1.3,0) -- (-1,0);
  \draw[->] (-1,0) -- node[above] {$l_1$} (-0.7,0.3);
  \draw (-0.8,0.2) -- (-0.5,0.5);
  \draw(-0.5,0.5) -- (0.5,0.5);
  \draw[->](1,0) -- node[above] {$l_3$} (0.7,0.3);
  \draw(0.8,0.2) -- (0.5,0.5);
  \draw[->](1,0) -- node[below] {$l_4$} (0.7,-0.3);
  \draw(0.8,-0.2) -- (0.5,-0.5);
   \draw[->](-1,0) -- node[below] {$l_2$} (-0.7,-0.3);
  \draw(-0.8,-0.2) -- (-0.5,-0.5);
  \draw (-0.5,0.5) -- (-0.5,-0.5);
  \draw (-0.5,-0.5) -- (0.5,0);
  \draw (-0.5,0) -- (-0.1,-0.2);
  \draw (0.1,-0.3) -- (0.5,-0.5);
   \draw (0.5,0.5) -- (0.5,-0.5);
  \draw[->] (1,0) -- node[above] {$q$} (1.4,0);
  \draw (1.3,0) -- (1.6,0);
\end{scope}

\end{tikzpicture}
\end{center}
    \caption{All possible topologies that contribute to five-point form factor square}
\label{2ptdiagram4loop}
\end{figure}
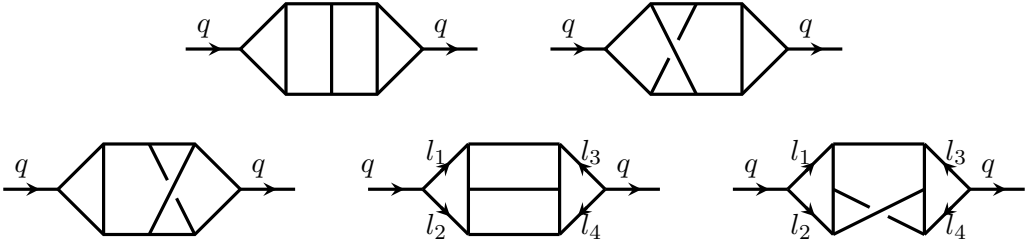

Unlike pervious simple lower loop cases, one needs to pay a special attention to the kinematic numerators. The numerators for the last two topologies have dependence on the loop momenta as induced from the form factor numerators \cite{Boels:2012ew}. After sewing the external legs, the form factor external legs should be taken off-shell. Moreover, the numerators should respect the symmetry of the two-point function graphs. Taking these into account, we modify the form factor numerators as
\begin{align}
& n_1 = n_2 = n_3 = (q^2)^3 \,, \\
& n_4 = - n_5 = {(q^2)^2 \over2} \big( \ell_1 \cdot \ell_3 - \ell_1 \cdot \ell_4 + \ell_2 \cdot \ell_4 - \ell_2 \cdot \ell_3 - q^2 \big) \,,
\end{align}
which correspond the numerators for the five diagrams in Figure \ref{2ptdiagram4loop}. 
Putting $\ell_3$ and $\ell_4$ on-shell, these numerators are equivalent to the form factor numerators in  \cite{Boels:2012ew}.
Now, one can consider the five-particle cut of the two-point integrands, there are 210 cut contributions. 
The expression of the form factor square is a bit lengthy and is given in the ancillary file.
We have checked that such obtained square of five-point form factor is numerically equivalent to a direct form factor summation
\begin{equation}
\int \prod_{i=1}^5 d^4 \eta_i \, \Big[ {\cal F}_{{\cal O}_{12},5}^{ {\rm MHV}}\, {\cal F}_{{\cal O}_{34},5}^{\textrm{N$^3$MHV}} + {\cal F}_{{\cal O}_{12},5}^{\textrm{N$^3$MHV}} \, {\cal F}_{{\cal O}_{34},5}^{{\rm MHV}} + {\cal F}_{{\cal O}_{12},5}^{ {\rm NMHV}}\, {\cal F}_{{\cal O}_{34},5}^{\textrm{N$^2$MHV}} + {\cal F}_{{\cal O}_{12},5}^{\textrm{N$^2$MHV}} \, {\cal F}_{{\cal O}_{34},5}^{{\rm NMHV}}  \Big] .
\end{equation}
As before, the result has cyclic permutational symmetry corresponding to the square of color-ordered form factors with ordering choose as ${\cal F}_{{\cal O}_{12},5}(1,2,3,4,5)$ and ${\cal F}_{{\cal O}_{34},5}(5,4,3,2,1)$.

Finally, we would like to comment on the color factors. In the above computation, we have considered only the color-ordered form factor product as 
\begin{equation}
{\cal F}_{{\cal O}_{12},n}(1,2,\ldots, n)\, {\cal F}_{{\cal O}_{34},n}(n, \ldots,2,1) \,.
\end{equation}
In this way we obtain the leading color contribution. For the form factor square up to five points, the result is actually complete, since there is no sub-leading color contribution. This can be understood from the color factor of the two-point function. For example, the color factors for the diagrams in Figure \ref{2ptdiagram4loop} are all proportional to $N_c^4$ and similarly for lower loop cases. However, starting from the six-point form factor square (relating to the four-loop Sudakov form factor in the above construction), there will be non-planar corrections due to the quadratic Casimir color factors \cite{Boels:2012ew}.
Given the known full-color four-loop and five-loop Sudakov form factor results \cite{Boels:2012ew, Yang:2016ear}, the above procedure can be also used to extract also the six- and seven-point form factor squares including both planar and non-planar corrections.

\begin{figure}[htbp]
    \centering
    \includegraphics[width=0.7\linewidth]{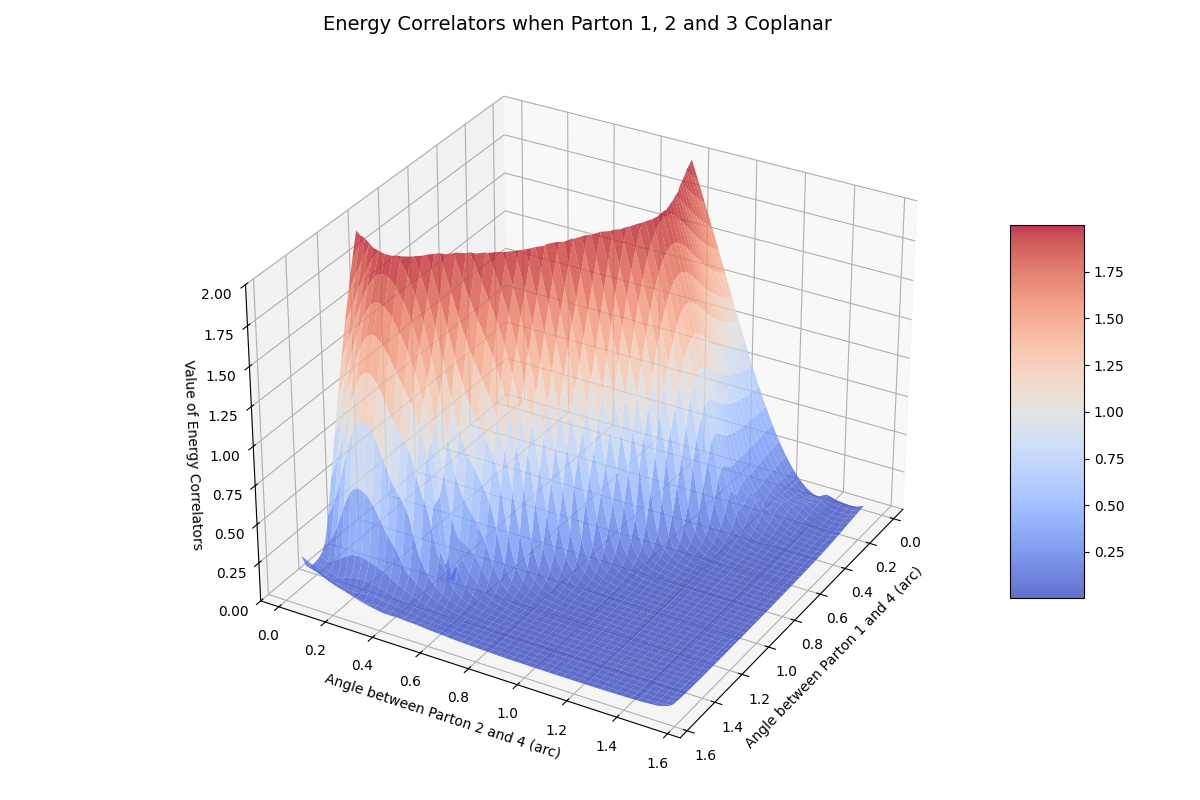}
    \caption{\centering{Four-Point Energy Correlators plot when Parton 1, 2 and 3 are coplanar.} We have chosen $z_2=2,\,z_3=1+2^{1\over4}i,\,\bar{z_3}=1-2^{1\over4}i$. This means there are two remaining variables $z_4$ and $\bar{z_4}$, which can be fixed by the angle between parton 1,4 and parton 2,4. }
    \label{fig:NumericalCop}
\end{figure}

\begin{figure}[htbp]
    \begin{minipage}{0.49\linewidth}
    \centering
    \includegraphics[width=1\linewidth]{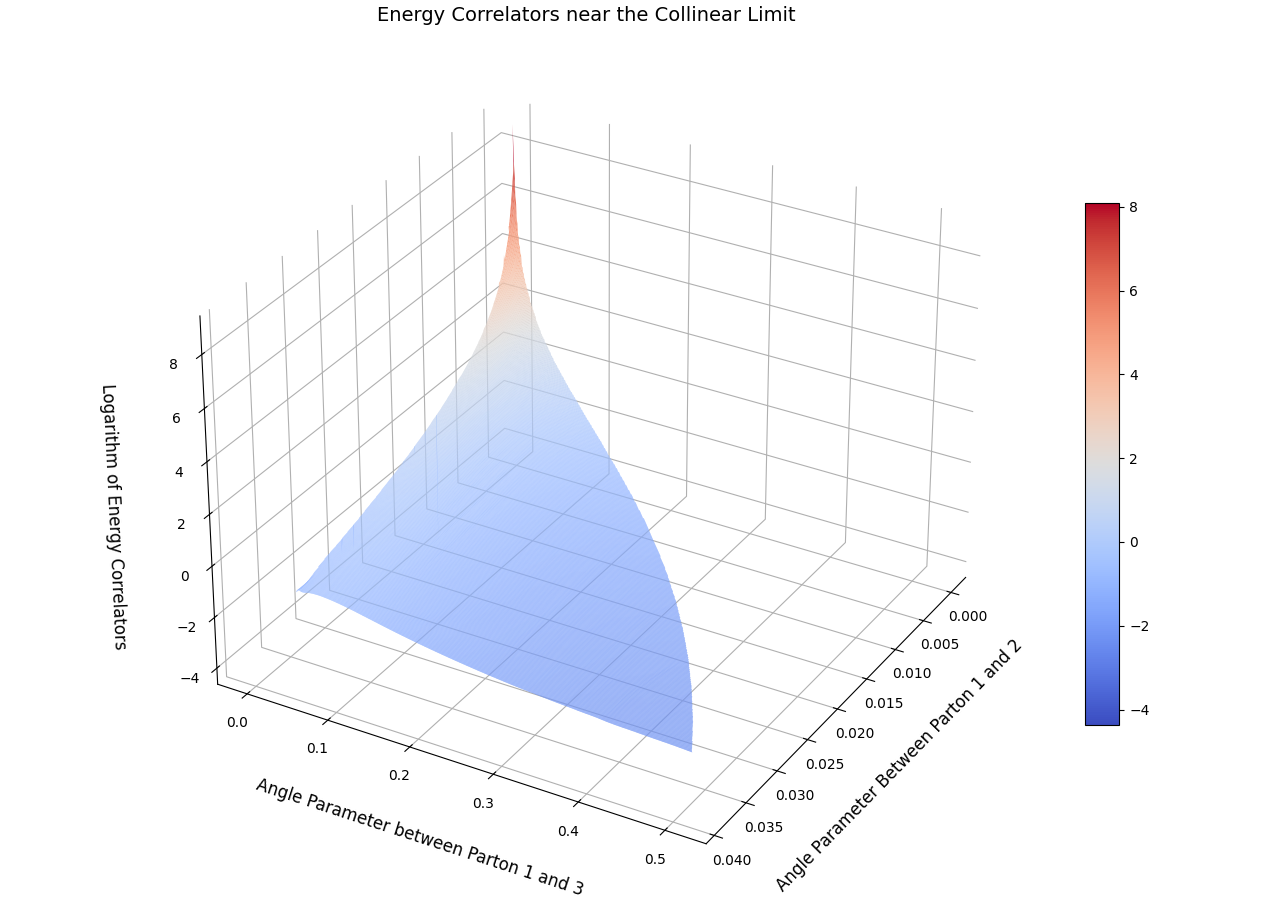}
    \end{minipage}
    \begin{minipage}{0.49\linewidth}
    \centering
    \includegraphics[width=1\linewidth]{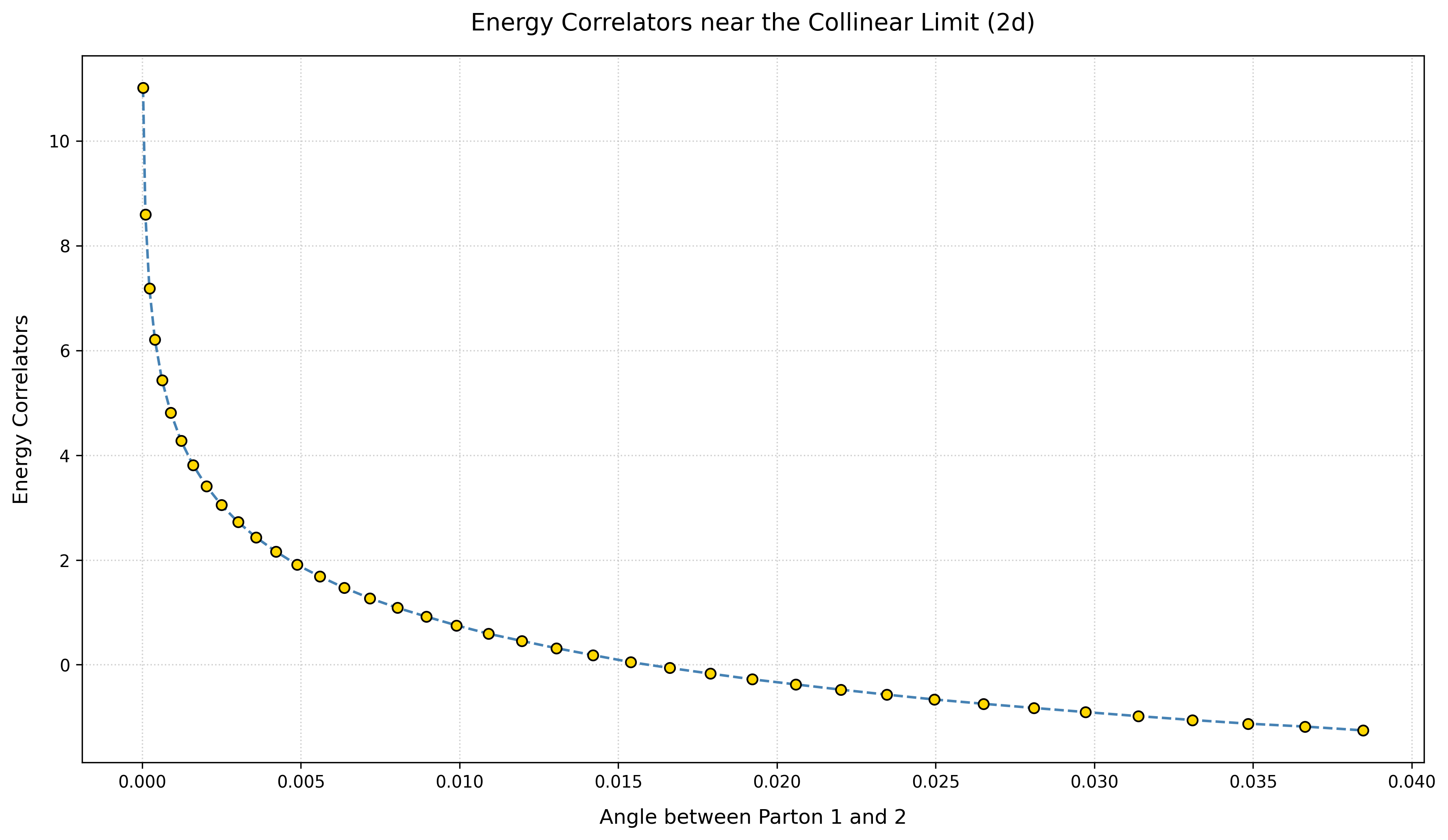}
    \end{minipage}
    \caption{\centering{Four-Point Energy Correlators plot near the multi-collinear limit.} The left figure shows the limit behavior when $z_2\rightarrow0$ with only fix $z_4=1+2i$ and $z_3$ changes, and the right figure shows the limit behavior when $z_2\rightarrow0$ with a fixed $z_3=2+i,\,z_4=1+2i$.} 
    \label{fig:NumericalCol}
\end{figure}

We first calculated four-point energy correlators numerically with the form factor square obtained from the above method. We choose some specific parameters to see what will happen for the coplanar limit (Figure. \ref{fig:NumericalCop}) and the multi-collinear limit (Figure. \ref{fig:NumericalCol}) in the four-point case. One should be reminded that for the multi-collinear limit, we require all the $\zeta_{ij}$ to $0$. Therefore, we perform $z_3\rightarrow z_2 z_3,\,\bar{z_3}\rightarrow z_2\bar{z_3}$ and $z_4\rightarrow z_2 z_4,\,\bar{z_4}\rightarrow z_2\bar{z_4}$ since $z_2$ is always real, and the new parametrization is

\begin{align*}
  &\zeta_{12}=\frac{z_2^2}{1+z_2^2},\quad \zeta_{13}=\frac{z_2^2|z_3|^2}{1+z_2^2|z_3|^2},\quad  \zeta_{14}=\frac{z_2^2|z_4|^2}{1+z_2^2|z_4|^2},\qquad\zeta_{23}=\frac{z_2^2|1-z_3|^2}{(1+z_2^2)(1+z_2^2|z_3|^2)}\\
  &\qquad\qquad\quad\zeta_{24}=\frac{z_2^2|1-z_4|^2}{(1+z_2^2)(1+z_2^2|z_4|^2)},\quad \zeta_{34}=\frac{z_2^2|z_3-z_4|^2}{(1+z_2^2|z_3|^2)(1+z_2^2|z_4|^2)}.
\end{align*}

\section{3-point Energy Correlator}
Energy correlator integrals are a type of unconventional Feynman integrals. On the one hand, they have some good properties, such as being finite and scalar integrals in energy fraction. This brings us some advantages for the calculation, we are free of regulator because of finiteness,  and we don't need to find another suitable parameter representation since they are already scalar integrals. On the other hand, it also challenges us to develop new IBP and DE techniques for these unconventional integrals.  Energy correlator integrals are somehow more like traditional Feynman integrals for the 3 point case, and from the 4 point on, one will realize their peculiarities and complexity.

We illustrate our methods and algorithms in the simpler 3 point case here.

\subsection{$\mathrm{E^3C}$ Integrals}
\paragraph{Symmetry $S_3$}
Considering the conservation of momentum , we can interpret $p_4$ as $q-p_1-p_2-p_3$. This 3-point energy correlator corresponds to \eqref{E^nC} for $n=3$. In this case, the form factor square $F_4^2$ \eqref{F_4} remains $S_3$ symmetry (permute $p_1$, $p_2$ and $p_3$). For simplicity, we shall consider only the resulting expressions after modding $F_4^2$ out by $S_3$ symmetry, say $F_4^2/S_3$, 
\begin{equation}
    \frac{(q^2)^2}{s_{12}s_{13}s_{24}s_{34}}+ \frac{(q^2)^2}{s_{12}s_{13}s_{124}s_{134}}+
    \frac{4(q^2)^2}{s_{12}s_{34}s_{124}s_{134}}+\frac{4(q^2)^2}{s_{13}s_{24}s_{123}s_{124}}+\frac{2(q^2)^2}{s_{23}s_{24}s_{123}s_{124}}+
   \frac{(q^2)^2}{s_{24}s_{34}s_{124}s_{134}}.
    \label{F4/S3}
\end{equation}
\paragraph{Integrand Simplification}
Via parameterization (\ref{energyPara}) and (\ref{anglePara}), we can express $s_{ij}$ and $s_{ijk}$ appear in (\ref{F4/S3}) as linear functions about $x_1$, $x_2$ and $x_3$ up to additional terms which will vanish at integral level with measure $\delta(1-x_1-x_2-x_3+\zeta_{12}x_1 x_2 +\zeta_{23} x_2 x_3 + \zeta_{13} x_1 x_3)$ in (\ref{E^nC}). Parameterize the first term in (\ref{F4/S3}) and time by $(x_1x_2x_3)^2$, we get a scalar integral,
\begin{equation}
    \frac{1}{\zeta_{12}\zeta_{13}}\int_{0}^{1}\mathrm{d}x_1\mathrm{d}x_2\mathrm{d}x_3\,\frac{\delta(1-x_1-x_2-x_3+\zeta_{12}x_1 x_2 +\zeta_{23} x_2 x_3 + \zeta_{13} x_1 x_3)}{(-1+x_1\zeta_{13}+x_2\zeta_{23})(-1+x_1\zeta_{12}+x_3\zeta_{23})}.
\end{equation}
One can find, for $i,j\in\{1,2,3\}$, the factor $s_{ij}=x_i x_j \zeta_{ij}$ on integrand denominator only contributes $\zeta_{ij}$, while $x_i x_j$ will be canceled by $(x_1x_2x_3)^2$. Therefore the most complicated parameter expression in \eqref{F4/S3} is the last term, its corresponding integral is,
\begin{equation}
    \int_{0}^{1}\mathrm{d}x_1\mathrm{d}x_2\mathrm{d}x_3\,\frac{\delta(1-x_1-x_2-x_3+\zeta_{12}x_1 x_2 +\zeta_{23} x_2 x_3 + \zeta_{13} x_1 x_3)}{(-1+x_1\zeta_{13}+x_2\zeta_{23})(-1+x_1\zeta_{12}+x_3\zeta_{23})(-1+x_3)(-1+x_2)}.
    \label{lastterm}
\end{equation}
All possible propagators in $\mathrm{E^3C}$ are,
\begin{align}
    &{\cal D}_1=\frac{s_{134}}{q^2}= -1+x_2,\, {\cal D}_2=\frac{s_{124}}{q^2}= -1+x_3, \,
    {\cal D}_3=\frac{s_{123}}{q^2}= -1+x_1+x_2+x_3, \nonumber\\
 &{\cal D}_4=\frac{s_{34}}{q^2 x_3}= -1+x_1 \zeta_{13}+x_2 \zeta_{23}, \, {\cal D}_5=\frac{s_{24}}{q^2 x_2} =  -1+x_1 \zeta_{12}+x_3 \zeta_{23}.
 \label{prop3point}
\end{align}
The four propagators in \eqref{lastterm} are linearly or algebraically dependent generally speaking, so we can perform the partial fraction decomposition to reduce this kind of integrand.

Partial fraction decomposition is a standard technique for simplifying an integrand and to decease the denominator number for the integration. The standard package {\sc MultiApart} \cite{Heller:2021qkz} or {\sc pfd-parallel} \cite{Bendle:2021ueg} is not directly used in this work, since the energy correlator integrand contains Dirac delta functions. Instead, we introduce a modified version of the {\sc MultiApart} algorithm \cite{Heller:2021qkz} for this purpose and the procedure is summarized as follows,
\begin{enumerate}
    \item Let $D_1,\ldots D_k$ be the irreducible denominators in integrand and $x_1, \ldots x_n$ be the variables. Denote the Dirac delta function as $\delta(F)$. Define $k$ free variables $q_1,\ldots q_k$.
    \item Rewrite the integrand as $P \cdot \delta(F)$ with the replacements $D_i^{-1} \to q_i$, $i=1,\ldots k$. Here $P$ is a polynomial,
        \begin{equation}
            P=q_1^{\beta_1} \ldots  q_k^{\beta_k} N(x_1,\ldots, x_n) \,.
        \end{equation}
    \item Define an ideal $I=\langle q_1 D_k-1, \ldots q_k D_k-1,F\rangle$. Calculate the Gr\"obner basis of $I$, $G(I)$ in the block ordering,
    \begin{equation}
        [q_1,\ldots q_k] \succ [x_1,\ldots, x_n]
    \end{equation}
    \item Divide $P$ towards the Gr\"obner basis $G(I)$ with the above block ordering. The monomials in the remainder become the decomposed integrands after the replacements, $q_i \to D_i^{-1}$ and the mulitplication with $\delta(F)$.
\end{enumerate}
This algorithm minimized the number of propagators by the elimination of $q_i$'s.
Therefore the integration computation for the $\mathrm{E^n C}$  can be simplified.
Comparing with the original {\sc MultiApart} algorithm, the new ingredient in this modified algorithm is to include the Dirac delta function's argument in the ideal. Then in the Gr\"obner basis computation, the $\delta(F)$ constraint is included and makes the remainder shorter. Thus
much simpler integrands are generated. This method is more convenient than to integrate out the Dirac delta function before the partial fraction decomposition. We propose our algorithm as a general preparation step for future $\mathrm{E^n C}$ computations. In the practice, this algorithm is powered by the computational algebraic geometry software {\sc Singular} \cite{DGPS}.

\paragraph{Power Counting and Finite Integrals}
There are six parts of $\mathrm{E^3 C}$ with respect to \eqref{F4/S3}, we will see that every part is a finite integrals then. An integral is finite if it is finite in every potentially divergent region. The so-called potential divergent regions correspond to the regions leading the propagators to zero. Considering the measure $\delta({\cal{D}}_\delta)\equiv\delta(1-x_1-x_2-x_3+\zeta_{12}x_1 x_2 +\zeta_{23} x_2 x_3 + \zeta_{13} x_1 x_3)$, the following three potential divergent regions are,
\begin{align}
  && \text{region 1}  &&\{\, x_1\rightarrow 1+(\zeta_{12}+\zeta_{13}-2)c x_1, \, x_2 \rightarrow c x_2, \, x_3 \rightarrow c x_3\, \}|_{c \rightarrow0}\, , \nonumber\\
   &&\text{region 2}  &&\{\, x_1\rightarrow c x_1, \, x_2 \rightarrow 1+(\zeta_{12}+\zeta_{23}-2)c x_2, \, x_3 \rightarrow c x_3\, \}|_{c \rightarrow0}\, , \nonumber\\
   &&\text{region 3}  &&\{\, x_1\rightarrow c x_1, \, x_2 \rightarrow c x_2, \, x_3 \rightarrow 1+(\zeta_{13}+\zeta_{23}-2)c x_3\, \}|_{c \rightarrow0}\, .
\end{align}
We give a criterion for judging finite integrals by power counting about $c$. The power counting of the integral measure part is always $2$ for the above three regions, 
\begin{equation}
    \text{d}x_1\text{d}x_2\text{d}x_3 \delta({\cal{D}}_\delta)\xrightarrow[\text{counting}]{\text{power}} 2
\end{equation}
The power counting of propagators would be different for different regions, 
\begin{align}
   \left\{ {\cal D}_1,\, {\cal D}_2,\, {\cal D}_3, \, {\cal D}_4, \, {\cal D}_5 \right\}\xrightarrow[\text{counting}]{\text{power}}\Biggl\{ 
    \begin{aligned}
        &\text{region 1}& \left\{0,\,0,\,1,\,0,\,0 \right\} \\
        &\text{region 2}& \left\{1,\,0,\,1,\,0,\,0\right\} \\
        &\text{region 3}& \left\{0,\,1,\,1,\,0,\,0\right\}
    \end{aligned}.
\end{align}
We call a propagator finite if its degree of divergence of any region above is $0$. Here, ${\cal{D}}_4$ and ${\cal{D}}_5$ are finite, and the reminders are divergent.
If the propagator appears on the denominator, its $c$ powers are negative; otherwise, it is positive. An integral is finite if the overall power count about $c$ is positive. For example, the following integral has only one potential divergent region, say $\{x\rightarrow c x, y \rightarrow c y\}$, when $c$ goes to $0$.
\begin{align}
   I(x_0,y_0) &= \int_0^{x_0} \text{d}x \int_0^{y_0} \text{d}y\frac{1}{x+y}=\left(x_0+y_0\right) \log \left(x_0+y_0\right)-x_0 \log \left(x_0\right)-y_0 \log \left(y_0\right)
   \label{eq:power_counting_example}
\end{align}
We show the power counting process of this integral below,
\begin{equation}
  I(x_0,y_0) \rightarrow  \int_0^{x_0} \text{d}(cx) \int_0^{y_0} \text{d}(cy)\frac{1}{cx+cy}\xrightarrow[\text{counting}]{\text{power}} 1>0
\end{equation}
By counting its power of $c$, we know it's finite, agree with eq.(\ref{eq:power_counting_example})

The $c$ powers of the initial six parts of $\mathrm{E^3C}$ correspond to \eqref{F4/S3} are positive, so all of them are finite integrals. This is a strong hint that makes us expect that all simplified integrals after PDF are still finite. They are indeed finite, except for two integrals whose combination turns to be finite finally.

\subsection{Integration by Parts}

\paragraph{Family Classification}
From the PDF result, we can divide the 3-point energy correlator integrand into three families with the same delta function $\delta(D_\delta)$ , where $D_\delta=1-x_1-x_2-x_3+\zeta_{12}x_1 x_2 +\zeta_{23} x_2 x_3 + \zeta_{13} x_1 x_3$. These three families are,
\begin{align}
    &&\text{family 1} &&D_1=-1+x_2, \, D_2=-1+x_3, \, D_3=-1+x_1+x_2+x_3, \nonumber\\
    &&\text{family 2} &&D_1=-1+x_2, \, D_2=-1+x_1+x_2+x_3, \, D_3=-1+x_1 \zeta_{12}+x_3 \zeta_{23}, \nonumber\\
    &&\text{family 3} &&D_1=-1+x_3, \, D_2=-1+x_1 \zeta_{13}+x_2 \zeta_{23}, \, D_3=-1+x_1 \zeta_{12}+x_3 \zeta_{23}.
\end{align}
We denote integrals as,
\begin{align}
 \text{Int}[n_1, n_2, n_3, 1]=\int \text{d}x_1 \text{d}x_2 \text{d}x_3 \frac{\delta(D_{\delta})}{D_1^{n_1} D_2^{n_2}D_3^{n_3}}=\int \text{d}x_1 \text{d}x_2 \text{d}x_3 \frac{1}{D_1^{n_1} D_2^{n_2}D_3^{n_3} D_{\delta}}\,\Big|_{\text{cut}(D_{\delta})}
\end{align}
For convenience for IBP, we regard delta function as a propagator $D_\delta$ with cut.
The target integrals we need to reduce are,
\begin{align}
    \text{family 1} \quad & \text{Int}[1, 1, 0, 1], \,
    \text{Int}[0, 1, 1, 1, \{x_2\}]=\text{Int}[-1, 1, 1, 1]+\text{Int}[0, 1, 1, 1];\label{divergent2finite}
    \\
    \text{family 2} \quad &\text{Int}[-1,1,1,1],\, \text{Int}[0,1,1,1],\, \text{Int}[1,-1,1,1],\, \text{Int}(1,0,1,1),\nonumber\\
    &  \text{Int}[0,0,1,1],\, \text{Int}[-1,1,0,1], \,\text{Int}[0,1,-1,1], \,\text{Int}[0,1,0,1], \nonumber\\
    &\text{Int}[1,0,-1,1],\, \text{Int}[1,-1,0,1],\, \text{Int}[1,0,0,1],\, \text{Int}[0,0,0,1]; \nonumber\\
    \text{family 3} \quad &\text{Int}[-1,1,1,1],\, \text{Int}[0,1,1,1],\, \text{Int}[1,1,-1,1],\, \text{Int}[1,1,0,1], \nonumber\\
    &\text{Int}[-1,0,1,1] , \, \text{Int}[0,0,1,1], \, \text{Int}[-1,1,0,1], \,\text{Int}[0,1,-1,1], \, \text{Int}[0,1,0,1],\nonumber \\
    & \text{Int}[1,-1,0,1], \, \text{Int}[1,0,-1,1], \, \text{Int}[1,0,0,1], \, \text{Int}[0,0,0,1] \nonumber
\end{align}
$\text{Int}[0,1,1,1,\{x_2\}]$ in eqn.\eqref{divergent2finite} means  $\int \text{d}x_1 \text{d}x_2 \text{d}x_3 (\delta(D_{\delta})x_2)/(D_2 D_3)$, which introduces a reducible numerator $x_2=D_1+1$ for finite consider.
\paragraph{Finite IBP}
\label{finite IBP}
We aim to build a finite IBP system for $\mathrm{E^3C}$ integrals. There are two points that we need to consider,
\begin{enumerate}
    \item do not increase the divergent power,
    \item keep the form of the delta function.
\end{enumerate}
The natural way to translate these two conditions to constraints on IBP relations is using computational algebraic geometry language. For an IBP operator,
\begin{equation}
{\cal{O}}_{\text{IBP}}= \sum_{i=1}^3 \frac{\partial}{\partial x_i}\left(a_{i}\, \cdot \,\,\right), 
\end{equation}
these conditions imply the so-called syzygy equations \cite{Gluza_2011,Schabinger_2012,B_hm_2018,Wu:2023upw},
\begin{equation}
   \sum_{i=1}^3 a_{i} \frac{\partial}{\partial x_i} D_j-b_j D_j=0,   \quad j\in \text{divergent set}\cup  \{\delta\}
   \label{syzygy equation}
\end{equation}
for $a_{i}, \, b_j \in Q(\zeta_{12},\zeta_{23},\zeta_{13})[x_1,x_2,x_3]$, where the divergent set corresponds to the ID set of divergent propagators.
When syzygy equations are satisfied, the powers of propagator in the "divergent set" do not increase and the delta function is unchanged. Therefore the IBP relations contain only finite integrals. This syzygy method is similar to the method used for computing $1/\epsilon$ coefficients of Feynman integrals \cite{Henn:2022vqp}.

Here we take the family 3 as an example. the first propagator $D_1=-1+x_3$ goes to $0$ at region 3, and the $D_1$ is the only one divergent propagator. So $j$ in \eqref{syzygy equation} can be $1$ and $\delta$ for finite IBP operator in family $3$.

We show this IBP operator act on an energy correlator integral as,
\begin{equation}
     {\cal{O}}_{\text{IBP}} \, \text{Int}[n_1,n_2,n_3,1] \equiv \int \text{d}x_1 \text{d}x_2 \text{d}x_3 \left({\cal{O}}_{\text{IBP}}\frac{1}{D_1^{n_1} D_2^{n_2}D_3^{n_3} D_{\delta}}\right)\Big|_{\text{cut}(D_{\delta})}
\end{equation}
\paragraph{Boundary IBP}
IBP identities for energy correlators have nonzero boundary terms, because of it's nontrivial integration boundary,
\begin{equation}
    {\cal{O}}_{\text{IBP}} \, \text{Int}[n_1,n_2,n_3,1] = \sum_{i=1}^3(\text{BT}_{x_i= 1}-\text{BT}_{x_i= 0}).
\end{equation}
For $x_i= 1$, boundary terms are $0$. When $x_i = 0$, boundary terms degenerate to 2-fold integrals which are  similar to 2-point energy correlator integrals. For family $3$, its boundary term integrals can be divided to 3 subfamily integrals.
\begin{align}
    \text{subfamily 1  }\qquad & D_1=-1+x_1 \zeta_{12}, \, D_2=-1+x_1 \zeta_{13}+x_2\zeta_{23}, \, D_{\delta}=1-x_1-x_2+x_1 x_2 \zeta_{12}\,,\nonumber \\
    \text{subfamily 2  }\qquad & D_1=-1+x_3 \zeta_{13}, \, D_2=-1+x_1 \zeta_{13}, \, D_{\delta}=1-x_1 -x_3+ x_1 x_3 \zeta_{13}\,,\nonumber \\
    \text{subfamily 3  }\qquad & D_1=-1+x_3\zeta_{23}, \, D_2=-1+x_2 \zeta_{23}, \, D_{\delta}=1-x_2-x_3+x_2 x_3 \zeta_{23}\,.
\end{align}
We denote those 2-fold integrals in boundary terms as,
\begin{equation}
    \text{Int}_2[sf,\{n_1,n_2,1\}]=\int \text{d}x_1 \hat{\text{d}x_i} \text{d}x_3 \frac{\delta(D_{\delta})}{D_1^{n_1} D_2^{n_2}},
\end{equation}
where $sf$ means the ID of subfamily, $\hat{\text{d}x_i}$ means $\text{d}x_i$ is absent.
For those 2-fold integrals in boundary terms, we can generate finite IBP systems just like what we did in 3-fold integral case. Combine all those IBP systems together, we perform IBP reductions by {\sc FiniteFlow} \cite{Peraro_2019}.

As an example, for family $3$, there are ten master integrals, contain five 3-fold integrals, four 2-fold integrals and one constant number $1$,
\begin{align}
    &\{ \text{Int}[1,1,-1,1],\,\text{Int}[-1,1,1,1],\,\text{Int}[1,1,0,1],\, \text{Int}[0,1,1,1],\, \text{Int}[1,0,0,1], \nonumber\\
    & \text{Int}_2[1,\{0,0,1\}], \, \text{Int}_2[2,\{0,1,1\}],\, \text{Int}_2[2,\{0,0,1\}], \, \text{Int}_2[3,\{0,0,1\}],\,1 \}
\end{align}

\subsection{Canonical Differential Equation}
A traditional differential operator $\frac{\partial}{\partial \zeta_{**}}$ ,in general, do not hold the form of energy correlator integrals, due to derivatives of Dirac delta function. Further, it is better not to increase the power of divergent propagators, to avoid the integrals with dotted divergent propagators. Instead, consider a differential operator 
\begin{equation}
{\cal{O}}_{\partial \zeta_{**}}=\frac{\partial}{\partial \zeta_{**}}+{\cal{O}}_{\text{IBP}}\equiv
\frac{\partial}{\partial \zeta_{**}}+ \sum_{i=1}^3  \frac{\partial}{\partial x_i} a_{i}
\end{equation}
such that, 
\begin{equation}
\frac{\partial}{\partial \zeta_{**}}D_j + \sum_{i=1}^3 a_{i} \frac{\partial}{\partial x_i} D_j-b_j D_j=0,\quad j\in \text{divergent set}\cup  \{\delta\},
\label{eq:DE_lift}
\end{equation}
where $a_{i}, \,b_j \in Q(\zeta_{12},\zeta_{23},\zeta_{13})[x_1,x_2,x_3]$. The divergent set corresponds to the ID set of divergent propagators. If the above condition holds, i.e., $a_{i}$'s and $b_j$'s exist, this differential operator is called "finite". The condition \eqref{eq:DE_lift}
guarantees that after the differential operator action, there is no Dirac delta derivative or the dotted propagators in the "divergent set". The goal is to write derivatives of a finite integral in terms of finite integrals. Mathematically, eqn.~\eqref{eq:DE_lift} is a lift equation in computational algebraic geometry and can be solved via the software {\sc Singular} \cite{DGPS}.

Two operators $\frac{\partial}{\partial \zeta_{**}}$ and ${\cal{O}}_{\partial \zeta_{**}}$ act on the same integral, the difference in the result would be,
\begin{equation}
     \frac{\partial}{\partial \zeta_{**}} \, \text{Int}[n_1,n_2,n_3,1]-{\cal{O}}_{\partial \zeta_{**}} \, \text{Int}[n_1,n_2,n_3,1]= \sum_{i=1}^3(\text{BT}_{x_i= 0}).
\end{equation}
So we can obtain the differential equations by finite differential operator  ${\cal{O}}_{\partial \zeta_{**}}$ up to some 2-fold boundary integrals. For those boundary integrals appeared in the differential equations about 3-fold integrals, we can get their differential equations similarly. One can realize that there are iterative structures in the processes of both IBPs and DEs. What's more, the canonical differential equation matrix should be strictly block upper triangular matrices. For 3-point case, canonical differential matrices are cubically nilpotent, which means the highest uniform transcendental weight is 2. We illustrate this form in the following diagram. Those weight-2 UT integrals correspond  to the dark green part, the light green part means weight-1 UT integrals and the yellow part is constant $1$.

\begin{figure}[htbp]
  \centering
  \begin{tikzpicture}[thick,font=\sffamily]

    \node[font=\fontsize{28}{28}\selectfont\color{block}] (d) at (-2.8,2.6) {$d$};
    \node[font=\fontsize{28}{28}\selectfont\color{block}] (eq) at (-0.6,2.5) {$=$};

    \draw[fill=w0] (-2.0,0)   rectangle (-1.3,0.675);   
    \draw[fill=w1] (-2.0,0.675)   rectangle (-1.3,2.5);   
    \draw[fill=w2] (-2.0,2.5)   rectangle (-1.3,4.9);  

    \begin{scope}[shift={(0,0)}]
    \draw (0,0) rectangle (5,5); 

    \foreach \x in {}{
    \draw (\x,4.95) -- (\x,5);           
    \node[font=\scriptsize,above] at (\x,5.1) {\x}; 
    }

    \foreach \y in {}{
    \draw (5,\y) -- (4.95,\y);           
    \node[font=\scriptsize,right] at (5.1,\y) {\y}; 
    }
    \foreach \x in {}{
    \draw (\x,0) -- (\x,0.05); }
    \foreach \y in {}{
    \draw (0,\y) -- (0.05,\y);}  
      \draw[dashed,block,line width=1.2pt] (0,5) -- (5,0);

      \fill[block] (2.5,2.5) rectangle (4.3,4.9);
      \node[text=w1,font=\bfseries] at (3.4,3.7) {$\int{w_1}$};

      \fill[block] (4.3,0.675) rectangle (5,2.5);
      \node[text=w0,font=\bfseries] at (4.65,1.6) {$\int{w_0}$};

      \node[anchor=west,align=left,font=\large\bfseries,block] 
            at (0.3,1.2) {cubically \\
            nilpotent};
    \end{scope}

    \begin{scope}[shift={(6,0)}]
      \draw[fill=w0] (0,0)  rectangle (0.7,0.675);
      \draw[fill=w1] (0,0.675)  rectangle (0.7,2.5);
      \draw[fill=w2] (0,2.5)  rectangle (0.7,4.9);

      \node[anchor=west,font=\Large\bfseries,text=w2] at (1.0,3.6) {$w_2$};
      \node[anchor=west,font=\Large\bfseries,text=w1] at (1.0,1.5) {$w_1$};
      \node[anchor=west,font=\Large\bfseries,text=w0] at (1.0,0.3) {$w_0$};

      \draw[->](4.1,0) -- (4.1,5)
            node[at start, right=0.6cm,rotate=90,align=center,
                 font=\Large\bfseries,text=w2] {Integrate iteratively};
    \end{scope}

  \end{tikzpicture}
  \label{fig:block-upper}
  \caption{The structures of canonical differential equations for three-point energy correlators.}
\end{figure}
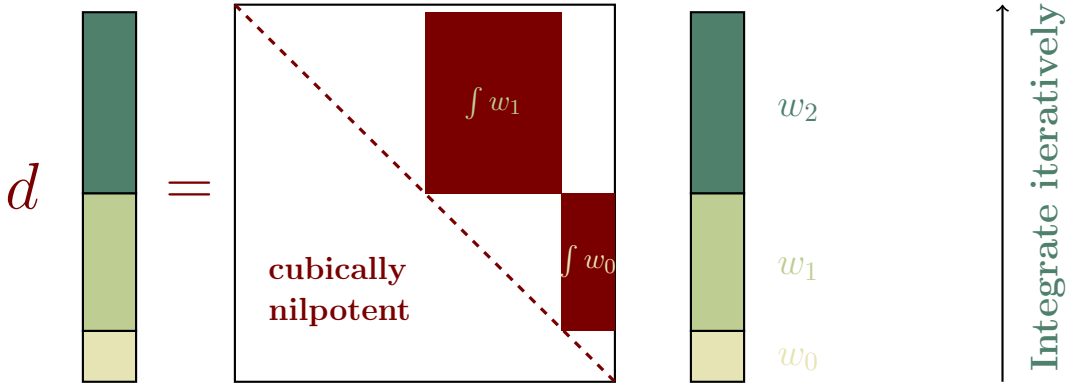

The canonical differential equation and UT integral files of these three families are in the ancillary directory.

\paragraph{{Analytic Result of $\mathrm{E^3C}$}}
To rationalize the square root in differential equation, we perform parameterization \eqref{para 2} for family 1 and family 2, parameterization \eqref{para 1} for family 3. 
We list the letters of these three families here,
\begin{align}
\text{family 1} \quad
    & s, \, x_1,\, x_2, \, 1-x_1,\, 1-x_2,\,1+s,\, s+x_1,\, s+x_2,\, \nonumber\\
    &   1+sx_1,\,s+x_1x_2,\, 1+s x_1 x_2 ,\,1-x_1 x_2,\, 1-x_1^2 x_2\,; \nonumber\\
   \text{family 2} \quad   & s, \, x_1,\, x_2, \, 1-x_1,\, 1-x_2,\,1+s,\, s+x_1,\, s+x_2,\, \nonumber\\
    &   1+sx_1,\,s+x_1x_2,\, 1+s x_1 x_2 ,\,1-x_1 x_2\,; \nonumber\\
   \text{family 3} \quad & z_2,\,z_3,\, \bar{z}_3,\,z_3-z_2,\,\bar{z}_3-z_2,\,\bar{z}_3-z_3,\,  z_2^2+1,\,z_2 z_3+1,\,z_2 \bar{z}_3+1,\,z_3\bar{z}_3+1\,. \nonumber
\end{align}
Family 3 is the simplest family, family 1 has one more letter, $1-x_1^2 x_2$, 
than family 2. One can also find the analytic results of UT integrals and propagators' definitions in the ancillary directory.

\section{4-point Energy Correlator}

Compared with 3-point EEC integrals, the difficulty of 4-point EEC calculation is not only the increase of integral variables, but also the new type of propagators, \textit{polynomial propagators}. The measure part of $\mathrm{E^4 C}$ is also more complicated than $\mathrm{E^3C}$, it's shows as a delta function in numerator,
\begin{equation}
    \delta(1-x_1-x_2-x_3-x_4+x_1x_2\zeta_{12}+x_1x_3\zeta_{13}+x_1x_4\zeta_{14}+x_2x_3\zeta_{23}+x_2x_4\zeta_{24}+x_3x_4\zeta_{34}). \label{eq:delta function}
\end{equation}
Some terms in $\mathrm{E^4 C}$ can contain $6$ propagators in the denominator, these complex integrals are hard to deal with. 
Polynomial propagators make things complicated in four ways. Firstly, 
those propagators are algebraic dependent in a way that is not so obvious, PDF plays a more important role in this case. Meanwhile, this method introduces some divergent integrals that we need to deal with, like the case \eqref{divergent2finite}, but not so easy to fix the divergent problem. We solve this issue in section \ref{combineDiv}.
Secondly, the representation of this kind of integrals would not be as straightforward as Baikov representation, because the relation between the integral variables and propagators is no longer a linear transformation for $\mathrm{E^4C}$. 
We will also define a standard form to represent these finite integrals in \ref{subsec:standard integral}. For the same reason, when we act an IBP operator on $\mathrm{E^4C}$ integrals, we need to perform polynomial division on numerator to represent an IBP relation about the standard form integrals.
Thirdly, in additional to IBP relations, there are still some additional relations about $\mathrm{E^4C}$ integrals at the integrand level that are necessary. We will give the details in section \ref{subsec:additional relation}.
Finally, the polynomial propagators make the integrals itself inevitably complicated even after PDF. There appear elliptic integrals and hyperelliptic integrals.

\subsection{$\mathrm{E^4C}$ Integrals}
\paragraph{Symmetries}
Similar with $\mathrm{E^3C}$, consider momentum conservation, $\mathrm{E^4C}$ remains $S_4$ symmetry.
\begin{equation}
    \mathrm{E^4C}(p_1,p_2,p_3,p_4)=\sum_{\sigma \in S_4} \sigma \, \mathcal{E}^4 \mathcal{C}(p_1,p_2,p_3,p_4) \label{S4symmetry}
\end{equation}
There are $43$ terms in $\mathrm{E}^4\mathrm{C}$.

\paragraph{Partial Fraction Decomposition}
To simplify $\mathrm{E}^4\mathrm{C}$ integrals, we apply PFD method and get fragment integrals in 62 families that contain at most $3$ of the following propagators,
\begin{align}
& \frac{s_{45}}{x_4}=1-x_1\zeta_{14}-x_2\zeta_{24}-x_3\zeta_{34}\, ,\quad \frac{s_{15}}{x_1}=1-x_2\zeta_{12}-x_3\zeta_{13}-x_4\zeta_{14}\, , \quad s_{2345}=1-x_1 \, ,\nonumber \\
&s_{1345}=1-x_2 \, ,\quad s_{1235}=1-x_4 \, , \quad s_{1234}=1-x_1-x_2-x_3-x_4 \, , 
    \quad \nonumber  \\
&   s_{345}=  1-x_1-x_2+x_1x_2\zeta_{12} \, , \,
s_{145}=1-x_2-x_3+x_2x_3 \zeta_{23}\, , \, s_{125}=1-x_3-x_4+x_3 x_4 \zeta_{34} \, , \nonumber \\
&  s_{123}=x_1 x_2 \zeta_{12}+x_1 x_3 \zeta_{13}+x_2 x_3 \zeta_{23}\, , \, s_{234}=x_2x_3\zeta_{23}+x_2x_4\zeta_{24}+x_3x_4\zeta_{34} \, , \, s_{1245}=1-x_3 \,.
\label{propagators}
\end{align}
The $c$ powers of the $43$ terms in $\mathrm{E}^4\mathrm{C}$ are positive, so all the initial complicated integrals are finite. This good property may not hold obviously after PFD, because we don't and it is difficult to perform the finite constraint for this algebraic calculation. However, we believe that the divergent parts would be canceled, since the initial integrals are finite. We introduce a systematic method for combining divergent integrals into finite integrals.


\paragraph{Standard Finite Integrals \label{subsec:standard integral}}

With the delta function constrain in \eqref{eq:delta function}, there are four possible regions that contribute the highest divergence degree when $c$ goes to $0$. They are,
\begin{align*}
    \text{region 1:}& \quad \{x_1 \rightarrow c x_1 \, ,\quad x_2\rightarrow cx_2 \, , \quad x_3\rightarrow cx_3 \, , \quad x_4\rightarrow 1+(\zeta_{14}+\zeta_{24}+\zeta_{34}-3)cx_4 \}  \\
    \text{region 2:}& \quad \{x_1 \rightarrow c x_1\, ,\quad x_2\rightarrow c x_2\, , \quad x_3\rightarrow 1+(\zeta_{13}+\zeta_{23}+\zeta_{34}-3)c x_3\, , \quad x_4\rightarrow c x_4\} \\
    \text{region 3:}& \quad \{x_1 \rightarrow c x_1\, ,\quad x_2\rightarrow 1+(\zeta_{12}+\zeta_{23}+\zeta_{24}-3)c x_2\, , \quad x_3\rightarrow c x_3\, , \quad x_4\rightarrow c x_4\}  \\
    \text{region 4:}& \quad \{x_1 \rightarrow 1+(\zeta_{12}+\zeta_{13}+\zeta_{14}-3)c x_1\, ,\quad x_2\rightarrow c x_2\, , \quad x_3\rightarrow cx_3 \, , \quad x_4\rightarrow c x_4\} 
\end{align*}
Propagators divergence degree in four regions:
\begin{align*}
    &\text{region 1: }\, \quad \{0, 0, 1, 0, 0, 1, 1, 0, 0, 1, 2,0\} \\
    &\text{region 2: }\, \quad \{0, 0, 0, 1, 0, 1, 1, 1, 0, 1, 1,1\} \\
    &\text{region 3: }\, \quad \{0, 0, 0, 0, 0, 1, 0, 1, 1, 1, 1,0\} \\
    &\text{region 4: }\, \quad \{0, 0, 0, 0, 1, 1, 0, 0, 1, 2, 1,0\} 
\end{align*}
Similarly with 3-point case, the $c$ degree of power counting of measure part is,
\begin{align}
dx_1dx_2dx_3dx_4\delta(D_\delta)\xrightarrow[\text{counting}]{\text{power}} 3.
\end{align}

Finite integrals can be confirmed by power counting. Consider the polynomial propagators, it's hard, sometimes impossible, to represent a finite integral with a specific numerator to an integral that can be described only by these propagators. We arrange the standard finite integrals numerator to be monomial for the convenience of expression. We denote the standard finite integral for 4-point as
\begin{equation}
    \mathrm{Int}[\{n_1,n_2,\cdots,n_{11},1\},N]=\int_0^1\frac{\text{d}x_1\cdots \text{d}x_4 N \delta(D_\delta)}{D_1^{n_1} D_2^{n_2}\cdots D_{11}^{n_{11}}D_{12}^{n_{12}}},
\end{equation}
where $n_i\geq0$. We characterize the numerator by $N$ and ask $N$ to always be monomial about $x_i$. Due to the simplification of PFD, every finite integral appears at most three propagators among $12$ propagators on the denominator.
\paragraph{Combine Divergent Integrals to Finite Integrals}
\label{combineDiv}
As what we analyzed before, divergent integrals appear from finite integrals because we applied PFD algorithm. For every divergent integral, we can regard it as the sum of finite part and divergent part. If we expect the divergent parts to be canceled, the most promising way to achieve this is representing those divergent parts as the same form, or minimize divergent types as less as possible.

We can cancel divergent integrals in $\mathcal{E}^4\mathcal{C}$ circuitously. Firstly, winnowing divergent integrals out and identifying which sector are they in. Then, seeding finite integrals in these sectors and doing polynomial reduction for integrand numerator with respect to the same monomial ordering as PFD algorithm. For example,
\begin{align}
  &\int\frac{x_2 \delta(D_\delta)}{D_1D_5D_{10}}=-\frac{\left(z_2^2+1\right) \left(z_3 \left(-\bar{z_3}\right)+z_3 \bar{z_4}+z_4 z_3 \bar{z_3} \bar{z_4}+z_4 \bar{z_3}\right)
   }{z_2 \left(z_3 \bar{z_3}+1\right) \left(z_2 z_4 \bar{z_4}+\bar{z_4}-z_2+z_4\right)}\int\frac{x_3\delta(D_\delta)}{D_1D_5D_{10}} \nonumber\\
  &\qquad \qquad+\frac{\left(z_2^2+1\right) z_4 \bar{z_4}}{z_2 \left(z_2 z_4
   \bar{z_4}+\bar{z_4}-z_2+z_4\right)} \int\frac{\delta(D_\delta)}{D_{10}}
   +\frac{\left(z_2^2+1\right) z_4 \bar{z_4} }{z_2 \left(z_2 z_4
   \bar{z_4}+\bar{z_4}-z_2+z_4\right)}\int\frac{\delta(D_\delta)}{D_1D_5} \nonumber\\
  &\qquad \qquad-\frac{\left(z_2^2+1\right)
   }{z_2 \left(z_2 z_4 \bar{z_4}+\bar{z_4}-z_2+z_4\right)}\int\frac{\delta(D_\delta)}{D_5D_{10}}-\frac{\left(z_2^2+1\right) }{z_2 \left(z_2 z_4
   \bar{z_4}
   +\bar{z_4}-z_2+z_4\right)}\int\frac{\delta(D_\delta)}{D_1D_5D_{10}} .\nonumber
\end{align}
For this identity, integrals appear in the first two lines are finite, whereas those in the last line are divergent. We will get some identities that mix finite integrals and divergent integrals together in this step. Then give finite integrals a prefer integral ordering and reduce those divergent integrals that appear in $\mathcal{E}^4\mathcal{C}$ by Gaussian elimination. It turns out that the divergent parts are eliminated by replacing the divergent integrals as the reduction result.

\subsection{Integration by Parts}
\paragraph{Family Classify}
To avoid the repetitive complication, we modulo the $S_4$ symmetry in \eqref{S4symmetry} at  $\mathrm{E^4C}$ observation level. When we go to the integral level, especially after PFD, there could still be some permutation symmetry among different families. It can also reduce the number of propagators in \eqref{propagators}, in fact, there are 26 families after symmetry reduction. The last propagator $s_{1245}$ can be reduced, and there are finally 11 propagators.
\begin{align}
    & D_1=-1+x_1\zeta_{14}+x_2\zeta_{24}+x_3\zeta_{34}\, ,\quad D_2=-1+x_2\zeta_{12}+x_3\zeta_{13}+x_4\zeta_{14}\, , \nonumber \\
&D_3=-1+x_1 \, ,\quad D_4=-1+x_2 \, ,\quad D_5=-1+x_4 \, , \quad D_6=-1+x_1+x_2+x_3+x_4 \, , 
    \quad \nonumber  \\
&   D_7=  -1+x_1+x_2-x_1x_2\zeta_{12} \, , \,
D_8=-1+x_2+x_3-x_2x_3 \zeta_{23}\, , \, D_9=-1+x_3+x_4-x_3 x_4 \zeta_{34} \, , \nonumber \\
&  D_{10}=x_1 x_2 \zeta_{12}+x_1 x_3 \zeta_{13}+x_2 x_3 \zeta_{23}\, , \, D_{11}=x_2x_3\zeta_{23}+x_2x_4\zeta_{24}+x_3x_4\zeta_{34} \, .
\label{4pProp}
\end{align}
\paragraph{Finite IBP}
Similarly with section \ref{finite IBP}, we establish our 4-point energy nenergy correlator IBP system by Syzygy method. here $D_1$ and $D_2$ are finite, and the reminders are divergent.

\begin{align}
     {\cal{O}}_{\text{IBP}} \, \text{Int}[\{n_1,n_2,\cdots,n_{11},1\},N] &\equiv \int_0^1 \text{d}x_1 \cdots \text{d}x_4 \left({\cal{O}}_{\text{IBP}}\frac{N}{D_1^{n_1} D_2^{n_2}\cdots D_{11}^{n_{11}} D_{\delta}}\right)\Big|_{\text{cut}(D_{\delta})} \nonumber\\
     &=\int_0^1 \text{d}x_1 \cdots \text{d}x_4   \frac{P}{D_1^{n_1'} D_2^{n_2'}\cdots D_{11}^{n_{11}} D_{\delta}} \Big|_{\text{cut}(D_{\delta})}
    \label{IBP4p}
\end{align}
Here, only the degree of propagator $D_1$ and $D_2$ can increase, since they are finite integrals. In \eqref{IBP4p}, $n_1' \geq n_1$, $n_2'\geq n_2$ and $n_i$ remains the same for $i \geq 3$. $P$ corresponds to the polynomial numerator in the result of acting $\mathcal{O}_\text{IBP}$ on integrand. If there is a linear transform between integral variables $x_i$ and propagators, we can rewrite this $P$ polynomial about $x_i$ as a polynomial about $D_i$ immediately just like IBP in Baikov representation. There is no linear relation between our polynomial propagators and integral variables. Some efforts can be made to simplify the IBP relation, rather than split the polynomial $P$ as the sum of monomials.

Since our IBPs are on $D_\delta$ cut, the polynomial $P$ is equivalent to the reminder $P'$ of $P$ modulo ideal $<D_\delta>$ at integral level. Represent $P'$ as the combination of propagators and necessary reminder polynomials. For the combination of propagators part, it can cancel some propagators and this relation show us the IBP relate top sector and sub sector integrals together. It is understandable that an additional reminder polynomial is sometimes necessary, since $P'$ could not be in the union ideal of the ideals generated by $D_i$ in denominator. Usually we can expect the reminder polynomial to have a lower degree than $P'$. 

These two ways to simplify IBP polynomial $P$ can be achieved by polynomial reduction with a specific monomial ordering. No matter what kind of monomial ordering we choose, there will be some high-degree monomials that cannot be reduced with respect to some three propagators and $D_\delta$ in a family. Our IBP system aims not only to relate high-sector and lower-sector integrals together, but also to represent integrals with a high-degree numerator to the combination of integrals with a lower-degree numerator.

\paragraph{Additional Relation at Integrand Level}
\label{subsec:additional relation}
To overcome the high-degree numerator that cannot be solved by polynomial reduction, we introduce some relations at integrand level.

\begin{align}
    \text{Int}[\{n_1,n_2,\dots,n_{11},1 \},N]&=\sum_{m\in \text{monomials in } D_i}c_i \text{Int}[\{n_1,\cdots,n_i+1,\cdots,n_{11},1 \},m N], i={1,2} \nonumber \\
    0&=\sum_{m\in \text{monomials in } D_\delta}c_i \text{Int}[\{n_1,n_2,\dots,n_{11},1\},m N] \nonumber
\end{align}
where $c_i$ is the coefficient correspond to monomial $m$ in $D_1$ or $D_2$ or in $D_\delta$.
These relation at integrand level can kill integrals with high degree numerator, such as
\begin{align}
    \text{Int}[\{2, 0, 0, 0, 1, 0, 0, 0, 0, 1, 0, 1\}, x_3^3],\,\text{Int}[\{2, 0, 2, 0, 0, 0, 0, 0,  0, 1, 0, 1\}, x_3x_4^3],\,  \cdots \nonumber
\end{align}
\paragraph{Boundary IBP}
The boundary terms of 4-point energy correlator integrals are some integrals similar to 3-point energy correlator. The boundary IBP story is also the same story with 3-point case, we don't go into details here.

\subsection{Finite Differential Equation}\label{sec:diffeqexample}
The method of finite differential equation is the same as 3-point case. For some families that contain only one polynomial propagator, for example $\{D_1,D_5,D_{10}\}$, we can get complete analytic differential equation. For some complicated families, it is difficult to get a complete analytic differential equations over parameter $z_i$ in \eqref{4pointPara}, not only due to the increase in parameter number, but also because of the difficulty of polynomial propagators. Like sector $\{D_{11},D_{12}\}$, we can get a semi-analytic differential equations with respect to a specific variable, say $z_2$ analytically, and others numerically. It still help a lot to conform the type of functions by Picard-Fuchs operators. We illustrate this part in section \ref{Picard-Fuchs Operators}.

To show the properties of $\mathrm{E^4C}$ functions, we give differential equations that truncated the boundary integrals. There are three examples for different types of functions, the complete analytic differential equations for family $\{D_1,D_5,D_{10}\}$ and the semi-analytic differential equations for sector $\{D_6,D_9\}$ and $\{D_{11},D_{12}\}$.

Family $\{D_1,D_5,D_{10}\}$ is a MPL family, sector $\{D_6,D_9\}$ is an elliptic sector and $\{D_{11},D_{12}\}$ is a higher-elliptic sector, their four-fold master integrals are
\begin{align}
  \{1,5,10\}: \quad & \text{Int}[\{0, 0, 0, 0, 0, 0, 0, 0, 0, 1, 0, 1\}, 1], \text{Int}[\{1, 0, 0, 0, 1, 0, 0, 0, 0, 0,0, 1\}, 1],  \nonumber\\
    & \text{Int}[\{1, 0, 0, 0, 1, 0, 0, 0, 0, 1, 0, 1\}, x_1], \text{Int}[\{1, 0, 0, 0, 1, 0, 0, 0, 0, 1, 0, 1\}, x_2]; \nonumber\\
    \{6,9\}:\qquad  &\text{Int}[\{0,0,0,0,0,1,0,0,1,0,0,1\},1],\text{Int}[\{0,0,0,0,0,1,0,0,1,0,0,1\},x_1],\nonumber\\
&\text{Int}[\{0,0,0,0,0,1,0,0,1,0,0,1\},x_2],\text{Int}[\{0,0,0,0,0,1,0,0,1,0,0,1\},x_3],\nonumber\\
&\text{Int}[\{0,0,0,0,0,1,0,0,1,0,0,1\},x_1^2]; \nonumber\\
\{11,12\}:\quad \,\, &\text{Int}[\{0,0,0,0,0,0,0,0,0,1,1,1\},x_2],
    \text{Int}[\{0,0,0,0,0,0,0,0,0,1,1,1\},x_3],\nonumber\\
    &\text{Int}[\{0,0,0,0,0,0,0,0,0,1,1,1\},x_2^2],
   \text{Int}[\{0,0,0,0,0,0,0,0,0,1,1,1\},x_3^2],\nonumber\\
   &\text{Int}[\{0,0,0,0,0,0,0,0,0,1,1,1\},x_1
   x_2],\text{Int}[\{0,0,0,0,0,0,0,0,0,1,1,1\},x_1
   x_3], \nonumber\\
    &\text{Int}[\{0,0,0,0,0,0,0,0,0,1,1,1\},x_1^2 x_2],
   \text{Int}[\{0,0,0,0,0,0,0,0,0,1,1,1\},x_1
   x_2^2],\nonumber\\
   &\text{Int}[\{0,0,0,0,0,0,0,0,0,1,1,1\},x_2 x_4].\nonumber
\end{align} 
We give the analytic differential equation matrix for $\{1,5,10\}$ and the semi-analytic differential equation matrix for $\{6,9\}$ and $\{11,12\}$ in the ancillary directory. The complexity of differential equations increases dramatically from MPL to higher-elliptic.


\section{Analysis of Master Integrals for 4-point Energy Correlators}

In the previous section, we have obtained irreducible integrals of the 4-point energy correlator in any angle scattering. In terms of these master integrals, some of the results are multiple polylogarithms (MPL), and the others are elliptic or even hyperelliptic integrals. We would first introduce the classifications of integrals and their forms. Then we move to the traditional cut method and leading singularity method to classify integrals. We also introduce Picard-Fuchs operators and perform the algorithm on the master integrals to ensure the validity of maximal cut or leading singularities. We will summarize the master integrals and their function types. For multiple polylogarithmic integrals, we introduce an approach, \textit{symbol integrations}, to calculate their symbols.

\subsection{Classification of Functions}

In order to fully calculate these master integrals, the first step is to classify them into multiple polylogarithms, elliptic integrals, or even beyond. A classic approach to classifying the type of functions of scattering amplitudes is to calculate the cuts \cite{Bosma:2017ens,Primo:2016ebd,Primo:2017ipr} or leading singularities \cite{Henn:2020lye,Bourjaily:2021vyj,Frellesvig:2021hkr} of the integrals, which can also be adapted in calculations of energy correlators. We assume that the number of propagators $k$ \footnote{Here we assume that the delta function is one of the propagator, since integration of a delta function is similar to take a residue with the propagator denoted by the function inside the delta function.} and the number of integration variables $n$ have the relation $k\geq n-1$. After cuts, the integrals are expected to have the form
\begin{equation}\label{eq:intdefine}
    \int {N(x)\mathrm{d}x\over\sqrt{P(x)}},
\end{equation}
where $N(x)$, $P(x)$ are some polynomials. If the degree of $P(x)$ is no more than two, the square roots can always be rationalized and the integration should be multiple polylogarithms. If the degree of $P(x)$ is more than two, the answer can be elliptic or beyond\footnote{When the degree of $P(x)$ is three, there are two cases. If the curve $y^2=P(x)$ is a rational curve, the answer should still be multiple polylogarithms. However, if the curve is irrational, which means that there are no singular points, the answer should be elliptic}. These functions are well studied in the Feynman diagrams and the scattering amplitudes aspect, see like \cite{Broedel:2017kkb,Huang:2013kh,Georgoudis:2015hca,Broedel:2018qkq,Bourjaily:2017bsb,Broedel:2019hyg}.

A simplification of this calculation is to solve this set of propagators in one variable and to see if there are degree-four polynomials in the square roots \cite{He:2024hbb}. For example, consider a master integral with propagators $D_7,\,D_9$ and $D_{\delta}$
\begin{equation}
    \frac{\delta(D_\delta)}{D_7D_9}={\delta(1-x_{1234}-s_{1234})\over {s_{125}s_{345}}}.
\end{equation}
The cut condition equations are
\begin{equation}
    \{1-x_{1234}-s_{1234}=0,\quad s_{125}=0,\quad s_{345}=0\}.
\end{equation}
After solving the equation as for any variable, let us choose $x_4$ without ambiguity, the solution involves a term
\begin{equation}
    x_1={{\cdots+\sqrt{(\zeta_{14}^2\zeta_{34}^2-2\zeta_{14}\zeta_{24}\zeta_{34}^2+4\zeta_{12}\zeta_{14}\zeta_{24}\zeta_{34}^2+\zeta_{24}^2\zeta_{34}^2)x_4^4+\cdots}}\over {2(-\zeta_{12}\zeta_{13}+x_4\zeta_{12}\zeta_{13}-x_4\zeta_{12}\zeta_{14}+x_4^2\zeta_{12}\zeta_{14}\zeta_{34})}}.
\end{equation}
This means if we directly calculate the integral (with a specific integration order)\footnote{We need to choose a correct integration order, since the propagator may be linearly dependent in one variable and integration over this variable is much easier than others.}, we will meet the integration like
\begin{equation}
    \int{{N(x_4)\,\mathrm{d}x_4}\over{\sqrt{(\zeta_{14}^2\zeta_{34}^2-2\zeta_{14}\zeta_{24}\zeta_{34}^2+4\zeta_{12}\zeta_{14}\zeta_{24}\zeta_{34}^2+\zeta_{24}^2\zeta_{34}^2)x_4^4+\cdots}}},
\end{equation}
where $N(x_4)$ is some weight two functions (without square roots or the same square root with the denominator) and this will be an elliptic integral. The real calculation matches our predictions, and thus this master integral should depend on an elliptic curve. For an example with four propagators
\begin{equation}
    \frac{\delta(D_\delta)}{D_1D_7D_9}={\delta(1-x_{1234}-s_{1234})x_4\over {s_{45}s_{125}s_{345}}},
\end{equation}
we should calculate all possible three propagator subsets (within delta function)
\begin{equation}
    \begin{split}
        &\{1-x_{1234}-s_{1234}=0,\quad s_{125}=0,\quad s_{345}=0\},\\
        &\{1-x_{1234}-s_{1234}=0,\quad s_{345}=0,\quad s_{45}=0\},\\
        &\{1-x_{1234}-s_{1234}=0,\quad s_{125}=0,\quad s_{45}=0\}.
    \end{split}
\end{equation}
Because there is a subtopology that contains the first example, which is an elliptic integral, this example would definitely be an elliptic integral. The algorithm can also be done numerically, in which we can set all the kinematics into some random numbers. This can enormously reduce the calculation time, but does not change the conclusion in most cases. One may counter the situation that there will be a cubic root after cuts, such as the sector including $D_{10}$ and $D_{11}$. An alternative method is to calculate the leading singularities of the integrals. In this way, one can also judge the type of functions and find that the sector that contains $D_{10}$ and $D_{11}$ is a hyperelliptic integral with a genus $2$ curve.


\subsection{Picard-Fuchs Operators}
\label{Picard-Fuchs Operators}
One powerful tool to determine whether a sector of master integrals is elliptic or MPL is the Picard-Fuchs operators \cite{Muller-Stach:2012tgj,Adams:2017tga}. These operators denote the differential equation that an integral satisfies and can accurately judge the type of functions.

A sector of master integrals can be written as
\begin{equation}
\vec{f}(x)=\left(f_1(x),f_2(x),...,f_n(x)\right)^T,
\end{equation}
and satisfies differential equations
\begin{equation}
\frac{d\vec{f}(x)}{d x}=\textbf{A}(x)\vec{f}(x),
\end{equation}
where $\mathbf{A}(x)$ is an $n\times n$ matrix function. We would like to find a higher-order differential equation of a single master integral, for instance $f_1(x)$. Its first derivative is
\begin{align}
f'_1(x)=&\sum_{i=1}^n\mathbf{A}_{1i}(x)f_i(x)
\\=&\alpha_{(1,1)}(x)f_1(x)+\alpha_{(1,2)}(x)f_2(x)+...\alpha_{(1,n)}(x)f_n(x),
\end{align}
where the first index of $\alpha_{(k,l)}$ indicates the order of derivative and the second for the expansion coefficient of $f_l(x)$. Differentiating with respect to $f^\prime_1(x)$ gives the second derivative
\begin{equation}
f^{''}_1(x)=\frac{d}{d x}f'_1(x)=\sum_{j=1}^n(\alpha'_{(1,j)}(x)f_j(x)+\alpha_{(1,j)}(x)f'_j(x)).
\end{equation}
For $f'_j(x)$, we can again use a similar substitution like $f'_1(x)$. Continuing the same process, we get a form like
\begin{equation}
f^{(m)}_1(x)=\alpha_{(m,1)}(x)f_1(x)+\alpha_{(m,2)}(x)f_2(x)+...\alpha_{(m,n)}(x)f_n(x).
\end{equation}

By treating these $\alpha_{(m,j)}(x)$ as a function of the coefficients, the derivatives of each order $f^{(m)}_1$ can be written together as
\begin{equation}
\vec{g}(x)
\;=\;
\begin{pmatrix}
f_{1}'(x)\\[6pt]
f_{1}''(x)\\
\vdots\\
f_{1}^{(r)}(x)
\end{pmatrix}
\;=\;
\mathbf{B}(x)\,
\begin{pmatrix}
f_{1}(x)\\
f_{2}(x)\\
\vdots\\
f_{n}(x)
\end{pmatrix}
\;=\;
\mathbf{B}(x)\,\vec{f}(x).
\end{equation}

In order to derive the Picard-Fuchs operator of $f_1(x)$, consider the first row of the equation $\vec{f}(x)=\mathbf{B}^{-1}(x)\vec{g}(x)$, we get
\begin{equation}
f_1(x)=\sum_{i=1}^n\mathbf{B}^{-1}_{1i}(x)g_i(x).
\end{equation}
So far, we have “converged” all dependencies on $f_2,...,f_n$ to a form containing only $f_1$ and its derivatives of each order, thus constructing a higher-order linear differential equation.
\begin{equation}
f_{1}^{(r)}(x)
\;-\;
\sum_{k=0}^{r-1} C_{k}(x)\,f_{1}^{(k)}(x)
\;=\;0,
\end{equation}
and the Picard-Fuchs operator should be 
\begin{equation}
   \hat{\mathcal{P}}= 1-\sum_{i=0}^{n-1}C_{i}(x)\frac{\mathrm{d}^i}{\mathrm{d}x^i}.
\end{equation}
If the matrix $\mathbf{B}$ is degenerate, one can calculate the null space of $\mathbf{B}$, i.e.
\begin{equation}
    \mathbf{N}_{k\times n}(x)\vec{g}_{n\times1}(x)=\mathbf{N}(x)\mathbf{B}(x)\vec{f}(x)=\vec{0}_{k},
\end{equation}
where $\mathbf{N}(x)$ is the null space matrix, and its scale depends on the rank of matrix $\mathbf{B}$. We can choose any row of $\mathbf{N}$ as the Picard-Fuchs operator
\begin{equation}
    \hat{\mathcal{P}}=\sum_{i=1}^n\mathbf{N}_{li}(x)\frac{\mathrm{d^{i-1}}}{\mathrm{d}x^{i-1}},\qquad \forall\,l\in\{1,2,\cdots,k\}
\end{equation}

After obtaining the Picard-Fuchs operator, one should check whether it is factorizable to determine the type of integrals. If the Picard-Fuchs operator can be completely factorized into a product of several first-order operators, then the corresponding solution can often be expressed in terms of polylogarithms, because the solution of a first-order equation is usually in the form of a simple exponential or logarithmic integral. If all equations are of first order, the overall solution can be iteratively expressed in terms of logarithmic or similar forms of closed form.
If Picard-Fuchs contains irreducible second-order derivative factors, it implies that the solution may contain elliptic integrals.

For example, we use Picard-Fuchs operators to determine whether the sector $\frac{\delta(D_\delta)}{D_7D_9}$ is elliptic or MPL, where there are nine master integrals. However, when analyzing the coefficient matrix, the first three integrals do not depend on the remaining integrals and this means one can treat the first three integrals as subsectors of the remaining six integrals. Therefore, it's reasonable to consider the first three master integrals
\begin{equation}
\vec{f}(z_2)=\left( \frac{\delta(D_\delta)}{D_7D_9}, \frac{x_1\delta(D_\delta)}{D_7D_9}, 
\frac{x_1^2\delta(D_\delta)}{D_7D_9}\right)^T.
\end{equation}
Its differential equations as to $z_2$ parameter are
\begin{equation}
\frac{d\vec{f}(z_2)}{d z_2}=\textbf{A}(z_2)\vec{f}(z_2).
\end{equation}
We take the second master integral $ f_2(z_2)=\frac{x_1\delta(D_\delta)}{D_7D_9}$ as an example, and we write the differential equation with respect to it as
\begin{equation}
\frac{d{f_2}(z_2)}{d z_2}=\textbf{A}_2(z_2)\vec{f}(z_2),
\end{equation}
where $\textbf{A}_2$ is the second row of matrix $\textbf{A}$. After calculating the higher derivative of $f_2(z_2)$, we get
\begin{equation}
\begin{pmatrix}
f_{2}'(z_2)\\[6pt]
f_{2}''(z_2)\\[6pt]
f_{2}^{(3)}(z_2)
\end{pmatrix}
\;=\;
\begin{pmatrix}
\textbf{A}_2\\[6pt]
\textbf{A}'_2+\textbf{A}_2.\textbf{A}\\[6pt]
\textbf{A}''_2+2(\textbf{A}'_2.\textbf{A})+(\textbf{A}_2.\textbf{A}')+(\textbf{A}_2.\textbf{A}.\textbf{A})
\end{pmatrix}
\begin{pmatrix}
f_{1}(z_2)\\[6pt]
f_{2}(z_2)\\[6pt]
f_{3}(z_2)
\end{pmatrix}
\;\triangleq\;
\mathbf{B}(z_2)\,\vec{f}(z_2).
\end{equation}
Inverse the matrix $\textbf{B}$ and the second row is the differetial equation corresponding to $f_2(z_2)$

\begin{equation}
f_{2}(z_2)-\textbf{B}^{-1}_2(z_2)\,
\begin{pmatrix}
f_{2}'(z_2)\\[6pt]
f_{2}''(z_2)\\[6pt]
f_{2}^{(3)}(z_2)
\end{pmatrix}
\;=\;0,
\end{equation}
where $\textbf{B}^{-1}_2$ is the second row of matrix $\textbf{B}^{-1}$, and the related Picard Fuchs operator is 
\begin{equation}
    \hat{\mathcal{P}}_{f_2}=1-\textbf{B}_{2}^{-1}\cdot\left(\frac{\mathrm{d}}{\mathrm{d}z_2},\frac{\mathrm{d}^2}{\mathrm{d}z_2^2},\frac{\mathrm{d}^3}{\mathrm{d}z_2^3}\right)^\textbf{T}.
\end{equation}

Using maple's DFactor function, the Picard-Fuchs operators of $f_2(z_2)$ can be factorized into the following form
\begin{equation}
\hat{\mathcal{P}}_{f_2}=\left(c_1\frac{\mathrm{d}^2}{\mathrm{d}z_2^2}+c_2\frac{\mathrm{d}}{\mathrm{d}z_2}+c_3\right)\left(\frac{\mathrm{d}}{\mathrm{d}z_2}+c_4\right).
\end{equation}
From the result, the third-order differential operator can be decomposed into a second-order and first-order differential operator. Therefore, the second integral in this family is an elliptic integral. Similar calculations can be performed for other master integrals in this family. 

Similarly, for the sector $\frac{\delta(D_\delta)}{D_6D_{11}}$, it also contains three master integrals
\begin{equation}
\vec{f}(z_2)=\left( \frac{x_2\delta(D_\delta)}{D_6D_{11}}, \frac{x_3\delta(D_\delta)}{D_6D_{11}}, 
\frac{x_4\delta(D_\delta)}{D_6D_{11}}\right)^T.
\end{equation}
The Picard-Fuchs operators of $f_1(z_2)$ can be factorized into the following form
\begin{equation}
\left(c_1\frac{\mathrm{d}}{\mathrm{d}z_2}+c_2\right)\left(\frac{\mathrm{d}}{\mathrm{d}z_2}+c_3\right)\left(\frac{\mathrm{d}}{\mathrm{d}z_2}+c_4\right).
\end{equation}
The result shows that it can be factored into a product of three first-order operators, so it is an MPL integral.

\subsection{Summary of Master Integrals}
In this section, we summarize all the master integrals of four-point energy correlators and their types of functions. There are 55 sectors for 4-point energy correlators in any angle scattering. For these sectors, there are also symmetries between each other. For simplicity, we give a list of sectors after modding the symmetries between sectors. There are two different types of elliptic integrals and one hyperelliptic integral of genus 2, and the other sectors are multiple polylogarithms. We use the propagators defined in (\ref{4pProp}). The reader should be reminded that there will be a delta function in the integrand of each master integral, and we omit the delta function here.
\begin{longtable}{c|c|c|c}
\hline
Propagators  & Count &  Numerators  &  Type of Functions\\
\hline
\endhead

\hline
\endfoot
        $D_{10}$ & 2 & $1,x_1$ & MPL \\
        $D_{9}$ & 3 & $1,x_1,x_2$ & MPL \\
        $D_{3}$  & 1 & 1 & MPL \\
        $D_{8},D_{11}$ & 6 & $1,x_1,x_2,x_2^2,x_3^2,x_2x_4$ & MPL \\
        $D_{4},D_{10}$ & 3 & $1,x_1,x_3$ & MPL \\
        $D_{4},D_{5}$ & 1 & 1 & MPL\\
        $D_{5},D_{8}$ & 3 & $1,x_2,x_3$ & MPL\\
        $D_6,D_{11}$ & 3 &  $x_2,x_3,x_4$ & MPL\\
        $D_3,D_6$ & 1 & 1 & MPL\\
        $D_6,D_9$ & 5 & $1,x_1,x_1^2,x_2,x_3$ & Elliptic\\
        $D_7,D_9$ & 9 & $1,x_1,x_1^2,x_2,x_3,x_1x_3,x_2x_3,x_4,x_1x_4$ & Elliptic \\
        $D_8,D_9$ & 4 & $1,x_1,x_2,x_4$ & MPL\\
        $D_2,D_6$ & 1 & 1 & MPL\\
        $D_{10},D_{11}$ & 9 & $x_2,x_1x_2,x_1^2x_2,x_2^2,x_1x_2^2,x_3,x_1x_3,x_3^2,x_2x_4$ & Hyperelliptic $g=2$\\
        $D_9,D_{10}$ & 5 & $x_1,x_2,x_3,x_1x_4,x_3x_4$ & MPL\\
        $D_2,D_{10}$ & 3 & $1,x_1,x_2$ & MPL \\
        $D_1,D_5$ & 1 & 1 & MPL \\
        $D_1,D_9$ & 3 & $1,x_1,x_2$ & MPL\\
        $D_1,D_2$ & 1 & 1 & MPL \\
        $D_5,D_6,D_{11}$ & 3 & $x_2,x_1x_2,x_3$ & MPL\\
        $D_2,D_5,D_6$ & 2 & $1,x_1$ & MPL\\
        $D_5,D_{10},D_{11}$ & 2 & $x_1x_2,x_2^2$& Hyperelliptic $g=2$\\
        $D_4,D_6,D_{10}$ & 2 & $x_1,x_2$ & MPL \\
        $D_5,D_8,D_{10}$ & 4 & $x_1,x_1^2,x_2,x_1x_2$& MPL \\
        $D_6,D_7,D_{10}$ & 6 & $x_1,x_1^2,x_1^3,x_1x_2,x_1x_3,x_2x_4$ & Elliptic \\
        $D_6,D_9,D_{10}$ & 3 & $x_1^2,x_2^2,x_1x_2$ & Elliptic \\
        $D_2,D_4,D_{10}$ & 2 & $1,x_1$ & MPL\\
        $D_2,D_5,D_{10}$ & 2 & $x_1,x_2$ & MPL \\
        $D_2,D_9,D_{10}$ & 4 & $x_1,x_2,x_3,x_3^2$ & MPL \\
        $D_1,D_8,D_{11}$& 4 & $1,x_1,x_2,x_4$ & MPL \\
        $D_1,D_4,D_{5}$ & 2 & $1,x_1$ & MPL \\
        $D_1,D_5,D_8$ & 4 & $1,x_1,x_1^2,x_2$ & MPL \\
        $D_1,D_6,D_{11}$& 5 & $x_2,x_1x_2,x_1^2x_2,x_3,x_4$ & MPL \\
        $D_1,D_6,D_8$ & 4 & $1,x_1,x_1^2,x_2$ & Elliptic \\
        $D_1,D_4,D_{9}$ & 2 & $1,x_1$ & MPL \\
        $D_1,D_7,D_9$ & 4 & $1,x_1,x_2,x_4$ & Elliptic\\
        $D_1,D_8,D_9$ & 4 & $1,x_1,x_2,x_4$ & MPL\\
        $D_1,D_2,D_{4}$ & 2 & $1,x_1$ & MPL\\
        $D_1,D_2,D_5$ & 2 & $1,x_1$ & MPL \\
        $D_1,D_2,D_9$ & 4 & $1,x_1,x_1^2,x_2$ & MPL \\
        $D_1,D_{10},D_{11}$ & 3 & $x_2,x_3,x_1x_4$ & Hyperelliptic g=2 \\
        $D_1,D_5,D_{10}$ & 2 & $x_1,x_2$ & MPL\\
        $D_1,D_9,D_{10}$ & 3 & $x_1,x_2,x_3$ & MPL\\
        $D_1,D_2,D_{10}$& 2 & $1,x_1$ & MPL\\
\hline

\end{longtable}

Using the package \textsc{SOFIA} \cite{Correia:2025yao}, we calculate the preliminary symbol alphabets of four-point energy correlators. The results involving only three points are shown here, and the complete results are in the appendix \ref{ap:symbol}. 
In order to compare with the result of $\mathrm{E^3C}$, we have used the parameterization \eqref{para 2}. 
\begin{align}
 s&, x_1, x_2, 1 - x_1, 1 - x_2, 1 + s, 1 - s, s + x_1, s + x_2, 
 1 + s x_1, 1 + s x_2, x_1 - x_2, s + x_1 x_2, \notag \\
1& - x_1 x_2,  1 + s x_1 x_2, 1 - x_1^2 x_2, 1 - x_1 x_2^2, s x_1 + s x_2 + x_1 x_2 - 2 s x_1 x_2 + s^2 x_1 x_2 + s x_1^2 x_2 + s x_1 x_2^2 \nonumber
\end{align}
It completely covers the results for 3-point energy correlator.  For the new items $1+s x_2, 1-x_1 x_2^2$, their appearance indicates that the $S_3$ symmetry moduled in the 3-point case is now recovered. The remaining new items $s x_1 + s x_2 + x_1 x_2 - 2 s x_1 x_2 + s^2 x_1 x_2 + s x_1^2 x_2 + s x_1 x_2^2$, $x_1-x_2$ and $1-s$ correspond to 
\begin{equation}
-1 + \zeta_{12} + \zeta_{13},\hspace{0.15cm} \zeta_{12}-\zeta_{13}, \hspace{0.15cm} \zeta_{12}^2 - 2 \zeta_{12} \zeta_{13} + \zeta_{13}^2 - 
 2 \zeta_{12} \zeta_{23} - 2 \zeta_{13} \zeta_{23} + 4 \zeta_{12} \zeta_{13} \zeta_{23} + \zeta_{23}^2\,, \nonumber
\end{equation}
which are absent in the case of $\rm{E^3}C$.

\subsection{Symbol Integrations}

In the last part of this section, we introduce a method for calculating symbols of some MPL integrals. Symbol alphabet of four-point energy correlators can be calculated by package \textsc{SOFIA} \cite{Correia:2025yao}, and we list the result in the appendix \ref{ap:symbol}. Some master integrals can be directly calculated by \texttt{HyperInt} \cite{Panzer:2014caa}, while most integrals will be stuck after integrating over one or two variables. Consequently, a compromised method is to calculate their symbols. We introduce a simple version of symbol integrations based on an extension of projection approach to symbol calculations \cite{Gong:2022erh} and tightly related to the Landau analysis \cite{Dennen:2015bet,Caron-Huot:2024brh,Fevola:2023kaw}. A complete version of symbol integration, one can see \cite{Zhenjie:2025}.

For integrals in the sector like 
\begin{equation}
\frac{\delta(D_\delta)}{D_8D_9},\hspace{1cm}\frac{\delta(D_\delta)}{D_4D_{9}},\hspace{1cm}\frac{\delta(D_\delta)}{D_5D_8},
\end{equation}
One can easily integrate over two variables, while the last integration is hard to perform. This means we know both the analyticity (symbols and poles) of the integrand\footnote{Here we slightly abuse the notation. The notation in the numerator means the numerator of the integrand has symbol $\sum_{S}S_1(x_k;\zeta_{ij})\otimes S_2(x_k;\zeta_{ij})$} and the integral contour
\begin{equation}\label{eq:intbegin}
    I(\zeta_{ij})=\int_{0}^1 \mathrm{d}x_k \frac{\sum_{S}S_1(x_k;\zeta_{ij})\otimes S_2(x_k;\zeta_{ij})}{P(x_k)},
\end{equation}
where $k$ labels the last variable to integrate and $P(x)$ is a polynomial and can be factorized into $P(x)=(x-x_1)(x-x_2)\cdots (x-x_n)$. From both Landau analysis and projection views, when the integrand is a rational function, analyticities after integration come from end-point singularities, which means $P(0)=0$ or $P(1)=0$. This is because when we deform the parameters, the contour also deforms, contributing to a $S^1$ contour around the pole. The integration over this circle is the residue and also the discontinuity of the integral. The symbol satisfies the following relation
\begin{equation}
    \mathcal{S}(\mathrm{Disc}(\mathcal{I}))=S_2\otimes\cdots\otimes S_n,
\end{equation}
if the original integral has the symbol $\mathcal{S}(\mathcal{I})=S_1\otimes S_2\otimes\cdots\otimes S_n$. For instance, a pure Log function 
\begin{equation}
    I=\int_0^\infty\frac{(r_2-r_1)\mathrm{d}x}{(x-r_1)(x-r_2)}=\log\left({r_1\over r_2}\right),
\end{equation}
whose symbol is $\otimes\,r_1/r_2$. This letter comes from $r_1$ or $r_2$ meet the endpoints of integral contour $0$ or $\infty$. The discontinuity of this integral is $2\pi i$, which does not contribute to any symbol.

However, in the $x_k$-plane of eq. (\ref{eq:intbegin}), in addition to the poles, the integrand also has a branch cut between $S_1=0$ and $S_1=\infty$. Note that when the contour deforms, the contour can also wind the branch point. This means that one can also have first entries that describe the coincidence of the branch points and the integral contour. For example, for the cases in (\ref{eq:intbegin}), assuming that $S_1(x_k;\zeta_{ij})=0$ and $S_1(x_k;\zeta_{ij})=\infty$ separately give solutions $x_k=x_0^{(1)}$ and $x_k=x_0^{(2)}$ (which are the positions of two branch points in the $x_k$-plane), the related first entires should be $x_0^{(1)}$, $x_0^{(1)}-1$, $x_0^{(2)}$ and $x_0^{(2)}-1$. Now, the discontinuites are contributed by the branch cut, but not the residue. Therefore, discontinuities can be expressed as an integration along the branch cut and set the original boundary as a new boundary of the integral. 
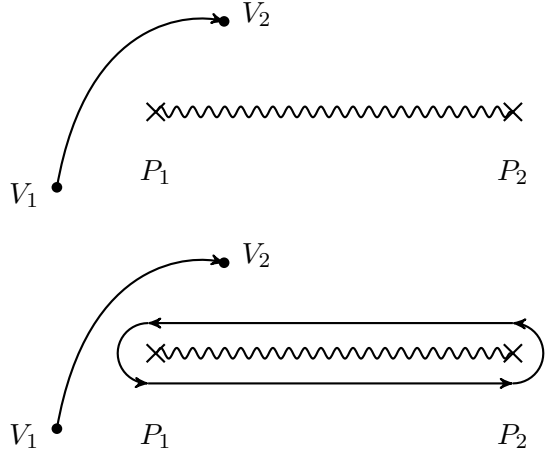
\begin{figure}[htbp]
  \centering
  \tikzset{ 
  photon/.style = {decorate, decoration={snake, amplitude=2pt, segment length=6pt}}, 
  cross/.style = {cross out, draw, thick, inner sep=1pt, minimum size=5pt},  midarrow/.style = { postaction = {decorate}, decoration = {markings, 
  mark = at position 0.5 with {\arrow{Stealth[length=6pt]}} } } }
  \begin{tikzpicture}[>={stealth'}]
    \begin{scope}
      \fill (-1,-1) circle (2pt);
      \fill (1.2,1.2) circle (2pt);
      \node[left] at (-1.1,-1.1) {$V_1$};
      \draw[photon,thick] (0.3,0) -- (5,0);
      \node[cross,minimum size=6pt,thick] at (0.3,0) {};
      \node[cross,minimum size=6pt,thick] at (5,0) {};
      \node[right] at (1.3,1.3) {$V_2$};
      \draw[->, thick] (-1.0,-1.0)
           .. controls (-0.6,1.3) and (0.8,1.3)
           .. (1.2,1.2);
      \node[below]  at (0.3,-0.5) {$P_1$};
      \node[below]  at (5,-0.5)   {$P_2$};
    \end{scope}

    \begin{scope}[yshift=-3.2cm]
      \fill (-1,-1) circle (2pt);
      \fill (1.2,1.2) circle (2pt);
      \node[left] at (-1.1,-1.1) {$V_1$};
      \draw[photon,thick] (0.3,0) -- (5,0);
      \node[cross,minimum size=6pt,thick] at (0.3,0) {};
      \node[cross,minimum size=6pt,thick] at (5,0) {};
     \draw[->,thick](0.2,-0.4) -- (5,-0.4);
     \draw[->, thick] (-1.0,-1.0)
           .. controls (-0.6,1.3) and (0.8,1.3)
           .. (1.2,1.2);
     \draw[->, thick, domain=180:0] plot ({5+0.4*sin(\x)},{0.4*cos(\x)});
      \node[right] at (1.3,1.3) {$V_2$};
      \draw[->, thick] (5,0.4) -- (0.2,0.4);
      \draw[->, thick, domain=0:180] plot ({0.2-0.4*sin(\x)},{0.4*cos(\x)});
      \node[below] at (0.3,-0.8) {$P_1$};
      \node[below] at (5,-0.8)   {$P_2$};
    \end{scope}
  \end{tikzpicture}
  \caption{Contour deformation if there is a branch cut in the plane}
\end{figure}

Therefore, the discontinuities of eq. (\ref{eq:intbegin}) have the symbol
\begin{equation}
    \mathcal{S}(\mathrm{Disc}(I(\zeta_{ij})))=\int_C\frac{\otimes S_2(x_k;\zeta_{ij})\mathrm{d}x_k}{P(x_k)}.
\end{equation}

In fact, we can generally write a recursive relation of symbol integrations. Consider an integral has the form
\begin{equation}
I=\int_{a}^{b}\frac{A_{1}(t)\otimes A_{2}(t)…\otimes A_{n}(t)}{t-c}\mathrm{d}t,
\end{equation}
after integration its symbol should be
\begin{align}
    \mathcal{S}(I)=&\frac{c-b}{c-a}\otimes A_{1}(c)\otimes A_{2}(c)…\otimes A_{n}(c)\notag\\
    &+\sum_{e}A(b)\otimes \left(\int_{e}^{b}\frac{1}{t-c} A_{2}(t)\otimes A_{3}(t)…\otimes A_{n}(t)\right)\\
    &+\sum_{e}A(a)\otimes \left(\int_{a}^{e}\frac{1}{t-c} A_{2}(t)\otimes A_{3}(t)…\otimes A_{n}(t)\right),\notag
\end{align}
where $A(a)$ and $A(b)$ indicate the condition that the point $a$ or $b$ coincide with the branch point of $A_1(t)$, and $e$ indicates the location of branch points. The first line shows the contribution of poles to the symbol and the second and third lines show the branch cut contribution.

Let us take a simple example. Assuming the numerator of the integrand is a logarithmic function, whose branch points are $t=d$ and $t=\infty$, and pole is $t=c$, and integral contour is from $a$ to $b$,
\begin{equation}
I_{\mathrm{e.g.}}=\int_{a}^{b}\frac{1}{t-c}\log(t-d) \mathrm{d}t
\end{equation}
Following the above analysis, its symbol can be written directly\footnote{Since one branch point is $t=\infty$, it contributes to the vanished symbol terms.},
\begin{align}
\mathcal{S}(I_{\mathrm{e.g.}})&=\frac{c-b}{c-a}\otimes(c-d)+(d-b)\otimes\left(\int_{d}^b\frac{\mathrm{d}t}{t-c}\right)+(d-a)\otimes\left(\int_{a}^d\frac{\mathrm{d}t}{t-c}\right)\notag\\
&=\frac{c-b}{c-a}\otimes(c-d)+(d-b)\otimes\frac{c-b}{c-d}+(d-a)\otimes\frac{c-d}{c-a}.
\end{align}
In four-point energy correlators, the final answer is weight-3 functions. Therefore, the master integrals that cannot be computed in the last step have the form\footnote{For integrands whose denominator is higher degree polynomials, they can always be written into forms in eq. (\ref{eq:intform}) by using relations like
\begin{equation*}
    \frac{1}{(x-a)(x-b)}=\frac{1}{a-b}\left(\frac{1}{x-b}-\frac{1}{x-a}\right).
\end{equation*}}
\begin{equation}\label{eq:intform}
I_{\mathrm{MI}}=\int_{a}^{b}\frac{1}{t-c} (t-d)\otimes(t-e) \mathrm{d}t,
\end{equation}
We can directly give a result in general
\begin{align}
&\mathcal{S}(I_{\mathrm{MI}})=\frac{c-b}{c-a}\otimes(a-d)\otimes(a-e)\\
&+(d-b)\otimes\frac{c-b}{c-d}\otimes(c-e)+(d-b)\otimes(e-b)\otimes\frac{c-b}{c-e}+(d-b)\otimes(e-d)\otimes\frac{c-e}{c-d}\notag\\
&+(d-a)\otimes\frac{c-d}{c-a}\otimes(c-e)+(d-a)\otimes(e-d)\otimes\frac{c-d}{c-e}+(d-a)\otimes(e-a)\otimes\frac{c-e}{c-a}\notag.
\end{align} 
This method can be performed for calculating symbols recursively. 

We should emphasize that there should not be square root branch cuts inside the symbols in the numerator of the integrand, because it will bring extra branch cuts. One direct idea is to first rationalize the square roots, and then one can do the symbol integration. However, this will bring more difficulties.

\section{Summary and Outlook}

In this paper, we introduced a systematic algorithm for computing the differential equations of energy correlators in any angle scattering. We mainly focus on three- and four-point energy correlators. For the three-point case, we calculated its canonical differential equations. The algorithm can be summarized in the following steps:
\begin{itemize}
    \item Calculating form factor squares of four points,
    \item Using symmetries to simplify the integrand,
    \item Using partial fraction decomposition to apart the integrand into some terms with monomial integrand and simplified propagators,
    \item Check the divergence of these integrals. For the divergent integrals, find the relations
    \begin{equation}
        \sum_n I_{n}^\infty=\sum_k I_k^\mathrm{finite}
    \end{equation}
    to make all the integrands finite,
    \item Finding integration by parts relations to master integrals,
    \item Recursively doing boundary IBPs,
    \item Calculating the canonical differential equations.
\end{itemize}
For the four-point energy correlators, there are some differences. We should emphasize that in addition to two extra parameters and one more integration variable in the integral, the propagators and the delta function measurement also get more complicated. In the three-point case, all the propagators are linear with integration variables except for the momentum conservation in the delta function. However, in the four-point case, there are five quadric propagators and there are integrals with three of them in the denominator. This will bring complicated integrals. Therefore, we present some methods, including maximal cut, leading singularities and Picard-Fuchs operators, to classify their type of functions after integration. Surprisingly, for the four-point any angle scattering, we have found that there are two different kinds of elliptic curves and one hyperelliptic curve that the master integrals depend on. For those integrals that do not depend on elliptic curves, we calculate their symbols and provide an intuition of a symbol integration method that can recursively calculate the full symbol of an integral.

There are still some remaining problems. One aspect is that the canonical differential equations of four-point any angle scattering energy correlators are difficult to compute. An advanced tool is necessary for further calculations. From the function type side, energy correlators are a new family of integrals like Feynman integrals. Since they already have information of elliptic and hyperelliptic curves, it is appealing to reveal the regularity of this type integrals, even Calabi-Yau in the future. Last but not least, energy correlators are the observable that they share the same set of master integrals as $\mathcal{N}=4$ sYM. We also hope that our master integrals can help for further calculations in QCD.

\section{Acknowledgement}
We thank Wen Chen and Huaxing Zhu on the early participant in this project. We also thank Mathieu Giroux, Ian Moult,  Lecheng Ren, Lorenzo Tancredi, Fabian Wagner and Chi Zhang for enlightening discussions.
This work is supported in part by the National Natural Science Foundation of China (No. 12357077). 
R.M. is funded by the Outstanding PhD Students Overseas Study Support Program of University of Science and Technology of China.
Y.Z. is supported from the NSF of China through Grant No. 12247103.


\appendix


\section{Symbol Alphabets for MPL Integrals}\label{ap:symbol}

Using the SOFIA package, we get the preliminary result of symbols for the MPL integrals. The obtained symbol are reparameterized with the following parameterization
\begin{equation}
(z_1,z_2,z_3,z_4)=\left(0,\lambda,\lambda z,\frac{\lambda (z-\omega)}{1-\omega}\right).
\end{equation}
Compared to the collinear limit of the four-point energy correlator, we introduce an additional $z_2=\bar{z}_2=\lambda$, which implies scaling the quadrilateral by a factor $\lambda$.  So we still have 5 parameters ($z, \bar{z}, \omega, \bar{\omega}, \lambda$) as before. 

\begin{figure}[htbp]
\begin{center}
\begin{tikzpicture}[scale=1.2, line cap=round, line join=round, 
                    >=latex, 
                    font=\small] 
  \coordinate (Z1) at (0,0);    
  \coordinate (Z2) at (4,0);   
  \coordinate (Z4) at (1,2);   
  \coordinate (Z3) at (3,2.5); 

  \draw[thick] (Z1) -- (Z2) -- (Z3) -- (Z4) -- cycle;

  \draw[thick] (Z1) -- (Z3);
  \draw[thick] (Z2) -- (Z4);

  \node[below left]  at (Z1) {$z_{1} = 0$};
  \node[below right] at (Z2) {$z_{2} = \lambda$};
  \node[above right] at (Z3) {$z_{3}= \lambda z $};
  \node[above left]  at (Z4) {$z_{4}= \frac{\lambda (z-\omega)}{1-\omega} $};

\end{tikzpicture}
\end{center}

 \caption{Parametrized in terms of complex parameters ($z, \omega$) and real parameter $\lambda$}
 \label{fig:E4Cpara2}
\end{figure}
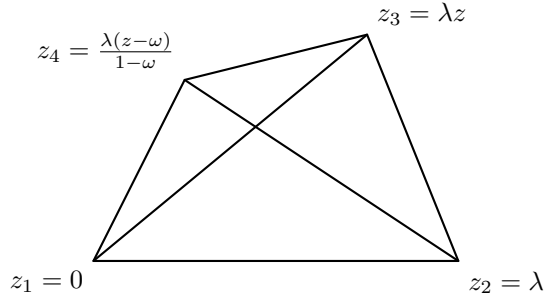


Considering the symmetry of exchanging two points of the quadrilateral, some symbol alphabets are equivalent. For these symbols, we only keep one of them. The final preliminary symbols by SOFIA are as follows:



\noindent
$\lambda, z, \bar{z}, w - z, \bar{z} - z, -1 + z, \bar{w} - \bar{z}, -1 + \bar{z}, 1 + \lambda^2, \
w - \bar{w} z, \bar{w} - \bar{z} w, -1 + \bar{z} z, 1 + \lambda^2 z, 1 + \bar{z} \lambda^2, \bar{z} w - \bar{w} z, 
\bar{w} w - \bar{z} z, 1 + \bar{z} \lambda^2 z, -1 + \bar{z} \lambda^4 z, 
\bar{w} w - \bar{z} w - \bar{w} z, -1 + w + \lambda^2 w - \lambda^2 z, -1 + \bar{z} + z + 
 \bar{z} \lambda^2 z, \bar{z} + z - \bar{z} z + \bar{z} \lambda^2 z, -1 + \bar{w} + \bar{w} \lambda^2 - 
 \bar{z} \lambda^2, -1 + w + \lambda^2 w - \bar{z} \lambda^2 z, -1 + \bar{w} + \bar{w} \lambda^2 - 
 \bar{z} \lambda^2 z, \bar{w} - w - \bar{z} \lambda^2 w + \bar{w} \lambda^2 z, -1 + w + \bar{z} \lambda^2 w - 
 \bar{z} \lambda^2 z, \bar{w} - z + \bar{w} \lambda^2 z - \bar{z} \lambda^2 z, -\bar{z} + w + \bar{z} \lambda^2 w - 
 \bar{z} \lambda^2 z, -1 + \bar{w} + \bar{w} \lambda^2 z - \bar{z} \lambda^2 z, -\bar{w} + \bar{z} + w - \bar{z} w - z + 
 \bar{w} z, 1 - \bar{w} w - \bar{w} \lambda^2 w + \bar{z} \lambda^2 z, 1 - \bar{w} z - \bar{w} \lambda^2 z + 
 \bar{z} \lambda^2 z, 1 - \bar{w} - w + \bar{z} w + \bar{w} z - \bar{z} z, -1 + \bar{z} w + \bar{z} \lambda^2 w - 
 \bar{z} \lambda^2 z, -1 + w + \lambda^2 w - 2 \lambda^2 z + \lambda^2 z^2, -1 + \bar{w} + \bar{w} \lambda^2 - 
 2 \bar{z} \lambda^2 + \bar{z}^2 \lambda^2, -\lambda^2 w - z + w z + 2 \lambda^2 w z - \lambda^2 z^2, \bar{z} - 
 \bar{w} \bar{z} + \bar{w} \lambda^2 - 2 \bar{w} \bar{z} \lambda^2 + \bar{z}^2 \lambda^2, -w + \bar{w} w - \bar{w} z + w z + 
 \bar{w} \lambda^2 w z - \bar{w} \lambda^2 z^2, -\bar{z} \lambda^2 w - z + w z + 2 \bar{z} \lambda^2 w z - 
 \bar{z} \lambda^2 z^2, \bar{z} - \bar{w} \bar{z} + \bar{w} \lambda^2 z - 2 \bar{w} \bar{z} \lambda^2 z + \bar{z}^2 \lambda^2 z, -\bar{w} +
  \bar{w} \bar{z} + \bar{w} w - \bar{z} w + \bar{w} \bar{z} \lambda^2 w - \bar{z}^2 \lambda^2 w,  -\bar{z} w + \bar{w} \bar{z} w + 
 \bar{w} z - \bar{w} \bar{z} z - \bar{w} w z + \bar{z} w z,  -\bar{w} w + \bar{z} w + \bar{w} z - \bar{w} \bar{z} z - 
 \bar{z} w z + \bar{w} \bar{z} w z,   \\[4pt]
 1 - \bar{w} - w + \bar{w} w + \bar{w} \lambda^2 w - \bar{z} \lambda^2 w - 
 \bar{w} \lambda^2 z + \bar{z} \lambda^2 z,  \\[4pt]
 1 - \bar{w} - w + \bar{w} w - \bar{w} \lambda^4 w + \bar{z} \lambda^4 w + 
 \bar{w} \lambda^4 z - \bar{z} \lambda^4 z,  \\[4pt]
 \bar{z} - \bar{z} w - \bar{z} \lambda^2 w - z + \bar{z} \lambda^2 z + w z + 
 \bar{z} \lambda^2 w z - \bar{z} \lambda^2 z^2,  \\[4pt]
 -\bar{z} w + \bar{z} w^2 + \bar{z} \lambda^2 w^2 + z - w z - 
 2 \bar{z} \lambda^2 w z + \bar{z} \lambda^2 z^2,  \\[4pt]
 1 - w - \lambda^2 w - \bar{z} \lambda^2 w - \bar{z} \lambda^4 w + 
 \lambda^2 z + \bar{z} \lambda^2 z + \bar{z} \lambda^4 z^2, \\[4pt]
 \bar{z} - \bar{w} \bar{z} - \bar{w} z + \bar{w}^2 z + 
 \bar{w}^2 \lambda^2 z - 2 \bar{w} \bar{z} \lambda^2 z + \bar{z}^2 \lambda^2 z,  \\[4pt]
 \bar{z} - \bar{w} \bar{z} - z + \bar{w} z + 
 \bar{w} \lambda^2 z - \bar{z} \lambda^2 z - \bar{w} \bar{z} \lambda^2 z + \bar{z}^2 \lambda^2 z,  \\[4pt]
 -1 + \bar{w} + \bar{w} \lambda^2 - 
 \bar{z} \lambda^2 + \bar{w} \lambda^2 z - \bar{z} \lambda^2 z + \bar{w} \lambda^4 z - \bar{z}^2 \lambda^4 z,  \\[4pt]
 -\bar{w} + \bar{z} - w + 
 \bar{w} w + \bar{w} \lambda^2 w - \bar{z} \lambda^2 w + z - \bar{z} z - \bar{w} \lambda^2 z + \bar{z} \lambda^2 z,  \\[4pt]
 1 - \bar{z} -
  \bar{w} w + \bar{z} w - \bar{w} \lambda^2 w + \bar{z} \lambda^2 w - z + \bar{w} z + \bar{w} \lambda^2 z - \bar{z} \lambda^2 z, \\[4pt]
\bar{z} w - \bar{w} \bar{z} w + \bar{w} z - \bar{w} w z - 2 \bar{w} \bar{z} \lambda^2 w z + \bar{z}^2 \lambda^2 w z + 
 \bar{w} \bar{z} \lambda^2 z^2, \\[4pt]
 -\bar{z}^2 + 2 \bar{z} z + 4 \bar{z} \lambda^2 z - 4 \bar{z}^2 \lambda^2 z - z^2 - 
 4 \bar{z} \lambda^2 z^2 + 4 \bar{z}^2 \lambda^2 z^2,  \\[4pt]
 -1 + \bar{w} + w - \bar{w} w + \bar{w} \bar{z} \lambda^4 w z - 
 \bar{z}^2 \lambda^4 w z - \bar{w} \bar{z} \lambda^4 z^2 + \bar{z}^2 \lambda^4 z^2,   \\[4pt]
\bar{z} w - \bar{w} \bar{z} w - \bar{w} z + 2 \bar{w} \bar{z} z - \bar{z}^2 z + \bar{w} w z - 2 \bar{z} w z + 
 \bar{z}^2 w z + \bar{z} z^2 - \bar{w} \bar{z} z^2,  \\[4pt]
 -\bar{w} + \bar{z} + w - \bar{z} w + \bar{z} \lambda^2 w - 
 \bar{z}^2 \lambda^2 w - z + \bar{w} z - \bar{w} \lambda^2 z + \bar{z}^2 \lambda^2 z + \bar{w} \lambda^2 z^2 - 
 \bar{z} \lambda^2 z^2,  \\[4pt]
\bar{z} w - \bar{w} \bar{z} w + \bar{w} z - \bar{z} z - \bar{w} w z + \bar{w} \bar{z} w z - \bar{w} \bar{z} \lambda^2 w z + 
 \bar{z}^2 \lambda^2 w z + \bar{w} \bar{z} \lambda^2 z^2 - \bar{z}^2 \lambda^2 z^2,  \\[4pt]
 -\bar{w} w + \bar{w} \bar{z} w + \bar{z} z -
  \bar{w} \bar{z} z + \bar{w} w z - \bar{z} w z + \bar{w} \bar{z} \lambda^2 w z - \bar{z}^2 \lambda^2 w z - 
 \bar{w} \bar{z} \lambda^2 z^2 + \bar{z}^2 \lambda^2 z^2,  \\[4pt]
 -w + \bar{w} w + \bar{w} \lambda^2 w - \bar{z} \lambda^2 w - 
 \bar{w} z - \bar{w} \lambda^2 z + w z + \bar{w} \lambda^2 w z + \bar{z} \lambda^2 w z + \bar{w} \lambda^4 w z - 
 \bar{w} \lambda^2 z^2 - \bar{w} \lambda^4 z^2,  \\[4pt]
 -\bar{w} + \bar{w} \bar{z} + \bar{w} w - \bar{z} w + \bar{w} \lambda^2 w - 
 \bar{z} \lambda^2 w + \bar{w} \bar{z} \lambda^2 w - \bar{z}^2 \lambda^2 w + \bar{w} \bar{z} \lambda^4 w - \bar{z}^2 \lambda^4 w - 
 \bar{w} \lambda^2 z + \bar{w} \bar{z} \lambda^2 z, \\[4pt]
 -\bar{z} w + \bar{w} \bar{z} w + \bar{w} \bar{z} \lambda^2 w - \bar{z}^2 \lambda^2 w + 
 \bar{w} z - \bar{w} \bar{z} z - \bar{w} w z + \bar{z} w z - \bar{w} \lambda^2 w z + \bar{z}^2 \lambda^2 w z + 
 \bar{w} \lambda^2 z^2 - \bar{w} \bar{z} \lambda^2 z^2,  \\[4pt]
\bar{w}^2 w - \bar{z} w - \bar{w} w^2 + \bar{z} w^2 - \bar{w} \bar{z} \lambda^2 w^2 + \bar{z}^2 \lambda^2 w^2 + 
 \bar{w} z - \bar{w}^2 z + \bar{w}^2 \lambda^2 w z - \bar{z}^2 \lambda^2 w z - \bar{w}^2 \lambda^2 z^2 + 
 \bar{w} \bar{z} \lambda^2 z^2,  \\[4pt]
 -\bar{w} w + \bar{z} w + \bar{w} w^2 - \bar{z} w^2 + \bar{w} \lambda^2 w^2 - 
 \bar{z} \lambda^2 w^2 - z + \bar{w} z + w z - \bar{w} w z - 2 \bar{w} \lambda^2 w z + 2 \bar{z} \lambda^2 w z + 
 \bar{w} \lambda^2 z^2 - \bar{z} \lambda^2 z^2,  \\[4pt]
 -\bar{z} + \bar{w} \bar{z} - \bar{w} w + \bar{w}^2 w + \bar{z} w - 
 \bar{w} \bar{z} w + \bar{w}^2 \lambda^2 w - 2 \bar{w} \bar{z} \lambda^2 w + \bar{z}^2 \lambda^2 w + \bar{w} z - \bar{w}^2 z - 
 \bar{w}^2 \lambda^2 z + 2 \bar{w} \bar{z} \lambda^2 z - \bar{z}^2 \lambda^2 z,   \\[4pt]
 \bar{w}^2 - 2 \bar{w} w - 
 4 \bar{w} \lambda^2 w + 2 \bar{w} \bar{z} \lambda^2 w + w^2 + 2 \bar{z} \lambda^2 w^2 + \bar{z}^2 \lambda^4 w^2 + 
 2 \bar{w}^2 \lambda^2 z + 2 \bar{w} \lambda^2 w z - 4 \bar{w} \bar{z} \lambda^2 w z - 2 \bar{w} \bar{z} \lambda^4 w z + 
 \bar{w}^2 \lambda^4 z^2,  \\[4pt]
 -w + \bar{w} w + \bar{w} \lambda^2 w - \bar{z} \lambda^2 w + z - \bar{w} z + \lambda^2 z - 
 \bar{w} \lambda^2 z - \lambda^2 w z + \bar{w} \lambda^2 w z + \bar{w} \lambda^4 w z - \bar{z} \lambda^4 w z - 
 \bar{w} \lambda^2 z^2 + \bar{z} \lambda^2 z^2 - \bar{w} \lambda^4 z^2 + \bar{z} \lambda^4 z^2,  \\[4pt]
 1 - \bar{w} - w + 
 \bar{w} w - \bar{w} \lambda^4 w + \bar{z} \lambda^4 w + \lambda^2 z - 2 \bar{w} \lambda^2 z + \bar{z} \lambda^2 z - 
 \lambda^2 w z + 2 \bar{w} \lambda^2 w z - \bar{z} \lambda^2 w z + 2 \bar{w} \lambda^4 w z - 2 \bar{z} \lambda^4 w z - 
 \bar{w} \lambda^4 z^2 + \bar{z} \lambda^4 z^2,  \\[4pt]
 -\bar{w} + \bar{z} + \bar{z} \lambda^2 - \bar{w} \bar{z} \lambda^2 + \bar{w} w - 
 \bar{z} w + \bar{w} \lambda^2 w - \bar{z} \lambda^2 w + \bar{w} \bar{z} \lambda^2 w - \bar{z}^2 \lambda^2 w + \bar{w} \bar{z} \lambda^4 w -
  \bar{z}^2 \lambda^4 w - \bar{w} \lambda^2 z + \bar{z}^2 \lambda^2 z - \bar{w} \bar{z} \lambda^4 z + 
 \bar{z}^2 \lambda^4 z, \\[4pt]
 1 - \bar{w} + \bar{z} \lambda^2 - \bar{w} \bar{z} \lambda^2 - w + \bar{w} w - 2 \bar{z} \lambda^2 w + 
 2 \bar{w} \bar{z} \lambda^2 w - \bar{w} \lambda^4 w + 2 \bar{w} \bar{z} \lambda^4 w - \bar{z}^2 \lambda^4 w + \bar{z} \lambda^2 z - 
 \bar{w} \bar{z} \lambda^2 z + \bar{w} \lambda^4 z - 2 \bar{w} \bar{z} \lambda^4 z + \bar{z}^2 \lambda^4 z,   \\[4pt]
 -\bar{w} + \bar{z} - w + 
 \bar{w} w - \bar{z} \lambda^2 w + \bar{w} \bar{z} \lambda^2 w + z - \bar{z} z - \bar{w} \lambda^2 z + 2 \bar{z} \lambda^2 z - 
 \bar{w} \bar{z} \lambda^2 z + \bar{w} \lambda^2 w z - \bar{z} \lambda^2 w z + \bar{w} \bar{z} \lambda^4 w z - 
 \bar{z}^2 \lambda^4 w z - \bar{w} \bar{z} \lambda^4 z^2 + \bar{z}^2 \lambda^4 z^2,  \\[4pt]
 -\bar{z} + \bar{w} \bar{z} + \bar{z} w - 
 \bar{w} \bar{z} w + \bar{z} \lambda^2 w - \bar{w} \bar{z} \lambda^2 w + z - \bar{w} z - \bar{w} \lambda^2 z + 
 2 \bar{w} \bar{z} \lambda^2 z - \bar{z}^2 \lambda^2 z - w z + \bar{w} w z + \bar{w} \lambda^2 w z - 
 2 \bar{z} \lambda^2 w z + \bar{z}^2 \lambda^2 w z + \bar{z} \lambda^2 z^2 - \bar{w} \bar{z} \lambda^2 z^2,  \\[4pt]
 -\bar{w} w + 
 \bar{w} \bar{z} w + \bar{z} z - \bar{w} \bar{z} z - \bar{w} \bar{z} \lambda^2 z + \bar{z}^2 \lambda^2 z + \bar{w} w z - 
 \bar{z} w z - \bar{z} \lambda^2 w z + 2 \bar{w} \bar{z} \lambda^2 w z - \bar{z}^2 \lambda^2 w z + 
 \bar{w} \bar{z} \lambda^4 w z - \bar{z}^2 \lambda^4 w z + \bar{z} \lambda^2 z^2 - \bar{w} \bar{z} \lambda^2 z^2 - 
 \bar{w} \bar{z} \lambda^4 z^2 + \bar{z}^2 \lambda^4 z^2,  \\[4pt]
 -1 + \bar{w} + \bar{z} - \bar{w} \bar{z} + w - \bar{w} w - 
 \bar{z} w + \bar{w} \bar{z} w + z - \bar{w} z - \bar{w} \bar{z} \lambda^2 z + \bar{z}^2 \lambda^2 z - w z + \bar{w} w z -
  \bar{z} \lambda^2 w z + 2 \bar{w} \bar{z} \lambda^2 w z - \bar{z}^2 \lambda^2 w z + \bar{w} \bar{z} \lambda^4 w z - 
 \bar{z}^2 \lambda^4 w z + \bar{z} \lambda^2 z^2 - \bar{w} \bar{z} \lambda^2 z^2 - \bar{w} \bar{z} \lambda^4 z^2 + 
 \bar{z}^2 \lambda^4 z^2,  \\[4pt]
 1 - 2 \bar{w} + \bar{w}^2 - 2 w + 4 \bar{w} w - 2 \bar{w}^2 w + 
 2 \bar{w} \lambda^2 w - 2 \bar{w}^2 \lambda^2 w + w^2 - 2 \bar{w} w^2 + \bar{w}^2 w^2 - 
 2 \bar{w} \lambda^2 w^2 + 2 \bar{w}^2 \lambda^2 w^2 + \bar{w}^2 \lambda^4 w^2 + 2 \bar{z} \lambda^2 z - 
 2 \bar{w} \bar{z} \lambda^2 z - 2 \bar{z} \lambda^2 w z + 2 \bar{w} \bar{z} \lambda^2 w z - 2 \bar{w} \bar{z} \lambda^4 w z + 
 \bar{z}^2 \lambda^4 z^2, \\[4pt]
 \bar{w} - \bar{w}^2 - \bar{w} w + \bar{w}^2 w - \bar{w} \lambda^2 w + \bar{w}^2 \lambda^2 w - z +
  \bar{w} z + 2 \bar{w} \lambda^2 z - 2 \bar{w}^2 \lambda^2 z - \bar{z} \lambda^2 z + \bar{w} \bar{z} \lambda^2 z + w z - 
 \bar{w} w z - \bar{w} \lambda^2 w z + \bar{w}^2 \lambda^2 w z + 2 \bar{z} \lambda^2 w z - 2 \bar{w} \bar{z} \lambda^2 w z +
  \bar{w}^2 \lambda^4 w z - 2 \bar{w} \bar{z} \lambda^4 w z + \bar{z}^2 \lambda^4 w z - \bar{z} \lambda^2 z^2 + 
 \bar{w} \bar{z} \lambda^2 z^2 - \bar{w}^2 \lambda^4 z^2 + 2 \bar{w} \bar{z} \lambda^4 z^2 - \bar{z}^2 \lambda^4 z^2,  \\[4pt]
 \bar{z} - 
 \bar{w} \bar{z} - w + \bar{w} w - \bar{z} w + \bar{w} \bar{z} w + \bar{w} \lambda^2 w - 2 \bar{z} \lambda^2 w + 
 \bar{w} \bar{z} \lambda^2 w + w^2 - \bar{w} w^2 - \bar{w} \lambda^2 w^2 + 2 \bar{z} \lambda^2 w^2 - 
 \bar{w} \bar{z} \lambda^2 w^2 - \bar{w} \bar{z} \lambda^4 w^2 + \bar{z}^2 \lambda^4 w^2 + \bar{z} \lambda^2 z - 
 2 \bar{w} \bar{z} \lambda^2 z + \bar{z}^2 \lambda^2 z - \bar{z} \lambda^2 w z + 2 \bar{w} \bar{z} \lambda^2 w z - 
 \bar{z}^2 \lambda^2 w z + 2 \bar{w} \bar{z} \lambda^4 w z - 2 \bar{z}^2 \lambda^4 w z - \bar{w} \bar{z} \lambda^4 z^2 + 
 \bar{z}^2 \lambda^4 z^2,  \\[4pt]
 1 - \bar{w} + \bar{z} \lambda^2 - \bar{w} \bar{z} \lambda^2 - w + \bar{w} w - 2 \bar{z} \lambda^2 w + 
 2 \bar{w} \bar{z} \lambda^2 w - \bar{w} \lambda^4 w + 2 \bar{w} \bar{z} \lambda^4 w - \bar{z}^2 \lambda^4 w + \lambda^2 z - 
 2 \bar{w} \lambda^2 z + \bar{z} \lambda^2 z + \bar{z} \lambda^4 z - 2 \bar{w} \bar{z} \lambda^4 z + \bar{z}^2 \lambda^4 z - 
 \lambda^2 w z + 2 \bar{w} \lambda^2 w z - \bar{w} \bar{z} \lambda^2 w z + 2 \bar{w} \lambda^4 w z - 
 2 \bar{z} \lambda^4 w z + \bar{w} \bar{z} \lambda^6 w z - \bar{z}^2 \lambda^6 w z - \bar{w} \lambda^4 z^2 + 
 \bar{z} \lambda^4 z^2 - \bar{w} \bar{z} \lambda^6 z^2 + \bar{z}^2 \lambda^6 z^2,  \\[4pt]
 1 - \bar{w} - \bar{w} \lambda^2 + 
 \bar{z} \lambda^2 - w + \bar{w} w - \lambda^2 w + \bar{w} \lambda^2 w - \bar{z} \lambda^2 w + \bar{w} \bar{z} \lambda^2 w - 
 \bar{z} \lambda^4 w + \bar{w} \bar{z} \lambda^4 w + \lambda^2 z - \bar{w} \lambda^2 z + \bar{z} \lambda^2 z - \bar{w} \bar{z} \lambda^2 z - 
 \bar{w} \lambda^4 z + \bar{z} \lambda^4 z - \bar{w} \bar{z} \lambda^4 z + \bar{z}^2 \lambda^4 z + \bar{w} \lambda^2 w z - 
 \bar{z} \lambda^2 w z + \bar{w} \lambda^4 w z - \bar{z} \lambda^4 w z + \bar{w} \bar{z} \lambda^4 w z - \bar{z}^2 \lambda^4 w z +
  \bar{w} \bar{z} \lambda^6 w z - \bar{z}^2 \lambda^6 w z + \bar{z} \lambda^4 z^2 - \bar{w} \bar{z} \lambda^4 z^2 - 
 \bar{w} \bar{z} \lambda^6 z^2 + \bar{z}^2 \lambda^6 z^2, \\[4pt]
 \bar{w}^2 - 2 \bar{w} \bar{z} + \bar{z}^2 - 2 \bar{w} w + 
 2 \bar{z} w + 2 \bar{w} \bar{z} w - 2 \bar{z}^2 w - 4 \bar{w} \lambda^2 w + 4 \bar{z} \lambda^2 w + 
 4 \bar{w} \bar{z} \lambda^2 w - 4 \bar{z}^2 \lambda^2 w + w^2 - 2 \bar{z} w^2 + \bar{z}^2 w^2 + 2 \bar{w} z - 
 2 \bar{w}^2 z - 2 \bar{z} z + 2 \bar{w} \bar{z} z + 4 \bar{w} \lambda^2 z - 4 \bar{z} \lambda^2 z - 
 4 \bar{w} \bar{z} \lambda^2 z + 4 \bar{z}^2 \lambda^2 z - 2 w z + 2 \bar{w} w z + 2 \bar{z} w z - 
 2 \bar{w} \bar{z} w z + 4 \bar{w} \lambda^2 w z - 4 \bar{z} \lambda^2 w z - 4 \bar{w} \bar{z} \lambda^2 w z + 
 4 \bar{z}^2 \lambda^2 w z + z^2 - 2 \bar{w} z^2 + \bar{w}^2 z^2 - 4 \bar{w} \lambda^2 z^2 + 
 4 \bar{z} \lambda^2 z^2 + 4 \bar{w} \bar{z} \lambda^2 z^2 - 4 \bar{z}^2 \lambda^2 z^2,  \\[4pt]
\bar{z}^2 w^2 - \bar{z}^3 w^2 + 2 \bar{w} \bar{z} w z - 4 \bar{w}^2 \bar{z} w z - 4 \bar{z}^2 w z + 
 6 \bar{w} \bar{z}^2 w z - 4 \bar{w} \bar{z} w^2 z + 4 \bar{w}^2 \bar{z} w^2 z + 3 \bar{z}^2 w^2 z - 
 4 \bar{w} \bar{z}^2 w^2 z + \bar{z}^3 w^2 z + 4 \bar{w}^2 \bar{z} \lambda^2 w^2 z - 
 8 \bar{w} \bar{z}^2 \lambda^2 w^2 z + 4 \bar{z}^3 \lambda^2 w^2 z + \bar{w}^2 z^2 - 4 \bar{w} \bar{z} z^2 + 
 3 \bar{w}^2 \bar{z} z^2 + 4 \bar{z}^2 z^2 - 4 \bar{w} \bar{z}^2 z^2 + 6 \bar{w} \bar{z} w z^2 - 
 4 \bar{w}^2 \bar{z} w z^2 - 4 \bar{z}^2 w z^2 + 2 \bar{w} \bar{z}^2 w z^2 - 
 8 \bar{w}^2 \bar{z} \lambda^2 w z^2 + 16 \bar{w} \bar{z}^2 \lambda^2 w z^2 - 8 \bar{z}^3 \lambda^2 w z^2 - 
 \bar{w}^2 z^3 + \bar{w}^2 \bar{z} z^3 + 4 \bar{w}^2 \bar{z} \lambda^2 z^3 - 8 \bar{w} \bar{z}^2 \lambda^2 z^3 + 
 4 \bar{z}^3 \lambda^2 z^3,  \\[4pt]
 -\bar{w} + \bar{w}^2 + \bar{z} - \bar{w} \bar{z} - w + 3 \bar{w} w - 2 \bar{w}^2 w - 
 \bar{z} w + \bar{w} \bar{z} w + 2 \bar{w} \lambda^2 w - 2 \bar{w}^2 \lambda^2 w - 2 \bar{z} \lambda^2 w + 
 2 \bar{w} \bar{z} \lambda^2 w + w^2 - 2 \bar{w} w^2 + \bar{w}^2 w^2 - 2 \bar{w} \lambda^2 w^2 + 
 2 \bar{w}^2 \lambda^2 w^2 + 2 \bar{z} \lambda^2 w^2 - 2 \bar{w} \bar{z} \lambda^2 w^2 + \bar{w}^2 \lambda^4 w^2 - 
 2 \bar{w} \bar{z} \lambda^4 w^2 + \bar{z}^2 \lambda^4 w^2 + z - \bar{w} z - \bar{z} z + \bar{w} \bar{z} z - 
 2 \bar{w} \lambda^2 z + 2 \bar{w}^2 \lambda^2 z + 2 \bar{z} \lambda^2 z - 2 \bar{w} \bar{z} \lambda^2 z - w z + 
 \bar{w} w z + \bar{z} w z - \bar{w} \bar{z} w z + 2 \bar{w} \lambda^2 w z - 2 \bar{w}^2 \lambda^2 w z - 
 2 \bar{z} \lambda^2 w z + 2 \bar{w} \bar{z} \lambda^2 w z - 2 \bar{w}^2 \lambda^4 w z + 4 \bar{w} \bar{z} \lambda^4 w z - 
 2 \bar{z}^2 \lambda^4 w z + \bar{w}^2 \lambda^4 z^2 - 2 \bar{w} \bar{z} \lambda^4 z^2 + \bar{z}^2 \lambda^4 z^2,  \\[4pt]
\bar{z}^2 w^2 - 2 \bar{w} \bar{z}^2 w^2 + \bar{w}^2 \bar{z}^2 w^2 - 2 \bar{w} \bar{z} w z + 
 2 \bar{w}^2 \bar{z} w z + 2 \bar{w} \bar{z}^2 w z - 2 \bar{w}^2 \bar{z}^2 w z + 2 \bar{w} \bar{z} w^2 z - 
 2 \bar{w}^2 \bar{z} w^2 z - 2 \bar{z}^2 w^2 z + 2 \bar{w} \bar{z}^2 w^2 z - 
 4 \bar{w}^2 \bar{z} \lambda^2 w^2 z + 4 \bar{w} \bar{z}^2 \lambda^2 w^2 z + 4 \bar{w}^2 \bar{z}^2 \lambda^2 w^2 z - 
 4 \bar{w} \bar{z}^3 \lambda^2 w^2 z + \bar{w}^2 z^2 - 2 \bar{w}^2 \bar{z} z^2 + \bar{w}^2 \bar{z}^2 z^2 - 
 2 \bar{w}^2 w z^2 + 2 \bar{w} \bar{z} w z^2 + 2 \bar{w}^2 \bar{z} w z^2 - 2 \bar{w} \bar{z}^2 w z^2 + 
 4 \bar{w}^2 \bar{z} \lambda^2 w z^2 - 4 \bar{w} \bar{z}^2 \lambda^2 w z^2 - 4 \bar{w}^2 \bar{z}^2 \lambda^2 w z^2 + 
 4 \bar{w} \bar{z}^3 \lambda^2 w z^2 + \bar{w}^2 w^2 z^2 - 2 \bar{w} \bar{z} w^2 z^2 + 
 \bar{z}^2 w^2 z^2 + 4 \bar{w}^2 \bar{z} \lambda^2 w^2 z^2 - 4 \bar{w} \bar{z}^2 \lambda^2 w^2 z^2 - 
 4 \bar{w}^2 \bar{z}^2 \lambda^2 w^2 z^2 + 4 \bar{w} \bar{z}^3 \lambda^2 w^2 z^2 - 
 4 \bar{w}^2 \bar{z} \lambda^2 w z^3 + 4 \bar{w} \bar{z}^2 \lambda^2 w z^3 + 4 \bar{w}^2 \bar{z}^2 \lambda^2 w z^3 - 
 4 \bar{w} \bar{z}^3 \lambda^2 w z^3,  \\[4pt]
 -\bar{w} w + \bar{w}^2 w + \bar{w} \bar{z} w - \bar{w}^2 \bar{z} w + \bar{w} w^2 - 
 \bar{w}^2 w^2 - \bar{w} \bar{z} w^2 + \bar{w}^2 \bar{z} w^2 + \bar{z} z - 2 \bar{w} \bar{z} z + \bar{w}^2 \bar{z} z + 
 \bar{w} w z - \bar{w}^2 w z - 2 \bar{z} w z + 3 \bar{w} \bar{z} w z - \bar{w}^2 \bar{z} w z + 
 2 \bar{w} \bar{z} \lambda^2 w z - 2 \bar{w}^2 \bar{z} \lambda^2 w z - 2 \bar{z}^2 \lambda^2 w z + 
 2 \bar{w} \bar{z}^2 \lambda^2 w z - \bar{w} w^2 z + \bar{w}^2 w^2 z + \bar{z} w^2 z - \bar{w} \bar{z} w^2 z - 
 2 \bar{w} \bar{z} \lambda^2 w^2 z + 2 \bar{w}^2 \bar{z} \lambda^2 w^2 z + 2 \bar{z}^2 \lambda^2 w^2 z - 
 2 \bar{w} \bar{z}^2 \lambda^2 w^2 z + \bar{w}^2 \bar{z} \lambda^4 w^2 z - 2 \bar{w} \bar{z}^2 \lambda^4 w^2 z + 
 \bar{z}^3 \lambda^4 w^2 z - 2 \bar{w} \bar{z} \lambda^2 z^2 + 2 \bar{w}^2 \bar{z} \lambda^2 z^2 + 
 2 \bar{z}^2 \lambda^2 z^2 - 2 \bar{w} \bar{z}^2 \lambda^2 z^2 + 2 \bar{w} \bar{z} \lambda^2 w z^2 - 
 2 \bar{w}^2 \bar{z} \lambda^2 w z^2 - 2 \bar{z}^2 \lambda^2 w z^2 + 2 \bar{w} \bar{z}^2 \lambda^2 w z^2 - 
 2 \bar{w}^2 \bar{z} \lambda^4 w z^2 + 4 \bar{w} \bar{z}^2 \lambda^4 w z^2 - 2 \bar{z}^3 \lambda^4 w z^2 + 
 \bar{w}^2 \bar{z} \lambda^4 z^3 - 2 \bar{w} \bar{z}^2 \lambda^4 z^3 + \bar{z}^3 \lambda^4 z^3,   \\[4pt]
4 \bar{w}^2 w^2 - 4 \bar{w}^3 w^2 - 4 \bar{w} \bar{z} w^2 + 4 \bar{w}^2 \bar{z} w^2 + \bar{z}^2 w^2 - 
 2 \bar{w} \bar{z}^2 w^2 + \bar{w}^2 \bar{z}^2 w^2 - 4 \bar{w}^2 w^3 + 4 \bar{w}^3 w^3 + 
 4 \bar{w} \bar{z} w^3 - 4 \bar{w}^2 \bar{z} w^3 - 4 \bar{w}^2 \bar{z} \lambda^2 w^3 + 
 4 \bar{w}^3 \bar{z} \lambda^2 w^3 + 4 \bar{w} \bar{z}^2 \lambda^2 w^3 - 4 \bar{w}^2 \bar{z}^2 \lambda^2 w^3 - 
 4 \bar{w}^2 w z + 4 \bar{w}^3 w z + 2 \bar{w} \bar{z} w z - 2 \bar{w}^2 \bar{z} w z + 
 2 \bar{w} \bar{z}^2 w z - 2 \bar{w}^2 \bar{z}^2 w z + 4 \bar{w}^2 w^2 z - 4 \bar{w}^3 w^2 z - 
 2 \bar{w} \bar{z} w^2 z + 2 \bar{w}^2 \bar{z} w^2 z - 2 \bar{z}^2 w^2 z + 2 \bar{w} \bar{z}^2 w^2 z - 
 4 \bar{w}^3 \lambda^2 w^2 z + 16 \bar{w}^2 \bar{z} \lambda^2 w^2 z - 8 \bar{w}^3 \bar{z} \lambda^2 w^2 z - 
 12 \bar{w} \bar{z}^2 \lambda^2 w^2 z + 8 \bar{w}^2 \bar{z}^2 \lambda^2 w^2 z + 4 \bar{w}^3 \lambda^2 w^3 z - 
 8 \bar{w}^2 \bar{z} \lambda^2 w^3 z + 4 \bar{w} \bar{z}^2 \lambda^2 w^3 z + 4 \bar{w}^3 \bar{z} \lambda^4 w^3 z - 
 8 \bar{w}^2 \bar{z}^2 \lambda^4 w^3 z + 4 \bar{w} \bar{z}^3 \lambda^4 w^3 z + \bar{w}^2 z^2 - 
 2 \bar{w}^2 \bar{z} z^2 + \bar{w}^2 \bar{z}^2 z^2 - 2 \bar{w}^2 w z^2 + 2 \bar{w} \bar{z} w z^2 + 
 2 \bar{w}^2 \bar{z} w z^2 - 2 \bar{w} \bar{z}^2 w z^2 + 4 \bar{w}^3 \lambda^2 w z^2 - 
 12 \bar{w}^2 \bar{z} \lambda^2 w z^2 + 4 \bar{w}^3 \bar{z} \lambda^2 w z^2 + 8 \bar{w} \bar{z}^2 \lambda^2 w z^2 - 
 4 \bar{w}^2 \bar{z}^2 \lambda^2 w z^2 + \bar{w}^2 w^2 z^2 - 2 \bar{w} \bar{z} w^2 z^2 + 
 \bar{z}^2 w^2 z^2 - 4 \bar{w}^3 \lambda^2 w^2 z^2 + 8 \bar{w}^2 \bar{z} \lambda^2 w^2 z^2 - 
 4 \bar{w} \bar{z}^2 \lambda^2 w^2 z^2 - 8 \bar{w}^3 \bar{z} \lambda^4 w^2 z^2 + 
 16 \bar{w}^2 \bar{z}^2 \lambda^4 w^2 z^2 - 8 \bar{w} \bar{z}^3 \lambda^4 w^2 z^2 + 
 4 \bar{w}^3 \bar{z} \lambda^4 w z^3 - 8 \bar{w}^2 \bar{z}^2 \lambda^4 w z^3 + 
 4 \bar{w} \bar{z}^3 \lambda^4 w z^3,   \\[4pt]
 -\bar{z}^2 w^2 + 2 \bar{w} \bar{z}^2 w^2 - \bar{w}^2 \bar{z}^2 w^2 + 
 2 \bar{w} \bar{z} \lambda^2 w^2 - 2 \bar{w}^2 \bar{z} \lambda^2 w^2 - 2 \bar{z}^2 \lambda^2 w^2 + 
 2 \bar{w} \bar{z}^2 \lambda^2 w^2 - \bar{w}^2 \lambda^4 w^2 + 2 \bar{w} \bar{z} \lambda^4 w^2 - \bar{z}^2 \lambda^4 w^2 + 
 2 \bar{w} \bar{z} w z - 2 \bar{w}^2 \bar{z} w z - 2 \bar{w} \bar{z}^2 w z + 2 \bar{w}^2 \bar{z}^2 w z + 
 2 \bar{w}^2 \lambda^2 w z - 4 \bar{w}^2 \bar{z} \lambda^2 w z + 2 \bar{w}^2 \bar{z}^2 \lambda^2 w z + 
 2 \bar{w}^2 \lambda^4 w z - 2 \bar{w} \bar{z} \lambda^4 w z - 2 \bar{w}^2 \bar{z} \lambda^4 w z + 
 2 \bar{w} \bar{z}^2 \lambda^4 w z - 2 \bar{w} \bar{z} w^2 z + 2 \bar{w}^2 \bar{z} w^2 z + 2 \bar{z}^2 w^2 z - 
 2 \bar{w} \bar{z}^2 w^2 z - 2 \bar{w}^2 \lambda^2 w^2 z - 4 \bar{w} \bar{z} \lambda^2 w^2 z + 
 8 \bar{w}^2 \bar{z} \lambda^2 w^2 z + 4 \bar{z}^2 \lambda^2 w^2 z - 4 \bar{w} \bar{z}^2 \lambda^2 w^2 z - 
 2 \bar{w}^2 \bar{z}^2 \lambda^2 w^2 z - 2 \bar{w} \bar{z} \lambda^4 w^2 z - 2 \bar{w}^2 \bar{z} \lambda^4 w^2 z + 
 2 \bar{z}^2 \lambda^4 w^2 z + 2 \bar{w} \bar{z}^2 \lambda^4 w^2 z + 4 \bar{w}^2 \bar{z}^2 \lambda^4 w^2 z - 
 4 \bar{w} \bar{z}^3 \lambda^4 w^2 z - 4 \bar{w}^2 \bar{z} \lambda^6 w^2 z + 4 \bar{w} \bar{z}^2 \lambda^6 w^2 z + 
 4 \bar{w}^2 \bar{z}^2 \lambda^6 w^2 z - 4 \bar{w} \bar{z}^3 \lambda^6 w^2 z - \bar{w}^2 z^2 + 
 2 \bar{w}^2 \bar{z} z^2 - \bar{w}^2 \bar{z}^2 z^2 - 2 \bar{w}^2 \lambda^2 z^2 + 4 \bar{w}^2 \bar{z} \lambda^2 z^2 - 
 2 \bar{w}^2 \bar{z}^2 \lambda^2 z^2 - \bar{w}^2 \lambda^4 z^2 + 2 \bar{w}^2 \bar{z} \lambda^4 z^2 - 
 \bar{w}^2 \bar{z}^2 \lambda^4 z^2 + 2 \bar{w}^2 w z^2 - 2 \bar{w} \bar{z} w z^2 - 2 \bar{w}^2 \bar{z} w z^2 + 
 2 \bar{w} \bar{z}^2 w z^2 + 2 \bar{w}^2 \lambda^2 w z^2 - 4 \bar{w}^2 \bar{z} \lambda^2 w z^2 + 
 2 \bar{w}^2 \bar{z}^2 \lambda^2 w z^2 + 2 \bar{w} \bar{z} \lambda^4 w z^2 + 2 \bar{w}^2 \bar{z} \lambda^4 w z^2 - 
 6 \bar{w} \bar{z}^2 \lambda^4 w z^2 - 2 \bar{w}^2 \bar{z}^2 \lambda^4 w z^2 + 4 \bar{w} \bar{z}^3 \lambda^4 w z^2 + 
 4 \bar{w}^2 \bar{z} \lambda^6 w z^2 - 4 \bar{w} \bar{z}^2 \lambda^6 w z^2 - 4 \bar{w}^2 \bar{z}^2 \lambda^6 w z^2 + 
 4 \bar{w} \bar{z}^3 \lambda^6 w z^2 - \bar{w}^2 w^2 z^2 + 2 \bar{w} \bar{z} w^2 z^2 - 
 \bar{z}^2 w^2 z^2 + 2 \bar{w} \bar{z} \lambda^2 w^2 z^2 - 2 \bar{w}^2 \bar{z} \lambda^2 w^2 z^2 - 
 2 \bar{z}^2 \lambda^2 w^2 z^2 + 2 \bar{w} \bar{z}^2 \lambda^2 w^2 z^2 + 4 \bar{w}^2 \bar{z} \lambda^4 w^2 z^2 - 
 \bar{z}^2 \lambda^4 w^2 z^2 - 2 \bar{w} \bar{z}^2 \lambda^4 w^2 z^2 - 5 \bar{w}^2 \bar{z}^2 \lambda^4 w^2 z^2 + 
 4 \bar{w} \bar{z}^3 \lambda^4 w^2 z^2 + 4 \bar{w}^2 \bar{z} \lambda^6 w^2 z^2 - 
 4 \bar{w} \bar{z}^2 \lambda^6 w^2 z^2 - 4 \bar{w}^2 \bar{z}^2 \lambda^6 w^2 z^2 + 
 4 \bar{w} \bar{z}^3 \lambda^6 w^2 z^2 - 4 \bar{w}^2 \bar{z} \lambda^4 w z^3 + 4 \bar{w} \bar{z}^2 \lambda^4 w z^3 + 
 4 \bar{w}^2 \bar{z}^2 \lambda^4 w z^3 - 4 \bar{w} \bar{z}^3 \lambda^4 w z^3 - 4 \bar{w}^2 \bar{z} \lambda^6 w z^3 + 
 4 \bar{w} \bar{z}^2 \lambda^6 w z^3 + 4 \bar{w}^2 \bar{z}^2 \lambda^6 w z^3 - 
 4 \bar{w} \bar{z}^3 \lambda^6 w z^3,  \\[4pt]
 1 - 2 \bar{w} + \bar{w}^2 - 2 w + 4 \bar{w} w - 2 \bar{w}^2 w + 
 2 \bar{w} \lambda^2 w - 2 \bar{w}^2 \lambda^2 w - 4 \bar{z} \lambda^2 w + 4 \bar{w} \bar{z} \lambda^2 w + 
 2 \bar{z}^2 \lambda^2 w - 2 \bar{w} \bar{z}^2 \lambda^2 w + w^2 - 2 \bar{w} w^2 + \bar{w}^2 w^2 - 
 2 \bar{w} \lambda^2 w^2 + 2 \bar{w}^2 \lambda^2 w^2 + 4 \bar{z} \lambda^2 w^2 - 4 \bar{w} \bar{z} \lambda^2 w^2 - 
 2 \bar{z}^2 \lambda^2 w^2 + 2 \bar{w} \bar{z}^2 \lambda^2 w^2 + \bar{w}^2 \lambda^4 w^2 - 
 4 \bar{w} \bar{z} \lambda^4 w^2 + 4 \bar{z}^2 \lambda^4 w^2 + 2 \bar{w} \bar{z}^2 \lambda^4 w^2 - 
 4 \bar{z}^3 \lambda^4 w^2 + \bar{z}^4 \lambda^4 w^2 - 4 \bar{w} \lambda^2 z + 4 \bar{w}^2 \lambda^2 z + 
 8 \bar{z} \lambda^2 z - 8 \bar{w} \bar{z} \lambda^2 z - 4 \bar{z}^2 \lambda^2 z + 4 \bar{w} \bar{z}^2 \lambda^2 z + 
 4 \bar{w} \lambda^2 w z - 4 \bar{w}^2 \lambda^2 w z - 8 \bar{z} \lambda^2 w z + 10 \bar{w} \bar{z} \lambda^2 w z - 
 2 \bar{w}^2 \bar{z} \lambda^2 w z + 2 \bar{z}^2 \lambda^2 w z - 2 \bar{w} \bar{z}^2 \lambda^2 w z - 
 4 \bar{w}^2 \lambda^4 w z + 14 \bar{w} \bar{z} \lambda^4 w z - 2 \bar{w}^2 \bar{z} \lambda^4 w z - 
 14 \bar{z}^2 \lambda^4 w z + 2 \bar{w} \bar{z}^2 \lambda^4 w z + 8 \bar{z}^3 \lambda^4 w z - 
 4 \bar{w} \bar{z}^3 \lambda^4 w z - 2 \bar{w} \bar{z} \lambda^2 w^2 z + 2 \bar{w}^2 \bar{z} \lambda^2 w^2 z + 
 2 \bar{z}^2 \lambda^2 w^2 z - 2 \bar{w} \bar{z}^2 \lambda^2 w^2 z - 2 \bar{w} \bar{z} \lambda^4 w^2 z + 
 4 \bar{w}^2 \bar{z} \lambda^4 w^2 z + 2 \bar{z}^2 \lambda^4 w^2 z - 8 \bar{w} \bar{z}^2 \lambda^4 w^2 z + 
 4 \bar{z}^3 \lambda^4 w^2 z + 2 \bar{w} \bar{z}^3 \lambda^4 w^2 z - 2 \bar{z}^4 \lambda^4 w^2 z + 
 2 \bar{w}^2 \bar{z} \lambda^6 w^2 z - 6 \bar{w} \bar{z}^2 \lambda^6 w^2 z + 4 \bar{z}^3 \lambda^6 w^2 z + 
 2 \bar{w} \bar{z}^3 \lambda^6 w^2 z - 2 \bar{z}^4 \lambda^6 w^2 z + 2 \bar{w} \lambda^2 z^2 - 
 2 \bar{w}^2 \lambda^2 z^2 - 4 \bar{z} \lambda^2 z^2 + 2 \bar{w} \bar{z} \lambda^2 z^2 + 
 2 \bar{w}^2 \bar{z} \lambda^2 z^2 + 4 \bar{z}^2 \lambda^2 z^2 - 4 \bar{w} \bar{z}^2 \lambda^2 z^2 + 
 4 \bar{w}^2 \lambda^4 z^2 - 14 \bar{w} \bar{z} \lambda^4 z^2 + 2 \bar{w}^2 \bar{z} \lambda^4 z^2 + 
 14 \bar{z}^2 \lambda^4 z^2 - 2 \bar{w} \bar{z}^2 \lambda^4 z^2 - 8 \bar{z}^3 \lambda^4 z^2 + 
 4 \bar{w} \bar{z}^3 \lambda^4 z^2 - 2 \bar{w} \lambda^2 w z^2 + 2 \bar{w}^2 \lambda^2 w z^2 + 
 4 \bar{z} \lambda^2 w z^2 - 2 \bar{w} \bar{z} \lambda^2 w z^2 - 2 \bar{w}^2 \bar{z} \lambda^2 w z^2 - 
 4 \bar{z}^2 \lambda^2 w z^2 + 4 \bar{w} \bar{z}^2 \lambda^2 w z^2 + 2 \bar{w}^2 \lambda^4 w z^2 + 
 2 \bar{w} \bar{z} \lambda^4 w z^2 - 8 \bar{w}^2 \bar{z} \lambda^4 w z^2 - 2 \bar{z}^2 \lambda^4 w z^2 + 
 8 \bar{w} \bar{z}^2 \lambda^4 w z^2 - 4 \bar{z}^3 \lambda^4 w z^2 + 2 \bar{w} \bar{z}^3 \lambda^4 w z^2 - 
 6 \bar{w}^2 \bar{z} \lambda^6 w z^2 + 18 \bar{w} \bar{z}^2 \lambda^6 w z^2 - 12 \bar{z}^3 \lambda^6 w z^2 - 
 6 \bar{w} \bar{z}^3 \lambda^6 w z^2 + 6 \bar{z}^4 \lambda^6 w z^2 + \bar{w}^2 \bar{z}^2 \lambda^4 w^2 z^2 - 
 2 \bar{w} \bar{z}^3 \lambda^4 w^2 z^2 + \bar{z}^4 \lambda^4 w^2 z^2 + 2 \bar{w}^2 \bar{z}^2 \lambda^6 w^2 z^2 - 
 4 \bar{w} \bar{z}^3 \lambda^6 w^2 z^2 + 2 \bar{z}^4 \lambda^6 w^2 z^2 + \bar{w}^2 \bar{z}^2 \lambda^8 w^2 z^2 - 
 2 \bar{w} \bar{z}^3 \lambda^8 w^2 z^2 + \bar{z}^4 \lambda^8 w^2 z^2 - 4 \bar{w}^2 \lambda^4 z^3 + 
 8 \bar{w} \bar{z} \lambda^4 z^3 + 4 \bar{w}^2 \bar{z} \lambda^4 z^3 - 8 \bar{z}^2 \lambda^4 z^3 - 
 4 \bar{w} \bar{z}^2 \lambda^4 z^3 + 8 \bar{z}^3 \lambda^4 z^3 - 4 \bar{w} \bar{z}^3 \lambda^4 z^3 + 
 4 \bar{w}^2 \bar{z} \lambda^6 z^3 - 12 \bar{w} \bar{z}^2 \lambda^6 z^3 + 8 \bar{z}^3 \lambda^6 z^3 + 
 4 \bar{w} \bar{z}^3 \lambda^6 z^3 - 4 \bar{z}^4 \lambda^6 z^3 - 4 \bar{w} \bar{z} \lambda^4 w z^3 + 
 2 \bar{w}^2 \bar{z} \lambda^4 w z^3 + 4 \bar{z}^2 \lambda^4 w z^3 + 2 \bar{w} \bar{z}^2 \lambda^4 w z^3 - 
 2 \bar{w}^2 \bar{z}^2 \lambda^4 w z^3 - 4 \bar{z}^3 \lambda^4 w z^3 + 2 \bar{w} \bar{z}^3 \lambda^4 w z^3 + 
 2 \bar{w}^2 \bar{z} \lambda^6 w z^3 - 6 \bar{w} \bar{z}^2 \lambda^6 w z^3 - 4 \bar{w}^2 \bar{z}^2 \lambda^6 w z^3 + 
 4 \bar{z}^3 \lambda^6 w z^3 + 10 \bar{w} \bar{z}^3 \lambda^6 w z^3 - 6 \bar{z}^4 \lambda^6 w z^3 - 
 2 \bar{w}^2 \bar{z}^2 \lambda^8 w z^3 + 4 \bar{w} \bar{z}^3 \lambda^8 w z^3 - 2 \bar{z}^4 \lambda^8 w z^3 + 
 \bar{w}^2 \lambda^4 z^4 - 2 \bar{w}^2 \bar{z} \lambda^4 z^4 + \bar{w}^2 \bar{z}^2 \lambda^4 z^4 - 
 2 \bar{w}^2 \bar{z} \lambda^6 z^4 + 6 \bar{w} \bar{z}^2 \lambda^6 z^4 + 2 \bar{w}^2 \bar{z}^2 \lambda^6 z^4 - 
 4 \bar{z}^3 \lambda^6 z^4 - 6 \bar{w} \bar{z}^3 \lambda^6 z^4 + 4 \bar{z}^4 \lambda^6 z^4 + 
 \bar{w}^2 \bar{z}^2 \lambda^8 z^4 - 2 \bar{w} \bar{z}^3 \lambda^8 z^4 + \bar{z}^4 \lambda^8 z^4,  \\[4pt]
\bar{z}^2 \lambda^4 w^2 - 2 \bar{z}^3 \lambda^4 w^2 + \bar{z}^4 \lambda^4 w^2 + 4 \bar{w} \bar{z} \lambda^2 w z - 
 4 \bar{w}^2 \bar{z} \lambda^2 w z - 2 \bar{z}^2 \lambda^2 w z - 2 \bar{w} \bar{z}^2 \lambda^2 w z + 
 4 \bar{w}^2 \bar{z}^2 \lambda^2 w z + 2 \bar{z}^3 \lambda^2 w z - 2 \bar{w} \bar{z}^3 \lambda^2 w z + 
 2 \bar{w} \bar{z} \lambda^4 w z - 4 \bar{w}^2 \bar{z} \lambda^4 w z - 2 \bar{z}^2 \lambda^4 w z + 
 2 \bar{w} \bar{z}^2 \lambda^4 w z + 4 \bar{w}^2 \bar{z}^2 \lambda^4 w z + 2 \bar{z}^3 \lambda^4 w z - 
 4 \bar{w} \bar{z}^3 \lambda^4 w z - 4 \bar{w} \bar{z} \lambda^2 w^2 z + 4 \bar{w}^2 \bar{z} \lambda^2 w^2 z + 
 2 \bar{z}^2 \lambda^2 w^2 z + 2 \bar{w} \bar{z}^2 \lambda^2 w^2 z - 4 \bar{w}^2 \bar{z}^2 \lambda^2 w^2 z - 
 2 \bar{z}^3 \lambda^2 w^2 z + 2 \bar{w} \bar{z}^3 \lambda^2 w^2 z - 4 \bar{w} \bar{z} \lambda^4 w^2 z + 
 8 \bar{w}^2 \bar{z} \lambda^4 w^2 z - 4 \bar{w} \bar{z}^2 \lambda^4 w^2 z - 8 \bar{w}^2 \bar{z}^2 \lambda^4 w^2 z + 
 4 \bar{z}^3 \lambda^4 w^2 z + 8 \bar{w} \bar{z}^3 \lambda^4 w^2 z - 4 \bar{z}^4 \lambda^4 w^2 z + 
 4 \bar{w}^2 \bar{z} \lambda^6 w^2 z - 6 \bar{w} \bar{z}^2 \lambda^6 w^2 z - 4 \bar{w}^2 \bar{z}^2 \lambda^6 w^2 z + 
 2 \bar{z}^3 \lambda^6 w^2 z + 6 \bar{w} \bar{z}^3 \lambda^6 w^2 z - 2 \bar{z}^4 \lambda^6 w^2 z + 
 \bar{z}^2 z^2 - 2 \bar{w} \bar{z}^2 z^2 + \bar{w}^2 \bar{z}^2 z^2 - 2 \bar{w} \bar{z} \lambda^2 z^2 + 
 2 \bar{w}^2 \bar{z} \lambda^2 z^2 + 2 \bar{z}^2 \lambda^2 z^2 - 2 \bar{w} \bar{z}^2 \lambda^2 z^2 + 
 \bar{w}^2 \lambda^4 z^2 - 2 \bar{w} \bar{z} \lambda^4 z^2 + \bar{z}^2 \lambda^4 z^2 - 2 \bar{z}^2 w z^2 + 
 4 \bar{w} \bar{z}^2 w z^2 - 2 \bar{w}^2 \bar{z}^2 w z^2 - 2 \bar{w} \bar{z} \lambda^2 w z^2 + 
 2 \bar{w}^2 \bar{z} \lambda^2 w z^2 - 2 \bar{z}^2 \lambda^2 w z^2 + 10 \bar{w} \bar{z}^2 \lambda^2 w z^2 - 
 8 \bar{w}^2 \bar{z}^2 \lambda^2 w z^2 - 4 \bar{z}^3 \lambda^2 w z^2 + 4 \bar{w} \bar{z}^3 \lambda^2 w z^2 + 
 2 \bar{w} \bar{z} \lambda^4 w z^2 - 4 \bar{w}^2 \bar{z} \lambda^4 w z^2 + 8 \bar{w} \bar{z}^2 \lambda^4 w z^2 - 
 2 \bar{w}^2 \bar{z}^2 \lambda^4 w z^2 - 8 \bar{z}^3 \lambda^4 w z^2 + 2 \bar{w} \bar{z}^3 \lambda^4 w z^2 + 
 2 \bar{z}^4 \lambda^4 w z^2 - 6 \bar{w}^2 \bar{z} \lambda^6 w z^2 + 10 \bar{w} \bar{z}^2 \lambda^6 w z^2 + 
 4 \bar{w}^2 \bar{z}^2 \lambda^6 w z^2 - 4 \bar{z}^3 \lambda^6 w z^2 - 6 \bar{w} \bar{z}^3 \lambda^6 w z^2 + 
 2 \bar{z}^4 \lambda^6 w z^2 + \bar{z}^2 w^2 z^2 - 2 \bar{w} \bar{z}^2 w^2 z^2 + 
 \bar{w}^2 \bar{z}^2 w^2 z^2 + 4 \bar{w} \bar{z} \lambda^2 w^2 z^2 - 4 \bar{w}^2 \bar{z} \lambda^2 w^2 z^2 - 
 8 \bar{w} \bar{z}^2 \lambda^2 w^2 z^2 + 8 \bar{w}^2 \bar{z}^2 \lambda^2 w^2 z^2 + 
 4 \bar{z}^3 \lambda^2 w^2 z^2 - 4 \bar{w} \bar{z}^3 \lambda^2 w^2 z^2 + 4 \bar{w} \bar{z} \lambda^4 w^2 z^2 - 
 8 \bar{w}^2 \bar{z} \lambda^4 w^2 z^2 - 2 \bar{w} \bar{z}^2 \lambda^4 w^2 z^2 + 
 14 \bar{w}^2 \bar{z}^2 \lambda^4 w^2 z^2 + 2 \bar{z}^3 \lambda^4 w^2 z^2 - 
 14 \bar{w} \bar{z}^3 \lambda^4 w^2 z^2 + 4 \bar{z}^4 \lambda^4 w^2 z^2 - 
 4 \bar{w}^2 \bar{z} \lambda^6 w^2 z^2 + 4 \bar{w} \bar{z}^2 \lambda^6 w^2 z^2 + 
 8 \bar{w}^2 \bar{z}^2 \lambda^6 w^2 z^2 - 12 \bar{w} \bar{z}^3 \lambda^6 w^2 z^2 + 
 4 \bar{z}^4 \lambda^6 w^2 z^2 + \bar{w}^2 \bar{z}^2 \lambda^8 w^2 z^2 - 2 \bar{w} \bar{z}^3 \lambda^8 w^2 z^2 + 
 \bar{z}^4 \lambda^8 w^2 z^2 + 2 \bar{w} \bar{z} \lambda^2 z^3 - 2 \bar{w}^2 \bar{z} \lambda^2 z^3 - 
 4 \bar{w} \bar{z}^2 \lambda^2 z^3 + 4 \bar{w}^2 \bar{z}^2 \lambda^2 z^3 + 2 \bar{z}^3 \lambda^2 z^3 - 
 2 \bar{w} \bar{z}^3 \lambda^2 z^3 - 2 \bar{w}^2 \lambda^4 z^3 + 2 \bar{w} \bar{z} \lambda^4 z^3 + 
 4 \bar{w}^2 \bar{z} \lambda^4 z^3 - 8 \bar{w} \bar{z}^2 \lambda^4 z^3 + 2 \bar{w}^2 \bar{z}^2 \lambda^4 z^3 + 
 4 \bar{z}^3 \lambda^4 z^3 - 2 \bar{w} \bar{z}^3 \lambda^4 z^3 + 2 \bar{w}^2 \bar{z} \lambda^6 z^3 - 
 4 \bar{w} \bar{z}^2 \lambda^6 z^3 + 2 \bar{z}^3 \lambda^6 z^3 - 2 \bar{w} \bar{z} \lambda^2 w z^3 + 
 2 \bar{w}^2 \bar{z} \lambda^2 w z^3 + 4 \bar{w} \bar{z}^2 \lambda^2 w z^3 - 4 \bar{w}^2 \bar{z}^2 \lambda^2 w z^3 - 
 2 \bar{z}^3 \lambda^2 w z^3 + 2 \bar{w} \bar{z}^3 \lambda^2 w z^3 - 4 \bar{w} \bar{z} \lambda^4 w z^3 + 
 8 \bar{w}^2 \bar{z} \lambda^4 w z^3 + 2 \bar{w} \bar{z}^2 \lambda^4 w z^3 - 14 \bar{w}^2 \bar{z}^2 \lambda^4 w z^3 - 
 2 \bar{z}^3 \lambda^4 w z^3 + 14 \bar{w} \bar{z}^3 \lambda^4 w z^3 - 4 \bar{z}^4 \lambda^4 w z^3 + 
 6 \bar{w}^2 \bar{z} \lambda^6 w z^3 - 6 \bar{w} \bar{z}^2 \lambda^6 w z^3 - 12 \bar{w}^2 \bar{z}^2 \lambda^6 w z^3 + 
 18 \bar{w} \bar{z}^3 \lambda^6 w z^3 - 6 \bar{z}^4 \lambda^6 w z^3 - 2 \bar{w}^2 \bar{z}^2 \lambda^8 w z^3 + 
 4 \bar{w} \bar{z}^3 \lambda^8 w z^3 - 2 \bar{z}^4 \lambda^8 w z^3 + \bar{w}^2 \lambda^4 z^4 - 
 4 \bar{w}^2 \bar{z} \lambda^4 z^4 + 2 \bar{w} \bar{z}^2 \lambda^4 z^4 + 4 \bar{w}^2 \bar{z}^2 \lambda^4 z^4 - 
 4 \bar{w} \bar{z}^3 \lambda^4 z^4 + \bar{z}^4 \lambda^4 z^4 - 2 \bar{w}^2 \bar{z} \lambda^6 z^4 + 
 2 \bar{w} \bar{z}^2 \lambda^6 z^4 + 4 \bar{w}^2 \bar{z}^2 \lambda^6 z^4 - 6 \bar{w} \bar{z}^3 \lambda^6 z^4 + 
 2 \bar{z}^4 \lambda^6 z^4 + \bar{w}^2 \bar{z}^2 \lambda^8 z^4 - 2 \bar{w} \bar{z}^3 \lambda^8 z^4 + 
 \bar{z}^4 \lambda^8 z^4, \\[4pt]
 \bar{z}^2 - 2 \bar{w} \bar{z}^2 + \bar{w}^2 \bar{z}^2 - 2 \bar{z}^2 w + 
 4 \bar{w} \bar{z}^2 w - 2 \bar{w}^2 \bar{z}^2 w + 2 \bar{w} \bar{z} \lambda^2 w - 2 \bar{w}^2 \bar{z} \lambda^2 w - 
 2 \bar{z}^2 \lambda^2 w + 2 \bar{w} \bar{z}^2 \lambda^2 w + \bar{z}^2 w^2 - 2 \bar{w} \bar{z}^2 w^2 + 
 \bar{w}^2 \bar{z}^2 w^2 - 2 \bar{w} \bar{z} \lambda^2 w^2 + 2 \bar{w}^2 \bar{z} \lambda^2 w^2 + 
 2 \bar{z}^2 \lambda^2 w^2 - 2 \bar{w} \bar{z}^2 \lambda^2 w^2 + \bar{w}^2 \lambda^4 w^2 - 2 \bar{w} \bar{z} \lambda^4 w^2 +
  \bar{z}^2 \lambda^4 w^2 - 2 \bar{z} z + 4 \bar{w} \bar{z} z - 2 \bar{w}^2 \bar{z} z - 2 \bar{w} \bar{z} \lambda^2 z + 
 2 \bar{w}^2 \bar{z} \lambda^2 z + 2 \bar{z}^2 \lambda^2 z - 2 \bar{w} \bar{z}^2 \lambda^2 z + 4 \bar{z} w z - 
 8 \bar{w} \bar{z} w z + 4 \bar{w}^2 \bar{z} w z + 2 \bar{w} \lambda^2 w z - 2 \bar{w}^2 \lambda^2 w z - 
 2 \bar{z} \lambda^2 w z - 4 \bar{w} \bar{z} \lambda^2 w z + 6 \bar{w}^2 \bar{z} \lambda^2 w z + 
 6 \bar{z}^2 \lambda^2 w z - 4 \bar{w} \bar{z}^2 \lambda^2 w z - 2 \bar{w}^2 \bar{z}^2 \lambda^2 w z - 
 2 \bar{z}^3 \lambda^2 w z + 2 \bar{w} \bar{z}^3 \lambda^2 w z - 2 \bar{w}^2 \lambda^4 w z + 
 8 \bar{w} \bar{z} \lambda^4 w z - 4 \bar{w}^2 \bar{z} \lambda^4 w z - 6 \bar{z}^2 \lambda^4 w z + 
 4 \bar{w}^2 \bar{z}^2 \lambda^4 w z + 4 \bar{z}^3 \lambda^4 w z - 4 \bar{w} \bar{z}^3 \lambda^4 w z - 
 2 \bar{z} w^2 z + 4 \bar{w} \bar{z} w^2 z - 2 \bar{w}^2 \bar{z} w^2 z - 2 \bar{w} \lambda^2 w^2 z + 
 2 \bar{w}^2 \lambda^2 w^2 z + 2 \bar{z} \lambda^2 w^2 z + 6 \bar{w} \bar{z} \lambda^2 w^2 z - 
 8 \bar{w}^2 \bar{z} \lambda^2 w^2 z - 8 \bar{z}^2 \lambda^2 w^2 z + 6 \bar{w} \bar{z}^2 \lambda^2 w^2 z + 
 2 \bar{w}^2 \bar{z}^2 \lambda^2 w^2 z + 2 \bar{z}^3 \lambda^2 w^2 z - 2 \bar{w} \bar{z}^3 \lambda^2 w^2 z - 
 4 \bar{w} \bar{z} \lambda^4 w^2 z + 2 \bar{w}^2 \bar{z} \lambda^4 w^2 z + 4 \bar{z}^2 \lambda^4 w^2 z + 
 4 \bar{w} \bar{z}^2 \lambda^4 w^2 z - 4 \bar{w}^2 \bar{z}^2 \lambda^4 w^2 z - 6 \bar{z}^3 \lambda^4 w^2 z + 
 4 \bar{w} \bar{z}^3 \lambda^4 w^2 z + 4 \bar{w}^2 \bar{z} \lambda^6 w^2 z - 8 \bar{w} \bar{z}^2 \lambda^6 w^2 z - 
 4 \bar{w}^2 \bar{z}^2 \lambda^6 w^2 z + 4 \bar{z}^3 \lambda^6 w^2 z + 8 \bar{w} \bar{z}^3 \lambda^6 w^2 z - 
 4 \bar{z}^4 \lambda^6 w^2 z + z^2 - 2 \bar{w} z^2 + \bar{w}^2 z^2 - 2 \bar{w} \lambda^2 z^2 + 
 2 \bar{w}^2 \lambda^2 z^2 + 2 \bar{z} \lambda^2 z^2 + 6 \bar{w} \bar{z} \lambda^2 z^2 - 
 8 \bar{w}^2 \bar{z} \lambda^2 z^2 - 8 \bar{z}^2 \lambda^2 z^2 + 6 \bar{w} \bar{z}^2 \lambda^2 z^2 + 
 2 \bar{w}^2 \bar{z}^2 \lambda^2 z^2 + 2 \bar{z}^3 \lambda^2 z^2 - 2 \bar{w} \bar{z}^3 \lambda^2 z^2 + 
 \bar{w}^2 \lambda^4 z^2 - 6 \bar{w} \bar{z} \lambda^4 z^2 + 4 \bar{w}^2 \bar{z} \lambda^4 z^2 + 
 5 \bar{z}^2 \lambda^4 z^2 - 4 \bar{w}^2 \bar{z}^2 \lambda^4 z^2 - 4 \bar{z}^3 \lambda^4 z^2 + 
 4 \bar{w} \bar{z}^3 \lambda^4 z^2 - 2 w z^2 + 4 \bar{w} w z^2 - 2 \bar{w}^2 w z^2 + 
 2 \bar{w} \lambda^2 w z^2 - 2 \bar{w}^2 \lambda^2 w z^2 - 2 \bar{z} \lambda^2 w z^2 - 
 4 \bar{w} \bar{z} \lambda^2 w z^2 + 6 \bar{w}^2 \bar{z} \lambda^2 w z^2 + 6 \bar{z}^2 \lambda^2 w z^2 - 
 4 \bar{w} \bar{z}^2 \lambda^2 w z^2 - 2 \bar{w}^2 \bar{z}^2 \lambda^2 w z^2 - 2 \bar{z}^3 \lambda^2 w z^2 + 
 2 \bar{w} \bar{z}^3 \lambda^2 w z^2 + 4 \bar{w}^2 \bar{z} \lambda^4 w z^2 - 8 \bar{w} \bar{z}^2 \lambda^4 w z^2 + 
 4 \bar{z}^3 \lambda^4 w z^2 - 8 \bar{w}^2 \bar{z} \lambda^6 w z^2 + 16 \bar{w} \bar{z}^2 \lambda^6 w z^2 + 
 8 \bar{w}^2 \bar{z}^2 \lambda^6 w z^2 - 8 \bar{z}^3 \lambda^6 w z^2 - 16 \bar{w} \bar{z}^3 \lambda^6 w z^2 + 
 8 \bar{z}^4 \lambda^6 w z^2 + w^2 z^2 - 2 \bar{w} w^2 z^2 + \bar{w}^2 w^2 z^2 - 
 2 \bar{w} \bar{z} \lambda^2 w^2 z^2 + 2 \bar{w}^2 \bar{z} \lambda^2 w^2 z^2 + 2 \bar{z}^2 \lambda^2 w^2 z^2 - 
 2 \bar{w} \bar{z}^2 \lambda^2 w^2 z^2 + 4 \bar{w} \bar{z} \lambda^4 w^2 z^2 - 
 4 \bar{w}^2 \bar{z} \lambda^4 w^2 z^2 - 4 \bar{z}^2 \lambda^4 w^2 z^2 + 
 5 \bar{w}^2 \bar{z}^2 \lambda^4 w^2 z^2 + 4 \bar{z}^3 \lambda^4 w^2 z^2 - 
 6 \bar{w} \bar{z}^3 \lambda^4 w^2 z^2 + \bar{z}^4 \lambda^4 w^2 z^2 - 4 \bar{w}^2 \bar{z} \lambda^6 w^2 z^2 + 
 8 \bar{w} \bar{z}^2 \lambda^6 w^2 z^2 + 4 \bar{w}^2 \bar{z}^2 \lambda^6 w^2 z^2 - 
 4 \bar{z}^3 \lambda^6 w^2 z^2 - 8 \bar{w} \bar{z}^3 \lambda^6 w^2 z^2 + 4 \bar{z}^4 \lambda^6 w^2 z^2 - 
 2 \bar{w} \bar{z} \lambda^2 z^3 + 2 \bar{w}^2 \bar{z} \lambda^2 z^3 + 2 \bar{z}^2 \lambda^2 z^3 - 
 2 \bar{w} \bar{z}^2 \lambda^2 z^3 + 4 \bar{w} \bar{z} \lambda^4 z^3 - 6 \bar{w}^2 \bar{z} \lambda^4 z^3 - 
 4 \bar{z}^2 \lambda^4 z^3 + 4 \bar{w} \bar{z}^2 \lambda^4 z^3 + 4 \bar{w}^2 \bar{z}^2 \lambda^4 z^3 + 
 2 \bar{z}^3 \lambda^4 z^3 - 4 \bar{w} \bar{z}^3 \lambda^4 z^3 + 4 \bar{w}^2 \bar{z} \lambda^6 z^3 - 
 8 \bar{w} \bar{z}^2 \lambda^6 z^3 - 4 \bar{w}^2 \bar{z}^2 \lambda^6 z^3 + 4 \bar{z}^3 \lambda^6 z^3 + 
 8 \bar{w} \bar{z}^3 \lambda^6 z^3 - 4 \bar{z}^4 \lambda^6 z^3 + 2 \bar{w} \bar{z} \lambda^2 w z^3 - 
 2 \bar{w}^2 \bar{z} \lambda^2 w z^3 - 2 \bar{z}^2 \lambda^2 w z^3 + 2 \bar{w} \bar{z}^2 \lambda^2 w z^3 - 
 4 \bar{w} \bar{z} \lambda^4 w z^3 + 4 \bar{w}^2 \bar{z} \lambda^4 w z^3 + 4 \bar{z}^2 \lambda^4 w z^3 - 
 6 \bar{w}^2 \bar{z}^2 \lambda^4 w z^3 - 4 \bar{z}^3 \lambda^4 w z^3 + 8 \bar{w} \bar{z}^3 \lambda^4 w z^3 - 
 2 \bar{z}^4 \lambda^4 w z^3 + 8 \bar{w}^2 \bar{z} \lambda^6 w z^3 - 16 \bar{w} \bar{z}^2 \lambda^6 w z^3 - 
 8 \bar{w}^2 \bar{z}^2 \lambda^6 w z^3 + 8 \bar{z}^3 \lambda^6 w z^3 + 16 \bar{w} \bar{z}^3 \lambda^6 w z^3 - 
 8 \bar{z}^4 \lambda^6 w z^3 + \bar{w}^2 \bar{z}^2 \lambda^4 z^4 - 2 \bar{w} \bar{z}^3 \lambda^4 z^4 + 
 \bar{z}^4 \lambda^4 z^4 - 4 \bar{w}^2 \bar{z} \lambda^6 z^4 + 8 \bar{w} \bar{z}^2 \lambda^6 z^4 + 
 4 \bar{w}^2 \bar{z}^2 \lambda^6 z^4 - 4 \bar{z}^3 \lambda^6 z^4 - 8 \bar{w} \bar{z}^3 \lambda^6 z^4 + 
 4 \bar{z}^4 \lambda^6 z^4$
 

There are a few integrals with 3 quadratic propagators (the $\delta$-function is also treated as a quadratic propagator) that do not yield results when run with SOFIA package,

\begin{equation*}
\frac{\delta(D_\delta)}{D_8D_{11}},\quad
\frac{\delta(D_\delta)}{D_1D_8D_{11}},\quad
\frac{\delta(D_\delta)}{D_2D_9D_{10}},\quad
\frac{\delta(D_\delta)}{D_5D_8D_{10}},\quad
\frac{\delta(D_\delta)}{D_8D_9},\quad
\frac{\delta(D_\delta)}{D_1D_8D_9}.
\end{equation*}
For these integrals, we leave the investigation for future work.

\bibliographystyle{JHEP}
\bibliography{projection}

\providecommand{\href}[2]{#2}\begingroup\raggedright\begin{thebibliography}{10}

\bibitem{Basham:1977iq}
C.~L. Basham, L.~S. Brown, S.~D. Ellis and S.~T. Love, \emph{{Electron - Positron Annihilation Energy Pattern in Quantum Chromodynamics: Asymptotically Free Perturbation Theory}}, \href{https://doi.org/10.1103/PhysRevD.17.2298}{\emph{Phys. Rev. D} {\bfseries 17} (1978) 2298}.

\bibitem{Basham:1978bw}
C.~L. Basham, L.~S. Brown, S.~D. Ellis and S.~T. Love, \emph{{Energy Correlations in electron - Positron Annihilation: Testing QCD}}, \href{https://doi.org/10.1103/PhysRevLett.41.1585}{\emph{Phys. Rev. Lett.} {\bfseries 41} (1978) 1585}.

\bibitem{Basham:1978zq}
C.~L. Basham, L.~S. Brown, S.~D. Ellis and S.~T. Love, \emph{{Energy Correlations in electron-Positron Annihilation in Quantum Chromodynamics: Asymptotically Free Perturbation Theory}}, \href{https://doi.org/10.1103/PhysRevD.19.2018}{\emph{Phys. Rev. D} {\bfseries 19} (1979) 2018}.

\bibitem{Basham:1979gh}
C.~L. Basham, L.~S. Brown, S.~D. Ellis and S.~T. Love, \emph{{Energy Correlations in Perturbative Quantum Chromodynamics: A Conjecture for All Orders}}, \href{https://doi.org/10.1016/0370-2693(79)90601-4}{\emph{Phys. Lett. B} {\bfseries 85} (1979) 297}.

\bibitem{Lee:1964is}
T.~D. Lee and M.~Nauenberg, \emph{{Degenerate Systems and Mass Singularities}}, \href{https://doi.org/10.1103/PhysRev.133.B1549}{\emph{Phys. Rev.} {\bfseries 133} (1964) B1549}.

\bibitem{Hofman:2008ar}
D.~M. Hofman and J.~Maldacena, \emph{{Conformal collider physics: Energy and charge correlations}}, \href{https://doi.org/10.1088/1126-6708/2008/05/012}{\emph{JHEP} {\bfseries 05} (2008) 012} [\href{https://arxiv.org/abs/0803.1467}{{\ttfamily 0803.1467}}].

\bibitem{Sveshnikov:1995vi}
N.~A. Sveshnikov and F.~V. Tkachov, \emph{{Jets and quantum field theory}}, \href{https://doi.org/10.1016/0370-2693(96)00558-8}{\emph{Phys. Lett. B} {\bfseries 382} (1996) 403} [\href{https://arxiv.org/abs/hep-ph/9512370}{{\ttfamily hep-ph/9512370}}].

\bibitem{Korchemsky:1999kt}
G.~P. Korchemsky and G.~F. Sterman, \emph{{Power corrections to event shapes and factorization}}, \href{https://doi.org/10.1016/S0550-3213(99)00308-9}{\emph{Nucl. Phys. B} {\bfseries 555} (1999) 335} [\href{https://arxiv.org/abs/hep-ph/9902341}{{\ttfamily hep-ph/9902341}}].

\bibitem{Lee:2006nr}
C.~Lee and G.~F. Sterman, \emph{{Momentum Flow Correlations from Event Shapes: Factorized Soft Gluons and Soft-Collinear Effective Theory}}, \href{https://doi.org/10.1103/PhysRevD.75.014022}{\emph{Phys. Rev. D} {\bfseries 75} (2007) 014022} [\href{https://arxiv.org/abs/hep-ph/0611061}{{\ttfamily hep-ph/0611061}}].

\bibitem{Belitsky:2013bja}
A.~V. Belitsky, S.~Hohenegger, G.~P. Korchemsky, E.~Sokatchev and A.~Zhiboedov, \emph{{Event shapes in $\mathcal{N} = 4$ super-Yang-Mills theory}}, \href{https://doi.org/10.1016/j.nuclphysb.2014.04.019}{\emph{Nucl. Phys. B} {\bfseries 884} (2014) 206} [\href{https://arxiv.org/abs/1309.1424}{{\ttfamily 1309.1424}}].

\bibitem{Belitsky:2013xxa}
A.~V. Belitsky, S.~Hohenegger, G.~P. Korchemsky, E.~Sokatchev and A.~Zhiboedov, \emph{{From correlation functions to event shapes}}, \href{https://doi.org/10.1016/j.nuclphysb.2014.04.020}{\emph{Nucl. Phys. B} {\bfseries 884} (2014) 305} [\href{https://arxiv.org/abs/1309.0769}{{\ttfamily 1309.0769}}].

\bibitem{Belitsky:2013ofa}
A.~V. Belitsky, S.~Hohenegger, G.~P. Korchemsky, E.~Sokatchev and A.~Zhiboedov, \emph{{Energy-Energy Correlations in N=4 Supersymmetric Yang-Mills Theory}}, \href{https://doi.org/10.1103/PhysRevLett.112.071601}{\emph{Phys. Rev. Lett.} {\bfseries 112} (2014) 071601} [\href{https://arxiv.org/abs/1311.6800}{{\ttfamily 1311.6800}}].

\bibitem{Dixon:2018qgp}
L.~J. Dixon, M.-X. Luo, V.~Shtabovenko, T.-Z. Yang and H.~X. Zhu, \emph{{Analytical Computation of Energy-Energy Correlation at Next-to-Leading Order in QCD}}, \href{https://doi.org/10.1103/PhysRevLett.120.102001}{\emph{Phys. Rev. Lett.} {\bfseries 120} (2018) 102001} [\href{https://arxiv.org/abs/1801.03219}{{\ttfamily 1801.03219}}].

\bibitem{Luo:2019nig}
M.-X. Luo, V.~Shtabovenko, T.-Z. Yang and H.~X. Zhu, \emph{{Analytic Next-To-Leading Order Calculation of Energy-Energy Correlation in Gluon-Initiated Higgs Decays}}, \href{https://doi.org/10.1007/JHEP06(2019)037}{\emph{JHEP} {\bfseries 06} (2019) 037} [\href{https://arxiv.org/abs/1903.07277}{{\ttfamily 1903.07277}}].

\bibitem{Henn:2019gkr}
J.~M. Henn, E.~Sokatchev, K.~Yan and A.~Zhiboedov, \emph{{Energy-energy correlation in $N$=4 super Yang-Mills theory at next-to-next-to-leading order}}, \href{https://doi.org/10.1103/PhysRevD.100.036010}{\emph{Phys. Rev. D} {\bfseries 100} (2019) 036010} [\href{https://arxiv.org/abs/1903.05314}{{\ttfamily 1903.05314}}].

\bibitem{Yan:2022cye}
K.~Yan and X.~Zhang, \emph{{Three-Point Energy Correlator in N=4 Supersymmetric Yang-Mills Theory}}, \href{https://doi.org/10.1103/PhysRevLett.129.021602}{\emph{Phys. Rev. Lett.} {\bfseries 129} (2022) 021602} [\href{https://arxiv.org/abs/2203.04349}{{\ttfamily 2203.04349}}].

\bibitem{chen2023nnllresummationprojectedthreepoint}
W.~Chen, J.~Gao, Y.~Li, Z.~Xu, X.~Zhang and H.~X. Zhu, \emph{Nnll resummation for projected three-point energy correlator},  2023.

\bibitem{Chicherin:2024ifn}
D.~Chicherin, I.~Moult, E.~Sokatchev, K.~Yan and Y.~Zhu, \emph{{The Collinear Limit of the Four-Point Energy Correlator in $\mathcal{N} = 4$ Super Yang-Mills Theory}},  \href{https://arxiv.org/abs/2401.06463}{{\ttfamily 2401.06463}}.

\bibitem{Kotikov:1990kg}
A.~V. Kotikov, \emph{{Differential equations method: New technique for massive Feynman diagrams calculation}}, \href{https://doi.org/10.1016/0370-2693(91)90413-K}{\emph{Phys. Lett. B} {\bfseries 254} (1991) 158}.

\bibitem{Henn_2013}
J.~M. Henn, \emph{Multiloop integrals in dimensional regularization made simple}, \href{https://doi.org/10.1103/physrevlett.110.251601}{\emph{Physical Review Letters} {\bfseries 110} (2013) }.

\bibitem{CHETYRKIN1981159}
K.~Chetyrkin and F.~Tkachov, \emph{Integration by parts: The algorithm to calculate $\beta$-functions in 4 loops}, \href{https://doi.org/https://doi.org/10.1016/0550-3213(81)90199-1}{\emph{Nuclear Physics B} {\bfseries 192} (1981) 159}.

\bibitem{Gert_2008}
G.~P. Gert-Martin~Greuel, \emph{A Singular Introduction to Commutative Algebra}. Springer Berlin, Heidelberg, 2008.

\bibitem{Gluza_2011}
J.~Gluza, K.~Kajda and D.~A. Kosower, \emph{Towards a basis for planar two-loop integrals}, \href{https://doi.org/10.1103/physrevd.83.045012}{\emph{Physical Review D} {\bfseries 83} (2011) }.

\bibitem{Larsen_2018}
K.~Larsen, J.~Bosma and Y.~Zhang, \emph{Differential equations for loop integrals without squared propagators},  in \emph{Proceedings of Loops and Legs in Quantum Field Theory — PoS(LL2018)}, LL2018, p.~064, Sissa Medialab, Oct., 2018, \href{https://doi.org/10.22323/1.303.0064}{DOI}.

\bibitem{Wu:2025aeg}
Z.~Wu, J.~B\"ohm, R.~Ma, J.~Usovitsch, Y.~Xu and Y.~Zhang, \emph{{Performing integration-by-parts reductions using NeatIBP 1.1 + Kira}},  \href{https://arxiv.org/abs/2502.20778}{{\ttfamily 2502.20778}}.

\bibitem{DGPS}
W.~Decker, G.-M. Greuel, G.~Pfister and H.~Sch\"onemann, ``{\sc Singular} {4-4-0} --- {A} computer algebra system for polynomial computations.'' \url{http://www.singular.uni-kl.de}, 2024.

\bibitem{Henn:2022vqp}
J.~Henn, R.~Ma, K.~Yan and Y.~Zhang, \emph{{Four-dimensional differential equations for the leading divergences of dimensionally-regulated loop integrals}}, \href{https://doi.org/10.1007/JHEP03(2023)162}{\emph{JHEP} {\bfseries 03} (2023) 162} [\href{https://arxiv.org/abs/2211.13967}{{\ttfamily 2211.13967}}].

\bibitem{Wu:2023upw}
Z.~Wu, J.~Boehm, R.~Ma, H.~Xu and Y.~Zhang, \emph{{NeatIBP 1.0, a package generating small-size integration-by-parts relations for Feynman integrals}}, \href{https://doi.org/10.1016/j.cpc.2023.108999}{\emph{Comput. Phys. Commun.} {\bfseries 295} (2024) 108999} [\href{https://arxiv.org/abs/2305.08783}{{\ttfamily 2305.08783}}].

\bibitem{Muller-Stach:2012tgj}
S.~M\"uller-Stach, S.~Weinzierl and R.~Zayadeh, \emph{{Picard-Fuchs equations for Feynman integrals}}, \href{https://doi.org/10.1007/s00220-013-1838-3}{\emph{Commun. Math. Phys.} {\bfseries 326} (2014) 237} [\href{https://arxiv.org/abs/1212.4389}{{\ttfamily 1212.4389}}].

\bibitem{Adams:2017tga}
L.~Adams, E.~Chaubey and S.~Weinzierl, \emph{{Simplifying Differential Equations for Multiscale Feynman Integrals beyond Multiple Polylogarithms}}, \href{https://doi.org/10.1103/PhysRevLett.118.141602}{\emph{Phys. Rev. Lett.} {\bfseries 118} (2017) 141602} [\href{https://arxiv.org/abs/1702.04279}{{\ttfamily 1702.04279}}].

\bibitem{Dlapa:2020cwj}
C.~Dlapa, J.~Henn and K.~Yan, \emph{{Deriving canonical differential equations for Feynman integrals from a single uniform weight integral}}, \href{https://doi.org/10.1007/JHEP05(2020)025}{\emph{JHEP} {\bfseries 05} (2020) 025} [\href{https://arxiv.org/abs/2002.02340}{{\ttfamily 2002.02340}}].

\bibitem{Dlapa:2022wdu}
C.~Dlapa, J.~M. Henn and F.~J. Wagner, \emph{{An algorithmic approach to finding canonical differential equations for elliptic Feynman integrals}}, \href{https://doi.org/10.1007/JHEP08(2023)120}{\emph{JHEP} {\bfseries 08} (2023) 120} [\href{https://arxiv.org/abs/2211.16357}{{\ttfamily 2211.16357}}].

\bibitem{Bosma:2017ens}
J.~Bosma, M.~Sogaard and Y.~Zhang, \emph{{Maximal Cuts in Arbitrary Dimension}}, \href{https://doi.org/10.1007/JHEP08(2017)051}{\emph{JHEP} {\bfseries 08} (2017) 051} [\href{https://arxiv.org/abs/1704.04255}{{\ttfamily 1704.04255}}].

\bibitem{Primo:2016ebd}
A.~Primo and L.~Tancredi, \emph{{On the maximal cut of Feynman integrals and the solution of their differential equations}}, \href{https://doi.org/10.1016/j.nuclphysb.2016.12.021}{\emph{Nucl. Phys. B} {\bfseries 916} (2017) 94} [\href{https://arxiv.org/abs/1610.08397}{{\ttfamily 1610.08397}}].

\bibitem{Primo:2017ipr}
A.~Primo and L.~Tancredi, \emph{{Maximal cuts and differential equations for Feynman integrals. An application to the three-loop massive banana graph}}, \href{https://doi.org/10.1016/j.nuclphysb.2017.05.018}{\emph{Nucl. Phys. B} {\bfseries 921} (2017) 316} [\href{https://arxiv.org/abs/1704.05465}{{\ttfamily 1704.05465}}].

\bibitem{Henn:2020lye}
J.~Henn, B.~Mistlberger, V.~A. Smirnov and P.~Wasser, \emph{{Constructing d-log integrands and computing master integrals for three-loop four-particle scattering}}, \href{https://doi.org/10.1007/JHEP04(2020)167}{\emph{JHEP} {\bfseries 04} (2020) 167} [\href{https://arxiv.org/abs/2002.09492}{{\ttfamily 2002.09492}}].

\bibitem{Bourjaily:2021vyj}
J.~L. Bourjaily, N.~Kalyanapuram, C.~Langer and K.~Patatoukos, \emph{{Prescriptive unitarity with elliptic leading singularities}}, \href{https://doi.org/10.1103/PhysRevD.104.125009}{\emph{Phys. Rev. D} {\bfseries 104} (2021) 125009} [\href{https://arxiv.org/abs/2102.02210}{{\ttfamily 2102.02210}}].

\bibitem{Frellesvig:2021hkr}
H.~Frellesvig, \emph{{On epsilon factorized differential equations for elliptic Feynman integrals}}, \href{https://doi.org/10.1007/JHEP03(2022)079}{\emph{JHEP} {\bfseries 03} (2022) 079} [\href{https://arxiv.org/abs/2110.07968}{{\ttfamily 2110.07968}}].

\bibitem{Broedel:2017kkb}
J.~Broedel, C.~Duhr, F.~Dulat and L.~Tancredi, \emph{{Elliptic polylogarithms and iterated integrals on elliptic curves. Part I: general formalism}}, \href{https://doi.org/10.1007/JHEP05(2018)093}{\emph{JHEP} {\bfseries 05} (2018) 093} [\href{https://arxiv.org/abs/1712.07089}{{\ttfamily 1712.07089}}].

\bibitem{Broedel:2018qkq}
J.~Broedel, C.~Duhr, F.~Dulat, B.~Penante and L.~Tancredi, \emph{{Elliptic Feynman integrals and pure functions}}, \href{https://doi.org/10.1007/JHEP01(2019)023}{\emph{JHEP} {\bfseries 01} (2019) 023} [\href{https://arxiv.org/abs/1809.10698}{{\ttfamily 1809.10698}}].

\bibitem{Bourjaily:2017bsb}
J.~L. Bourjaily, A.~J. McLeod, M.~Spradlin, M.~von Hippel and M.~Wilhelm, \emph{{Elliptic Double-Box Integrals: Massless Scattering Amplitudes beyond Polylogarithms}}, \href{https://doi.org/10.1103/PhysRevLett.120.121603}{\emph{Phys. Rev. Lett.} {\bfseries 120} (2018) 121603} [\href{https://arxiv.org/abs/1712.02785}{{\ttfamily 1712.02785}}].

\bibitem{Broedel:2019hyg}
J.~Broedel, C.~Duhr, F.~Dulat, B.~Penante and L.~Tancredi, \emph{{Elliptic polylogarithms and Feynman parameter integrals}}, \href{https://doi.org/10.1007/JHEP05(2019)120}{\emph{JHEP} {\bfseries 05} (2019) 120} [\href{https://arxiv.org/abs/1902.09971}{{\ttfamily 1902.09971}}].

\bibitem{Huang:2013kh}
R.~Huang and Y.~Zhang, \emph{{On Genera of Curves from High-loop Generalized Unitarity Cuts}}, \href{https://doi.org/10.1007/JHEP04(2013)080}{\emph{JHEP} {\bfseries 04} (2013) 080} [\href{https://arxiv.org/abs/1302.1023}{{\ttfamily 1302.1023}}].

\bibitem{Georgoudis:2015hca}
A.~Georgoudis and Y.~Zhang, \emph{{Two-loop Integral Reduction from Elliptic and Hyperelliptic Curves}}, \href{https://doi.org/10.1007/JHEP12(2015)086}{\emph{JHEP} {\bfseries 12} (2015) 086} [\href{https://arxiv.org/abs/1507.06310}{{\ttfamily 1507.06310}}].

\bibitem{Duhr:2011zq}
C.~Duhr, H.~Gangl and J.~R. Rhodes, \emph{{From polygons and symbols to polylogarithmic functions}}, \href{https://doi.org/10.1007/JHEP10(2012)075}{\emph{JHEP} {\bfseries 10} (2012) 075} [\href{https://arxiv.org/abs/1110.0458}{{\ttfamily 1110.0458}}].

\bibitem{Duhr:2019tlz}
C.~Duhr and F.~Dulat, \emph{{PolyLogTools \textemdash{} polylogs for the masses}}, \href{https://doi.org/10.1007/JHEP08(2019)135}{\emph{JHEP} {\bfseries 08} (2019) 135} [\href{https://arxiv.org/abs/1904.07279}{{\ttfamily 1904.07279}}].

\bibitem{Goncharov:2010jf}
A.~B. Goncharov, M.~Spradlin, C.~Vergu and A.~Volovich, \emph{{Classical Polylogarithms for Amplitudes and Wilson Loops}}, \href{https://doi.org/10.1103/PhysRevLett.105.151605}{\emph{Phys. Rev. Lett.} {\bfseries 105} (2010) 151605} [\href{https://arxiv.org/abs/1006.5703}{{\ttfamily 1006.5703}}].

\bibitem{Correia:2025yao}
M.~Correia, M.~Giroux and S.~Mizera, \emph{{SOFIA: Singularities of Feynman Integrals Automatized}},  \href{https://arxiv.org/abs/2503.16601}{{\ttfamily 2503.16601}}.

\bibitem{Panzer:2014caa}
E.~Panzer, \emph{{Algorithms for the symbolic integration of hyperlogarithms with applications to Feynman integrals}}, \href{https://doi.org/10.1016/j.cpc.2014.10.019}{\emph{Comput. Phys. Commun.} {\bfseries 188} (2015) 148} [\href{https://arxiv.org/abs/1403.3385}{{\ttfamily 1403.3385}}].

\bibitem{Bork:2010wf}
L.~V. Bork, D.~I. Kazakov and G.~S. Vartanov, \emph{{On form factors in N=4 sym}}, \href{https://doi.org/10.1007/JHEP02(2011)063}{\emph{JHEP} {\bfseries 02} (2011) 063} [\href{https://arxiv.org/abs/1011.2440}{{\ttfamily 1011.2440}}].

\bibitem{Bork:2011cj}
L.~V. Bork, D.~I. Kazakov and G.~S. Vartanov, \emph{{On MHV Form Factors in Superspace for N=4 SYM Theory}}, \href{https://doi.org/10.1007/JHEP10(2011)133}{\emph{JHEP} {\bfseries 10} (2011) 133} [\href{https://arxiv.org/abs/1107.5551}{{\ttfamily 1107.5551}}].

\bibitem{Penante:2014sza}
B.~Penante, B.~Spence, G.~Travaglini and C.~Wen, \emph{{On super form factors of half-BPS operators in N=4 super Yang-Mills}}, \href{https://doi.org/10.1007/JHEP04(2014)083}{\emph{JHEP} {\bfseries 04} (2014) 083} [\href{https://arxiv.org/abs/1402.1300}{{\ttfamily 1402.1300}}].

\bibitem{Brandhuber:2011tv}
A.~Brandhuber, O.~Gurdogan, R.~Mooney, G.~Travaglini and G.~Yang, \emph{{Harmony of Super Form Factors}}, \href{https://doi.org/10.1007/JHEP10(2011)046}{\emph{JHEP} {\bfseries 10} (2011) 046} [\href{https://arxiv.org/abs/1107.5067}{{\ttfamily 1107.5067}}].

\bibitem{vanNeerven:1985ja}
W.~L. van Neerven, \emph{{Infrared behaviour of on shell form factors in anN=4 supersymmetric Yang-Mills field theory}}, \href{https://doi.org/10.1007/BF01571808}{\emph{Zeitschrift f{\"u}r Physik C Particles and Fields} {\bfseries 30} (1986) 595}.

\bibitem{Boels:2012ew}
R.~H. Boels, B.~A. Kniehl, O.~V. Tarasov and G.~Yang, \emph{{Color-kinematic Duality for Form Factors}}, \href{https://doi.org/10.1007/JHEP02(2013)063}{\emph{JHEP} {\bfseries 02} (2013) 063} [\href{https://arxiv.org/abs/1211.7028}{{\ttfamily 1211.7028}}].

\bibitem{Yang:2016ear}
G.~Yang, \emph{{Color-kinematics duality and Sudakov form factor at five loops for N=4 supersymmetric Yang-Mills theory}}, \href{https://doi.org/10.1103/PhysRevLett.117.271602}{\emph{Phys. Rev. Lett.} {\bfseries 117} (2016) 271602} [\href{https://arxiv.org/abs/1610.02394}{{\ttfamily 1610.02394}}].

\bibitem{Heller:2021qkz}
M.~Heller and A.~von Manteuffel, \emph{{MultivariateApart: Generalized partial fractions}}, \href{https://doi.org/10.1016/j.cpc.2021.108174}{\emph{Comput. Phys. Commun.} {\bfseries 271} (2022) 108174} [\href{https://arxiv.org/abs/2101.08283}{{\ttfamily 2101.08283}}].

\bibitem{Bendle:2021ueg}
D.~Bendle, J.~Boehm, M.~Heymann, R.~Ma, M.~Rahn, L.~Ristau et~al., \emph{{pfd-parallel, a Singular/GPI-Space package for massively parallel multivariate partial fractioning}}, \href{https://doi.org/10.1016/j.cpc.2023.108942}{\emph{Comput. Phys. Commun.} {\bfseries 294} (2024) 108942} [\href{https://arxiv.org/abs/2104.06866}{{\ttfamily 2104.06866}}].

\bibitem{Schabinger_2012}
R.~M. Schabinger, \emph{A new algorithm for the generation of unitarity-compatible integration by parts relations}, \href{https://doi.org/10.1007/jhep01(2012)077}{\emph{Journal of High Energy Physics} {\bfseries 2012} (2012) }.

\bibitem{B_hm_2018}
J.~Böhm, A.~Georgoudis, K.~J. Larsen, M.~Schulze and Y.~Zhang, \emph{Complete sets of logarithmic vector fields for integration-by-parts identities of feynman integrals}, \href{https://doi.org/10.1103/physrevd.98.025023}{\emph{Physical Review D} {\bfseries 98} (2018) }.

\bibitem{Peraro_2019}
T.~Peraro, \emph{Finiteflow: multivariate functional reconstruction using finite fields and dataflow graphs}, \href{https://doi.org/10.1007/jhep07(2019)031}{\emph{Journal of High Energy Physics} {\bfseries 2019} (2019) }.

\bibitem{He:2024hbb}
S.~He, X.~Jiang, Q.~Yang and Y.-Q. Zhang, \emph{{From squared amplitudes to energy correlators}},  \href{https://arxiv.org/abs/2408.04222}{{\ttfamily 2408.04222}}.

\bibitem{Gong:2022erh}
J.~Gong and E.~Y. Yuan, \emph{{Towards analytic structure of Feynman parameter integrals with rational curves}}, \href{https://doi.org/10.1007/JHEP10(2022)145}{\emph{JHEP} {\bfseries 10} (2022) 145} [\href{https://arxiv.org/abs/2206.06507}{{\ttfamily 2206.06507}}].

\bibitem{Dennen:2015bet}
T.~Dennen, M.~Spradlin and A.~Volovich, \emph{{Landau Singularities and Symbology: One- and Two-loop MHV Amplitudes in SYM Theory}}, \href{https://doi.org/10.1007/JHEP03(2016)069}{\emph{JHEP} {\bfseries 03} (2016) 069} [\href{https://arxiv.org/abs/1512.07909}{{\ttfamily 1512.07909}}].

\bibitem{Caron-Huot:2024brh}
S.~Caron-Huot, M.~Correia and M.~Giroux, \emph{{Recursive Landau Analysis}},  \href{https://arxiv.org/abs/2406.05241}{{\ttfamily 2406.05241}}.

\bibitem{Fevola:2023kaw}
C.~Fevola, S.~Mizera and S.~Telen, \emph{{Landau Singularities Revisited: Computational Algebraic Geometry for Feynman Integrals}}, \href{https://doi.org/10.1103/PhysRevLett.132.101601}{\emph{Phys. Rev. Lett.} {\bfseries 132} (2024) 101601} [\href{https://arxiv.org/abs/2311.14669}{{\ttfamily 2311.14669}}].

\bibitem{Zhenjie:2025}
Z.~Li, L.~Ren and C.~Zhang, \emph{to appear}, {\emph{in progress} (2025) }.

\end{thebibliography}\endgroup

\end{document}